\documentclass[11pt]{article}
\input psfig.sty

\usepackage{graphics}
\usepackage{cite} % collapse adjacent citations

\psdraft
\psfigurepath{/home/httpd/html/mrenna/lc/}

\newenvironment{Eqnarray}%
     {\arraycolsep 0.14em\begin{eqnarray}}{\end{eqnarray}}

\makeatletter
\@addtoreset{equation}{section}

\def\BR{{\rm BR}}
\def\msusy{M_{\rm SUSY}}
\makeatother

\def\nicefrac#1#2{\hbox{${#1\over #2}$}}
\def\half{\nicefrac{1}{2}}
\def\hsm{h_{\rm SM}}

\newcommand{\be}{\begin{equation}}
\newcommand{\ee}{\end{equation}}
\newcommand{\bea}{\begin{Eqnarray}}
\newcommand{\eea}{\end{Eqnarray}}

\def\lsim{\mathrel{\raise.3ex\hbox{$<$\kern-.75em\lower1ex\hbox{$\sim$}}}}
\def\gsim{\mathrel{\raise.3ex\hbox{$>$\kern-.75em\lower1ex\hbox{$\sim$}}}}
\def\ifmath#1{\relax\ifmmode #1\else $#1$\fi}
\def\ls#1{\ifmath{_{\lower1.5pt\hbox{$\scriptstyle #1$}}}}

\begin{document}

\pagestyle{empty}
\begin{flushright}
FERMILAB-Pub-00/334-T  \\
SCIPP-01/25 \\
UCD-01-23 \\
hep-ph/0106116 \\
\end{flushright}

\renewcommand{\thefootnote}{\fnsymbol{footnote}}
\begin{center}
{\Large\bf Distinguishing a MSSM Higgs Boson from the SM Higgs Boson 
at a Linear Collider
}\\[1cm]
{\large Marcela Carena$^a$, Howard E. Haber$^b$,
Heather E. Logan$^a$ and Stephen Mrenna$^c$~\footnote{Electronic
addresses:
carena@fnal.gov, haber@scipp.ucsc.edu,
logan@fnal.gov, mrenna@physics.ucdavis.edu}
}\\[6pt]
{\it
$^a$ Theoretical Physics Department \\
   Fermi National Accelerator Laboratory, Batavia, IL 60510-0500, USA.\\
$^b$ Santa Cruz Institute for Particle Physics  \\
   University of California, Santa Cruz, CA 95064, USA. \\
$^c$ Davis Institute for High Energy Physics\\
   University of California, Davis, CA 95616, USA.}\\[.5cm]

{\bf Abstract}
\end{center}

\noindent

The decoupling properties of the Higgs sector in the Minimal
Supersymmetric Standard Model (MSSM) imply that a light CP-even Higgs
boson discovered at the Tevatron or LHC may closely resemble the
Standard Model (SM) Higgs boson.  In this paper, we investigate how
precision measurements of Higgs properties at a Linear Collider (LC)
can distinguish between a CP-even Higgs boson of the MSSM and the SM
Higgs boson.  We review the expected theoretical behavior of the
partial widths and branching ratios for decays of the neutral MSSM
Higgs bosons with significant couplings to the $W$ and $Z$ bosons,
including the leading radiative corrections to the mixing angle
$\alpha$ and $\tan\beta$-enhanced vertex corrections.  The general
expectation is that the Higgs couplings to $W^+W^-$, $ZZ$, $c\bar c$
and $t\bar t$ should quickly approach their SM values for increasing
CP-odd Higgs mass $m_A$, while the couplings to $b\bar b$ and
$\tau^+\tau^-$ do so more slowly.  Using the expected experimental and
theoretical accuracy in determining SM branching ratios and partial
widths, we demonstrate the sensitivity of measurements at the LC to
variations in the MSSM parameters, with particular attention to the
decoupling limit.  For a wide range of MSSM parameters, the LC is
sensitive to $m_A\sim 600$ GeV almost independently of $\tan\beta$.
For large values of $\tan\beta$ and some specific choices of MSSM
parameters [{\it e.g.}, $A_t\mu<0$ and $|A_t|\simeq|\mu|\simeq
\mathcal{O}(M_S)$], one of the CP-even Higgs bosons can be SM-like
independent of the value of $m_A$.  In the case of large deviations
from the SM, we present a procedure using Higgs coupling measurements
to extract the supersymmetric correction to the relation between the
$b$ quark mass and Yukawa coupling.

\vfill
\clearpage
%%%%%%%%%%%%%%%%%%%%%%%%%%%%%%%%%%%%%%%%%%%%%%%%%%%%%%%%%%%%%%%%%%%%%%%%

\renewcommand{\thefootnote}{\arabic{footnote}}
\setcounter{footnote}{0}

%%%%%%%%%%%%%%%%%%%%%%%%%%%%%%%%%%%%%%%%%%%%%%%%%%%%%%%%%%%%%%%%%%%%%%%%%%%%
\pagestyle{plain}

\section{Introduction}

The radiative corrections to
Higgs boson masses and couplings in the
minimal supersymmetric extension of the Standard Model (MSSM)
have been investigated thoroughly using different theoretical
approaches.  Derived quantities such as
Higgs boson production cross sections, partial widths and branching
ratios (BRs) are predicted to a high level of precision 
for any given set of MSSM parameters.
While at least one of the neutral Higgs bosons of the MSSM has
a coupling to $W$ and $Z$ bosons similar in magnitude to a 
Standard Model (SM) Higgs boson,
some of its properties can differ from those of the SM
Higgs boson of the same mass.  Nevertheless, over a significant
region of parameter space, the deviation of the couplings of 
this SM-like\footnote{In this paper, a
SM-like Higgs boson always refers to the neutral Higgs boson of the
MSSM with $g^2_{h_iVV}\geq 0.5 g^2_{h_{\rm SM}VV}$, 
where $V=W$ or $Z$.  
%(Recall that in multi-Higgs doublet models,
%$\sum_i g_{VVh_i}^2= [g^{\rm SM}_{VVh_i}]^2$.)   
The decay
properties of this Higgs boson may be quite different than those of a
SM Higgs boson in the parameter regime away from the decoupling limit.} 
Higgs boson from the corresponding couplings of the SM Higgs boson is
small and approaches zero in the so-called ``decoupling limit'' of
the model \cite{decoupling}.

Experiments at the Fermilab Tevatron \cite{HiggsWGrep}
and CERN LHC \cite{atlas_tdr} will be sensitive to the Higgs 
bosons of the SM and MSSM.  The Tevatron can discover a SM-like
Higgs boson for most choices of MSSM parameters if enough
data can be accumulated, the detectors perform as expected,
and systematic errors are demonstrably small.
The LHC can discover at least
one MSSM Higgs boson over all
of the MSSM parameter space \cite{atlas_tdr}, and  several Higgs bosons
are likely to be discovered in a significant region of the parameter space.
For moderate values of the ratio, $\tan\beta$, of the two Higgs vacuum
expectation values,\footnote{The absence of a Higgs boson
discovery at LEP implies that the range $0.5<\tan\beta<2.4$ is
excluded at 95\% confidence level \cite{LEPHIGGSgroup}.}
in the range $2.4\lsim \tan\beta\lsim 8$ [ $2.4\lsim \tan\beta\lsim 17$]
for a CP-odd Higgs mass of $m_A = 250$~GeV [$m_A=500$~GeV],
only one SM-like Higgs boson will be visible 
at the LHC \cite{atlas_tdr,Gianottiplot}.
If more than one
Higgs boson is observed at the hadron colliders,
then additional precision measurements of Higgs decay properties can
determine if the Higgs bosons originate from a 
two-Higgs doublet sector of a supersymmetric (SUSY) 
model (as in the MSSM) or from 
a different model.  If only
one Higgs boson is observed, such precision measurements could indicate whether
any additional Higgs structure exists.
Of course, the MSSM also contains supersymmetric particles, which can be
discovered at the Tevatron and/or the LHC.
However, these particles can be heavy,
so that only a part of the spectrum may ultimately
be observable at the LHC, or the
interpretation of the data as SUSY particle production may be ambiguous.
Hence, precision measurements of the properties of the Higgs
sector can provide crucial supporting evidence for the MSSM.
In this study, we explore the potential of 
a future $e^+ e^-$ linear collider (LC) to explore the MSSM Higgs sector
in regions
of MSSM parameter space with very different behaviors.

If a Higgs boson couples to $Z$ bosons with SM-like strength, then its
mass can be determined to high precision at the LC through measurements of
the recoil mass spectrum against a $Z$ boson.  With 500~fb$^{-1}$ of data
at $\sqrt{s} = 350$ GeV, a precision of $40$--$90$ MeV can be achieved for
Higgs masses between 120 and 180 GeV
\cite{Garcia-Abia,BattagliaDesch}.
%If a Higgs boson couples to $Z$ bosons, then with sufficient data
%its mass can be determined
%to within 50 MeV at the LC 
%through measurements of
%the recoil mass spectrum against a $Z$
%boson~\cite{Garcia-Abia,BattagliaDesch}.\footnote{Based on the 
%requirement of at least 50 signal events at the LC for 500 fb$^{-1}$
%of integrated luminosity
%at $\sqrt{s}=350$ GeV, a Higgs boson with $m_h=120$ GeV
%can be discovered if the coupling $g_{hZZ}\simeq 0.1g_{h_{\rm
%SM} ZZ}$,
%while a Higgs boson with SM-like couplings to $ZZ$
%is observable if $m_h\lsim 260$~GeV.
%The quoted accuracy for the mass measurement is for the
%SM Higgs boson.}
%
With the mass of the Higgs boson so constrained, the theoretical
predictions for many of the Higgs branching ratios and partial widths
({\it i.e.}, the Higgs couplings) will be known to great accuracy.
As a result, precision experimental measurements of Higgs branching
ratios and partial widths may allow one to discriminate between the
SM and new physics of electroweak symmetry breaking.
A number of recent studies have evaluated how precisely
the branching ratios
and couplings of the SM Higgs boson can be measured at the LC
\cite{BattagliaDesch,Battaglia,Gunion96,Sachwitz97,VanKooten,Brau,Boos}.
In this paper, we exploit the results of these studies to examine the
potential of Higgs boson BR and partial width measurements at the
LC to distinguish the SM-like Higgs boson  
of the MSSM from the SM Higgs boson.  We focus on
choices of MSSM parameters that exhibit a significant variation in
the approach to the decoupling limit.  This allows us to identify
parameter regions where hadron collider measurements
will most likely not be able to distinguish a MSSM from the SM Higgs
boson, and LC measurements can be essential for this purpose.  
In addition, one new observation is that 
%one of our most interesting results is that 
there are regions where decoupling
occurs at fairly low values of $m_A$ so that precision measurements
will not be able to reveal much about the MSSM parameter space structure.
In such cases, direct measurements of the properties of the heavier
Higgs bosons will be necessary to elucidate the true nature of the
Higgs sector.
Even if the existence of a MSSM Higgs boson is established,
it may be challenging to extract the underlying MSSM 
parameters \cite{ParamSets}.  Nevertheless, in some regions of MSSM
parameter space, we demonstrate that
the SUSY vertex corrections to the Higgs boson couplings to
bottom quark and tau pairs 
%(denoted $\Delta_b$) 
can be extracted from branching ratio 
and Higgs coupling measurements.\footnote{The distinctive effects of
SUSY vertex corrections to Higgs-fermion pair couplings on branching
ratio measurements have also been explored in Refs.~\cite{Babu,Hollik}.}

This paper is organized as follows.  In Sec.~\ref{sec:theory}, we
outline the features of the MSSM Higgs sector relevant for our
analysis.  In particular, we review and expand on the details of the radiative
corrections that contribute significantly to the Higgs couplings.  In
Sec.~\ref{sec:partialwidths}, we analyze the behavior of the Higgs
partial decay widths in various regions of MSSM parameter space.  We
consider three benchmark scenarios that lead to very different
behaviors in the MSSM Higgs sector.  In Sec.~\ref{sec:brmeas}, we
combine the expected experimental capabilities of the LC to measure
Higgs boson BRs and couplings with our theoretical results.  For the
benchmark scenarios, we quantify the ability of the LC to distinguish
the lightest MSSM Higgs boson from the SM Higgs boson.  
In Sec.~\ref{sec:Deltab}, we describe how the SUSY Yukawa correction to
the Higgs boson couplings to bottom quark pairs, $\Delta_b$, can be
extracted from Higgs measurements.  
In Sec.~\ref{sec:brmeas} and \ref{sec:Deltab} 
we also evaluate the impact of the 
theoretical uncertainties on our analysis.
Finally,
Sec.~\ref{sec:conclusions} contains our conclusions.  Some preliminary
results of this work were presented in Ref.~\cite{LoganMrenna}.

%------------------------------------------------------------------
\section{The MSSM Higgs sector}
\label{sec:theory}

In this section, we review those properties of the MSSM Higgs sector relevant
to our analysis.
At tree level, the masses and couplings of the MSSM Higgs bosons are determined
by two parameters, which are conveniently chosen to be
the mass of the CP-odd Higgs boson, $m_A$, and the ratio
of the vacuum expectation values of the two neutral Higgs fields, $\tan\beta$.
Radiative corrections to the MSSM Higgs sector
introduce significant dependence
on other MSSM parameters (for a review see Ref.~\cite{HiggsWGrep}).
These radiative corrections have
been analyzed extensively in the literature
\cite{Okada,Li,Brignole91,Zhang,
GunionTurski,HaberHempfling,Brignole92,
Berger,Chankowski92,PBMZ,HempflingHoang,Haber:1997fp,Heinemeyer,
CHWW,Barbieri,EspinosaQuiros,HaberHempfling93,CEQW,CHHHWW,
JoseRamon,Degrassi2001,HiggsWGrep,
Coarasa,HRS,COPW,Dabelsteinhbb,CMWpheno,HHLPRT,Heinemeyer00}.

The two main sources of radiative corrections to the couplings
of the MSSM Higgs bosons are: (i) the radiative corrections to the
Higgs squared-mass matrix
\cite{Okada,Li,Brignole91,Zhang,
GunionTurski,HaberHempfling,Brignole92,
Berger,Chankowski92,PBMZ,HempflingHoang,Haber:1997fp,Heinemeyer,
CHWW,Barbieri,EspinosaQuiros,HaberHempfling93,CEQW,CHHHWW,
JoseRamon,Degrassi2001},
%(for a review
%see Ref.~\cite{HiggsWGrep}),
which give rise to corrections to an effective CP-even Higgs mixing
angle $\alpha$,
and (ii) vertex corrections to the Higgs-fermion Yukawa couplings
\cite{Coarasa,HRS,COPW,PBMZ,Dabelsteinhbb,CMWpheno,HHLPRT,Heinemeyer00}.
In this paper we examine the effects of these two types of corrections in
the CP-conserving MSSM.\footnote{The MSSM Higgs sector 
automatically conserves CP
at tree-level, although non-trivial CP-violating effects can enter at
one-loop (due to complex MSSM parameters)
and be phenomenologically significant \cite{CEPW,CDL}.  
In this paper, we assume that the one-loop CP-violating effects are absent.
The  CP-violating case will be addressed elsewhere.}

The squared-mass matrix for the CP-even neutral MSSM Higgs bosons $h$
and $H$ (where $m_{h} < m_{H}$) is given by:
% can be expressed in the interaction basis as:
\begin{equation}
        \mathcal{M}^2 \equiv
        \left( \begin{array}{cc}
        \mathcal{M}^2_{11} & \mathcal{M}^2_{12} \\
        \mathcal{M}^2_{12} & \mathcal{M}^2_{22}
        \end{array} \right)
=
\left( \begin{array}{cc}
        m_A^2 s^2_{\beta} + m_Z^2 c^2_{\beta}
                & -(m_A^2 + m_Z^2) s_{\beta} c_{\beta} \\
        -(m_A^2 + m_Z^2) s_{\beta} c_{\beta}
                & m_A^2 c^2_{\beta} + m_Z^2 s^2_{\beta}
        \end{array} \right) +\delta{\cal M}^2\,,
        \label{eq:massmatrix}
\end{equation}
where $\delta{\cal M}^2$ is a consequence of the radiative corrections.
At tree level, one obtains $m_{h} \leq m_Z |\cos 2\beta | \leq
m_Z$.  
Such
a light $h$ is essentially ruled out by searches at LEP2.\footnote{The
current MSSM Higgs mass limits are $m_h > 91.0$ GeV and
$m_A > 91.9$ GeV \cite{LEPHIGGSgroup}.}
However, once radiative corrections to the squared-mass 
matrix are included, the theoretical upper bound on
$m_{h}$ is raised substantially.  For a fixed value of $\tan\beta$
and a specified set of MSSM parameters,
$m_{h}$ grows with increasing $m_A$ and
reaches an asymptotic value $m_{h}^{\rm max}(\tan\beta)$ in the limit of
large $m_A$.
%(see Ref.~\cite{HiggsWGrep} for details).
If $\tan\beta$ is now
allowed to vary (while holding all other free parameters fixed), 
$m_{h}^{\rm max}(\tan\beta)$ increases with
$\tan\beta$ and typically\footnote{In some regions of MSSM parameter
space at large $\tan\beta$, radiative corrections to the
Higgs-bottom quark Yukawa coupling can yield 
large negative loop corrections to
$m_h^{\rm max}(\tan\beta)$, so that the latter begins to decrease
for $\tan\beta\gsim 10$.} 
reaches an asymptotic value $m_{h}^{\rm max}$ for
$\tan\beta \gsim 10$.
For large values of $\tan \beta$, 
$m_{h} \simeq m_{h}^{\rm max}$ and
$m_{H} \simeq m_A$ for $m_A >
m_{h}^{\rm max}$.   Conversely, if $m_A < m_{h}^{\rm max}$ then
$m_{h} \simeq m_A$ and $m_{H} \simeq m_{h}^{\rm max}$.

At the LC, a light SM-like Higgs boson will be produced through
Higgsstrahlung [$e^+e^- \to Z^* \to Zh$], $WW$ fusion
[$e^+e^- \to W^*W^*\nu \bar \nu \to h\nu \bar \nu$] and $ZZ$ fusion
[$e^+e^-\to Z^*Z^* e^+e^- \to he^+e^-$].  The cross sections for all the
above processes
depend on the couplings of the Higgs boson to vector boson pairs.
In the MSSM,
the couplings of $h$ [$H$] to vector boson pairs are given by
$\sin(\beta - \alpha)$ [$\cos(\beta - \alpha)$] times the corresponding
SM Higgs coupling.  The decoupling limit corresponds to $m_A\gg m_Z$,
in which case $\sin(\beta-\alpha)\simeq 1$, and the properties of $h$
approach those of the SM Higgs boson.
The CP-even Higgs squared-masses obey the sum rule \cite{Carena:2000bh,EspCom}
\begin{equation}
        m_{H}^2 \cos^2(\beta - \alpha)
        + m_{h}^2 \sin^2(\beta - \alpha)
        = \left[m_{h}^{\rm max}(\tan\beta) \right]^2.
\end{equation}
In particular, combining this sum rule with the large $\tan\beta$
behavior of the Higgs masses noted above implies that
$H$ is the SM-like Higgs boson for
$m_A < m_{h}^{\rm max}$ and large values of $\tan \beta$,
while $h$ is the SM-like Higgs boson for $m_A > m_{h}^{\rm max}$.
Note that the decoupling limit implies that the latter holds 
for any value of $\tan \beta$.
Most of the analysis of this paper will focus on the case where
$h$ is the SM-like Higgs boson and $m_A > m_{h}^{\rm max}$.

Along with $\tan\beta$, the CP-even Higgs mixing angle $\alpha$ determines the
Higgs boson couplings to fermions.  In particular, 
relative to their SM values, the couplings
of $h$ [$H$] to down-type fermions are multiplied by 
$-\sin\alpha/\cos\beta$ [$\cos\alpha/\cos\beta$],
and those of $h$ [$H$] to up-type fermions are multiplied by
$\cos\alpha/\sin\beta$ [$\sin\alpha/\sin\beta$].
Thus radiative corrections to $\alpha$ can have significant effects on
the Higgs boson couplings to fermions.
The mixing angle $\alpha$ which diagonalizes the mass matrix in
Eq.~\ref{eq:massmatrix} can be expressed as:
\begin{equation}
        s_{\alpha} c_{\alpha} = \frac{\mathcal{M}^2_{12}}
        {\sqrt{({\rm Tr}\mathcal{M}^2)^2 - 4 \, {\rm det} \mathcal{M}^2}}\,,
\ \ \ \
        c^2_{\alpha} - s^2_{\alpha} =
        \frac{\mathcal{M}^2_{11} - \mathcal{M}^2_{22}}
        {\sqrt{({\rm Tr}\mathcal{M}^2)^2 - 4 \, {\rm det} \mathcal{M}^2}}\,,
        \label{eq:alpha}
\end{equation}
where $s_{\alpha} \equiv \sin\alpha$ and $c_{\alpha} \equiv \cos\alpha$.
Note that if $\mathcal{M}^2_{12} \to 0$, then either $\sin\alpha \to 0$
(if $\mathcal{M}_{11}^2 > \mathcal{M}_{22}^2$) or
$\cos\alpha \to 0$ (if $\mathcal{M}_{11}^2 < \mathcal{M}_{22}^2$).
At tree level (see Eq.~\ref{eq:massmatrix}),
$\mathcal{M}^2_{12}$ is small for small
$m_A$ and/or large $\tan\beta$, but it cannot vanish.
This is no longer true after including radiative corrections,
which can be of the same order as the tree level value
for small values of $m_A$ and large $\tan\beta$.
In particular, the radiatively-corrected value of $\mathcal{M}^2_{12}$ 
exhibits a widely varying behavior as a function 
of the MSSM parameters.
The radiative corrections to
${\cal M}^2$, including dominant corrections 
coming from the one-loop top and bottom quark and
top and bottom squark contributions plus the two-loop
leading logarithmic contributions, are given to $\mathcal{O}(h_t^4, h_b^4)$ by
\cite{Haber:1997fp,CMWpheno}
\bea
        \delta \mathcal{M}^2_{11} &\simeq&
        - \bar \mu^2 x_t^2 \frac{h_t^4 v^2}{32 \pi^2}
        s^2_{\beta}
        \left[ 1 + c_{11}
        \ln \left( \frac{M_S^2}{m_t^2} \right) \right] 
        - \bar \mu^2 a_b^2 \frac{h_b^4 v^2}{32 \pi^2}
        s^2_{\beta}
        \left[ 1 + c_{12}
        \ln \left( \frac{M_S^2}{m_t^2} \right) \right], \nonumber \\[5pt]
        \delta \mathcal{M}^2_{22} &\simeq&
        \frac{3 h_t^4 v^2}{8 \pi^2} s^2_{\beta} 
        \ln \left( \frac{M_S^2}{m_t^2} \right)
        \left[ 1 + \half c_{21}
        \ln \left( \frac{M_S^2}{m_t^2} \right) \right] \nonumber \\[5pt]
        &&
        + \frac{h_t^4 v^2}{32 \pi^2} s^2_{\beta} x_t a_t (12 - x_t
        a_t)
        \left[ 1 + c_{21}
        \ln \left( \frac{M_S^2}{m_t^2} \right) \right] 
        - \bar \mu^4 \frac{h_b^4 v^2}{32 \pi^2} s^2_{\beta}
        \left[ 1 + c_{22}
        \ln \left( \frac{M_S^2}{m_t^2} \right) \right]\,,\nonumber \\[5pt]
        \delta \mathcal{M}^2_{12} &\simeq&
        - \bar \mu x_t \frac{h_t^4 v^2}{32 \pi^2} (6 - x_t a_t)
        s^2_{\beta}
%        - \frac{3 h_t^2 m_Z^2}{32 \pi^2} \right]
        \left[ 1 + c_{31}
        \ln \left( \frac{M_S^2}{m_t^2} \right) \right] 
        + \bar \mu^3 a_b \frac{h_b^4 v^2}{32 \pi^2}
        s^2_{\beta}
        \left[ 1 + c_{32}
        \ln \left( \frac{M_S^2}{m_t^2} \right) \right]\,,
        \label{eq:M12approx}
\eea
where $s_{\beta} \equiv \sin\beta$,
$c_{\beta} \equiv \cos\beta$, and the coefficients $c_{ij}$ are:
\begin{equation}
c_{ij}\equiv {t_{ij}h_t^2+b_{ij}h_b^2-32g_3^2\over 32\pi^2}\,,
\end{equation}
with $(t_{11},t_{12},t_{21},t_{22},t_{31},t_{32})=(12,-4,6,-10,9,-7)$
and $(b_{11},b_{12},b_{21},b_{22},b_{31},b_{32})=(-4,12,2,18,-1,15)$.
Above, $h_t$ and $h_b$ are the top and bottom quark Yukawa couplings
[see Eqs.~\ref{yuklag}--\ref{tyukmassrel}], 
$g_3$ is the strong QCD coupling,
$v = 246$ GeV is the SM Higgs
vacuum expectation value, and $M_S^2 = \half(M^2_{\tilde t_1} + 
M^2_{\tilde t_2})$ is the average squared top squark 
mass.\footnote{Eq.~\ref{eq:M12approx} is derived under the assumption that
$|M^2_{\tilde t_1} - M^2_{\tilde t_2}|/(M^2_{\tilde t_1} + M^2_{\tilde t_2})
\ll 1$.  The approximate forms of Eq.~\ref{eq:M12approx} are sufficient
to provide insight on the dependence of the radiatively-corrected
Higgs masses and couplings on the MSSM parameters.  
The numerical work of this paper employs
more exact expressions for the Higgs squared-mass matrix elements
as noted below.}
%which is the case in the benchmark scenarios considered in this
%paper.
The $\delta\mathcal{M}_{ij}^2$ also depend on the MSSM parameters $A_t$,
$A_b$ and $\mu$ that enter the off-diagonal top-squark and
bottom-squark squared-mass matrices.
We employ the following notation:
$\bar \mu \equiv \mu / M_S$,
$a_t\equiv  A_t / M_S$, $a_b \equiv A_b / M_S$ and 
$x_t \equiv X_t / M_S$, where $X_t \equiv A_t - \mu \cot\beta$.
Note that the leading radiative corrections
to $\mathcal{M}^2_{12}$  depend strongly on
the sign of $\mu X_t$ and the magnitude of $A_t$.
For the scenarios we shall consider, with $a_t^2 \lsim 6$,
the combination $A_t\mu<0$ [$A_t\mu>0$] 
can lead to a suppression [enhancement] of
$\mathcal{M}^2_{12}=-(m_A^2+m_Z^2)s_\beta c_\beta+\delta 
\mathcal{M}^2_{12}$, and hence, to a suppression [enhancement] of
the coupling of the SM-like Higgs boson to $b$ quarks and
$\tau$ leptons.

%The contributions of bottom squarks in the loops,
%which can be large at
%large $\tan\beta$, are not included in Eq.~\ref{eq:M12approx} for
%compactness, although they are included in our numerical
%calculations.
Our numerical calculation of the radiative corrections to the Higgs
masses and mixing angle is based on the results of Ref.~\cite{CHHHWW}
and incorporates the renormalization group improved one-loop effective
potential, plus the non-logarithmic two-loop contributions
of the Yukawa vertex corrections for top and bottom quarks.
The Yukawa vertex corrections modify the effective 
Lagrangian that describes the coupling of 
the Higgs bosons to the third generation 
quarks:\footnote{The Higgs couplings to leptons and to first and
second generation quarks can be treated similarly.}
\begin{equation} \label{yuklag}
	-\mathcal{L}_{\rm eff} = \epsilon_{ij} \left[
	(h_b + \delta h_b) \bar b_R H_d^i Q_L^j 
	+ (h_t + \delta h_t) \bar t_R Q_L^i H_u^j \right]
        + \Delta h_t \bar t_R Q_L^k H_d^{k \ast} 
        + \Delta h_b \bar b_R Q_L^k H_u^{k \ast} 
	+ {\rm h.c.}\,,
\end{equation}
resulting in a modification of the tree-level relation between 
$h_t$ [$h_b$] and $m_t$ [$m_b$] as follows:
\bea 
        m_b &=& \frac{h_b v}{\sqrt{2}} \cos\beta 
	\left(1 + \frac{\delta h_b}{h_b} 
        + \frac{\Delta h_b \tan\beta}{h_b} \right)
        \equiv\frac{h_b v}{\sqrt{2}} \cos\beta 
	(1 + \Delta_b)\,, \label{byukmassrel} \\[5pt]
        m_t &=& \frac{h_t v}{\sqrt{2}} \sin\beta 
	\left(1 + \frac{\delta h_t}{h_t} + \frac{\Delta
        h_t\cot\beta}{h_t} \right)
        \equiv\frac{h_t v}{\sqrt{2}} \sin\beta 
	(1 + \Delta_t)\,. \label{tyukmassrel} 
\eea
%Note that $\Delta_b$ contains a term that is
%$\tan\beta$-enhanced; this is true in general for
%the down-type Yukawa vertex corrections.
The dominant contributions to $\Delta_b$ are $\tan\beta$-enhanced,
with $\Delta_b\simeq (\Delta h_b/h_b)\tan\beta$; for
$\tan\beta\gg 1$, $\delta h_b/h_b$ provides a small correction to
$\Delta_b$.   In the same limit, $\Delta_t\simeq\delta h_t/h_t$, with
the additional contribution of $(\Delta h_t/h_t)\cot\beta$ providing a
small correction.\footnote{Because the one-loop
corrections $\delta h_b$, $\Delta h_b$, $\delta h_t$ and $\Delta h_t$
depend only on Yukawa and gauge couplings and SUSY particle masses, they
contain no hidden $\tan\beta$ enhancements \cite{Carena:2001uj}.}
Explicitly, one finds that   
for $\tan\beta \gg 1$ \cite{PBMZ,HRS,COPW}
\bea
        \Delta_b 
%\simeq {\Delta h_b\tan\beta\over h_b} 
        &\simeq&\left[ 
        \frac{2 \alpha_s}{3 \pi} \mu M_{\tilde g} \,
        I(M^2_{\tilde b_1}, M^2_{\tilde b_2}, M^2_{\tilde g})
        + \frac{h_t^2}{16 \pi^2} \mu A_t \,
        I(M^2_{\tilde t_1}, M^2_{\tilde t_2}, \mu^2)\right]\tan\beta\,,\\[5pt]
        \label{eq:Deltab}
        \Delta_t &\simeq& 
        -\frac{2 \alpha_s}{3 \pi} A_t M_{\tilde g} I(M^2_{\tilde t_1},
        M^2_{\tilde t_2}, M^2_{\tilde g})
        - \frac{h_b^2}{16 \pi^2} \mu^2 I(M^2_{\tilde b_1}, 
        M^2_{\tilde b_2}, \mu^2), \label{eq:Deltat} 
\eea
where $\alpha_s\equiv g_3^2/4\pi$,
$M_{\tilde g}$ is the gluino mass,
$M_{\tilde b_{1,2}}$ are the
bottom squark masses, and smaller electroweak
corrections have been ignored.
The loop integral $I(a^2,b^2,c^2)$ is of order $1/{\rm
max}(a^2,b^2,c^2)$ when at least one of its arguments is large
compared to $m_Z^2$;
for the explicit expression see Ref. \cite{COPW}.
These one-loop Yukawa vertex corrections enter indirectly as two-loop
effects in the Higgs squared-mass matrix elements via the dependence on
$h_t$ and $h_b$.
%%%

We have noted earlier that in the decoupling limit
$\sin(\beta-\alpha)=1$ [or equivalently  
$\cos(\beta-\alpha)=0$], in which case the couplings of $h$ are
identical to those of the SM Higgs boson.  This limit is achieved when
$m_A\gg m_Z$.  This behavior, which is easy to verify for the
tree-level expressions, continues to hold when radiative corrections
are included.  However, the onset of decoupling can be significantly
affected by the radiative corrections, as we now discuss.  From
Eq.~\ref{eq:alpha}, one easily obtains:
\bea \label{eq:cosbma}
\cos(\beta-\alpha)&=&{(\mathcal{M}_{11}^2-\mathcal{M}_{22}^2)\sin 2\beta
-2\mathcal{M}_{12}^2\cos 2\beta\over
2(m_H^2-m_h^2)\sin(\beta-\alpha)} \nonumber \\[5pt]
&=&{m_Z^2\sin 4\beta+
({\delta\mathcal{M}}_{11}^2-{\delta\mathcal{M}}_{22}^2)\sin 2\beta
-2{\delta\mathcal{M}}_{12}^2\cos 2\beta}\over
2(m_H^2-m_h^2)\sin(\beta-\alpha)\,.
\eea
Since $\delta\mathcal{M}^2_{ij}\sim {\mathcal O}(m_Z^2)$, 
and $m_H^2-m_h^2=m_A^2+\mathcal{O}(m_Z^2)$, one obtains 
for $m_A\gg m_Z$
\begin{equation} \label{cosbmadecoupling}
	\cos(\beta-\alpha)=c\left[{m_Z^2\sin 4\beta\over
	2m_A^2}+\mathcal{O}\left(m_Z^4\over m_A^4\right)\right]\,,
\end{equation}
where
\begin{equation} \label{cdef}
	c\equiv 1+{{\delta\mathcal{M}}_{11}^2-{\delta\mathcal{M}}_{22}^2\over
	2m_Z^2\cos 2\beta}-{{\delta\mathcal{M}}_{12}^2\over m_Z^2\sin
	2\beta}\,.
\end{equation}
Eq.~\ref{cosbmadecoupling} exhibits the expected decoupling behavior 
for $m_A\gg m_Z$.
However, Eq.~\ref{eq:cosbma} exhibits another way in which
$\cos(\beta-\alpha)=0$ can be achieved---simply choose the
supersymmetric parameters (that govern the Higgs mass radiative
corrections) such that the numerator of Eq.~\ref{eq:cosbma} vanishes.
That is,\footnote{Eq.~\ref{eq:tanbetadecoup} is
equivalent to the condition $c=0$ [see Eqs.~\ref{cosbmadecoupling}
and \ref{cdef}].}
\begin{equation}
        2 m_Z^2 \sin 2\beta =
        2\, \delta \mathcal{M}^2_{12} 
        - \tan 2\beta
        \left(\delta \mathcal{M}^2_{11} - \delta \mathcal{M}^2_{22} \right)\,.
        \label{eq:tanbetadecoup}
\end{equation}
Note that Eq.~\ref{eq:tanbetadecoup} is independent of the value of $m_A$.
For a typical choice of MSSM parameters,
Eq.~\ref{eq:tanbetadecoup} yields a solution at large $\tan\beta$---by
approximating $\tan 2\beta\simeq -\sin 2\beta \simeq -2/ \tan \beta$, 
one can determine
the value of $\beta$ at which the decoupling occurs:
\begin{equation} \label{earlydecoupling}
\tan \beta\simeq \frac{2m_Z^2-
\delta\mathcal{M}_{11}^2+\delta \mathcal{M}_{22}^2}
{ \delta\mathcal{M}_{12}^2}\,.
\end{equation}
The explicit expressions for $\delta\mathcal{M}_{ij}^2$
quoted in Eq.~\ref{eq:M12approx} confirm that the
assumption of $\tan\beta\gg 1$ used to derive this result is a consistent
approximation because $\delta \mathcal{M}^2_{12}$ is typically small.
We conclude that for the value of $\tan\beta$ specified
in Eq.~\ref{earlydecoupling}, $\cos(\beta-\alpha)=0$ independently of
the value of $m_A$.  We shall refer to this phenomenon as
$m_A$-independent decoupling.
From Eq.~\ref{eq:M12approx}, it follows that explicit solutions to
Eq.~\ref{eq:tanbetadecoup} depend on ratios of SUSY parameters and  
%on $M_S$, 
so are insensitive to the overall SUSY mass scale,
modulo a mild logarithmic dependence on $M_S/m_t$.

The introduction of the radiatively-corrected value for 
the CP-even Higgs mixing angle $\alpha$
affects the MSSM Higgs boson couplings to all down-type 
fermions (and likewise to all up-type fermions) in the same way.
In particular,\footnote{We use the notation
$\BR(b)\equiv \BR(h\to b\bar b)$ and $\Gamma(b)\equiv \Gamma(h\to
b\bar b)$, and similarly for other Higgs decay final states.}
$\BR(b)/\BR(\tau)=\Gamma(b)/\Gamma(\tau) 
\propto g^2_{hbb}/g^2_{h\tau\tau}\propto
m_b^2/m_\tau^2$ \cite{Babu,Hollik,CMWpheno,Wells}.
However, the Yukawa vertex corrections
enter directly in the couplings of fermions to the Higgs
bosons.  These corrections can be understood as a modification
of the relation between the fermion Yukawa coupling and its mass, as
exhibited in Eqs.~\ref{byukmassrel}--\ref{tyukmassrel}.
After including the dominant corrections, the CP-even Higgs boson couplings
to $b$ and $\tau$ are modified relative to the
SM coupling, $g_{\hsm ff}=gm_f/2m_W$, as follows \cite{CMWpheno}:
\bea
        g_{hbb} &=& -\frac{g m_b \sin\alpha}{2 m_W \cos\beta}\,
        \frac{1}{1 + \Delta_b} \left[ 1 - \Delta_b\cot\alpha \cot\beta
        +{\delta h_b\over h_b}\left(1+\cot\alpha\cot\beta\right) \right]\,,
         \nonumber \\[6pt]
        g_{Hbb} &=& \phantom{-}\frac{g m_b \cos\alpha}{2 m_W \cos\beta}\,
        \frac{1}{1 + \Delta_b} \left[ 1 + \Delta_b \tan\alpha\cot\beta
        +{\delta h_b\over h_b}\left(1-\tan\alpha\cot\beta\right)\right]\,.
%	 \nonumber \\[5pt]
%        g_{h\tau\tau} &=& -\frac{g m_{\tau} \sin\alpha}{2 m_W \cos\beta}
%        \frac{1}{1 + \Delta_\tau} 
%          \left[ 1 - \frac{\Delta_\tau}{\tan\alpha \tan\beta}
%        \right] \nonumber \\[5pt]
%        g_{H\tau\tau} &=& \frac{g m_{\tau} \cos\alpha}{2 m_W \cos\beta}
%        \frac{1}{1 + \Delta_\tau} \left[ 1 + \frac{\Delta_\tau
%        \tan\alpha}{\tan\beta} \right]\,.
        \label{eq:couplingsDeltab}
\eea
At large $\tan\beta$, terms involving
$\Delta_b\propto\tan\beta$ [Eq.~\ref{eq:Deltab}]
provide the dominant corrections to the neutral Higgs couplings to
$b\bar b$.  The corrections proportional to $\delta h_b/h_b$ [see
Eqs.~\ref{yuklag}--\ref{byukmassrel} and the discussion that follows]
are never $\tan\beta$-enhanced and are therefore numerically unimportant.
The $\tau$ couplings are obtained from Eq.~\ref{eq:couplingsDeltab}
by replacing $m_b$, $\Delta_b$ and $\delta h_b$
with $m_{\tau}$, $\Delta_{\tau}$ and $\delta h_\tau$, respectively.  Note that
$\Delta_\tau$ and $\delta h_\tau$
arise in analogy with $\Delta_b$ and $\delta h_b$ from SUSY particle loops 
involving the leptonic sector.  In particular, at large $\tan\beta$
 \cite{Hollik,PBMZ,HRS,COPW}:
\begin{equation} \label{eq:deltatau}
	\Delta_\tau \simeq \left[{\alpha_1 \over 4\pi} M_1\mu
	I(M^2_{\tilde\tau_1},
	M^2_{\tilde\tau_2},M^2_1) - {\alpha_2 \over 4\pi} M_2\mu \,
	I(M^2_{\tilde\nu_\tau},M^2_2,\mu^2)\right]\tan\beta\,,
\end{equation}
where $\alpha_2\equiv g^2/4\pi$ and $\alpha_1\equiv g^{\prime 2}/4\pi$
are the electroweak gauge couplings.  
Since corrections to $h_\tau$ are proportional to $\alpha_1$ and
$\alpha_2$, we expect $|\Delta_\tau|\ll |\Delta_b|$.
Nevertheless,
we shall formally keep the $\Delta_\tau$ corrections in our analysis,
although they will have negligible effect in our numerical results.
%After including $\Delta_b$
%the SM relation between the bottom and $\tau$ branching ratios 
%is violated, except in the decoupling
%limit $\tan\alpha\tan\beta\to -1$.

To see how the decoupling limit is achieved for the $hbb$ 
(and $h\tau \tau$) couplings, note that we can write:
\begin{equation}
-{\sin\alpha\over\cos\beta}=\sin(\beta-\alpha)
-\tan\beta\cos(\beta-\alpha)\,.
\end{equation}
Working to first order in $\cos(\beta-\alpha)$, and
using\footnote{In the decoupling limit ($m_A\gg m_Z$),
Eq.~\ref{cosbmadecoupling} implies that that $|(\tan\beta+\cot\beta)
\cos(\beta-\alpha)|\ll 1$ for all values of $\tan\beta$.}
\begin{equation}
	\tan\alpha\tan\beta=-1+(\tan\beta+\cot\beta)\cos(\beta-\alpha)
	+ \mathcal{O}\left(\cos^2(\beta-\alpha)\right)\,,
	\label{eq:tanatanb}
\end{equation}
it follows that
\begin{equation} \label{hbbcorrection}
%	g_{hbb}\simeq g_{\hsm bb}\left[1 - \tan\beta\cos(\beta-\alpha)
%	\left(1 - \frac{1}{\sin^2\beta}\frac{\Delta_b}{1+\Delta_b}\right)
%	\right]\,.
	g_{hbb}\simeq g_{\hsm bb}\left[1-
          (\tan\beta+\cot\beta) \cos(\beta-\alpha)
	\left(\sin^2\beta-{\Delta_b-\delta h_b/h_b\over 1+\Delta_b}
        \right)\right]\,.
\end{equation}
Two points are particularly noteworthy.  First, for $m_A\gg m_Z$,
decoupling is achieved since $\cos(\beta-\alpha)\sim
\mathcal{O}(m_Z^2/m_A^2)$ [Eq.~\ref{cosbmadecoupling}].  However, because
$\Delta_b\propto\tan\beta$,
decoupling can be delayed until $m_A^2\gg
m_Z^2\tan\beta$ \cite{HHLPRT}.  Second, as noted above, if
Eq.~\ref{eq:tanbetadecoup} is satisfied (or equivalently if $c=0$
[Eq.~\ref{cdef}]), decoupling is achieved independently of the value
of $m_A$.  One also obtains a similar expression for $g_{h\tau\tau}$
by replacing $\Delta_b$ and $\delta h_b$
with $\Delta_\tau$ and $\delta h_\tau$ in Eq.~\ref{hbbcorrection}.
Since $|\Delta_\tau|\ll |\Delta_b|$, it follows that the SM
expectation, $g_{hbb}^2/g_{h\tau\tau}^2\propto m_b^2/m_\tau^2$ is
violated except in the decoupling limit \cite{Babu,Hollik,CMWpheno,Wells}.

Another limiting case of interest is one where $\sin(\beta-\alpha)$ 
is close to zero.  This limit can be
reached for values of $m_A\lsim m_h^{\rm max}$ and large $\tan\beta$.
In this limit, the $H$ couplings to the $W$ and $Z$
bosons approach their Standard Model values and become relevant for
our analysis.
However, this is not a decoupling limit, and the $H$ couplings to
fermion pairs can deviate from the corresponding Standard Model couplings.
This can be confirmed by observing that 
\begin{equation}
{\cos\alpha\over\cos\beta}=\cos(\beta-\alpha)+\tan\beta\sin(\beta-\alpha)\,.
\end{equation}
But, for large $\tan\beta$, it is possible to have
$\tan\beta\sin(\beta-\alpha)\sim\mathcal{O}(1)$ even in the limit of
small $\sin(\beta-\alpha)$.  
Working to first order in $\sin(\beta-\alpha)$ and using
\begin{equation}
	\cot\alpha\tan\beta=1+(\tan\beta+\cot\beta)\sin(\beta-\alpha)
	+ \mathcal{O}\left(\sin^2(\beta-\alpha)\right)\,,
	\label{eq:cotatanb}
\end{equation}
it follows that 
\begin{equation} \label{Hbbcorrection}
	g_{Hbb}\simeq g_{\hsm bb}\left[1-
          (\tan\beta+\cot\beta) \sin(\beta-\alpha)
	\left(\cos^2\beta-
        {1+\delta h_b/h_b \over 1+\Delta_b}\right)\right]\,.
\end{equation}
Note that if
$|(\tan\beta+\cot\beta)\sin(\beta-\alpha)|\ll 1$, then the $Hbb$
coupling approaches the Standard Model value, even when the Yukawa
vertex corrections are included.  

We next consider 
the CP-even Higgs boson couplings to top quark pairs.  The analysis is
similar to the one given above, and one obtains
%(the charm quark couplings are obtained by replacing $t$ with $c$):
\begin{eqnarray}
	g_{htt} &=& \frac{g m_t \cos\alpha}{2 m_W \sin\beta}
	\left[ 1 - \frac{1}{1 + \Delta_t} 
	\frac{\Delta h_t}{h_t}(\cot\beta + \tan\alpha) \right]\,, 
	\nonumber \\[6pt]
	g_{Htt} &=& \frac{g m_t \sin\alpha}{2 m_W \sin\beta}
	\left[ 1 - \frac{1}{1 + \Delta_t}
	\frac{\Delta h_t}{h_t}(\cot\beta - \cot\alpha) \right]\,.
%	g_{htt} &=& \frac{g m_t \cos\alpha}{2 m_W \sin\beta}
%	\frac{\left[1 + \delta h_t/h_t 
%		- \Delta h_t/h_t \cot\alpha \right]}
%	{\left[1 + \delta h_t/h_t + \Delta h_t/h_t \tan\beta
%	\right]} 
%	\simeq \frac{g m_t \cos\alpha}{2 m_W \sin\beta}
%	\nonumber \\
%	g_{Htt} &=& \frac{g m_t \sin\alpha}{2 m_W \sin\beta}
%	\frac{\left[1 + \delta h_t/h_t
%		+ \Delta h_t/h_t \tan\alpha \right]}
%	{\left[1 + \delta h_t/h_t + \Delta h_t/h_t \tan\beta
%	\right]}
%	\simeq \frac{g m_t \sin\alpha}{2 m_W \sin\beta}\,,
\label{eq:couplingsDeltat}
\end{eqnarray}
%where we use the notation of Eq.~\ref{tyukmassrel} for the Yukawa vertex
%corrections.  
Here, it is more convenient to express our results in
terms of $\Delta_t$ and $\Delta h_t/h_t$, since $\Delta_t\simeq \delta
h_t/h_t$, while the corresponding contribution of $\Delta h_t/h_t$ is 
$\tan\beta$ suppressed [Eq.~\ref{tyukmassrel}].
The Higgs couplings to charm quark pairs are obtained from 
Eq.~\ref{eq:couplingsDeltat} by replacing $m_t$, $\Delta_t$ 
and $\Delta h_t$ with
$m_c$, $\Delta_c$ and $\Delta h_c$, respectively.  Using
$\cot\beta+\tan\alpha \simeq \cos(\beta-\alpha)/\sin^2\beta$
[see Eq.~\ref{eq:tanatanb}],
it follows that the SUSY vertex corrections to 
$g_{htt}$ and $g_{hcc}$ are suppressed in the decoupling limit (with no 
enhancement in the limit of large $\tan\beta$), and so $g_{htt}$
and $g_{hcc}$ approach their Standard Model values.
In the opposite limit in which
$\sin(\beta-\alpha)$ is close to zero, we use 
$\cot\beta-\cot\alpha \simeq -\sin(\beta-\alpha)/\sin^2\beta$
[see Eq.~\ref{eq:cotatanb}] to conclude that the
SUSY vertex corrections to $g_{Htt}$ and $g_{Hcc}$ are
suppressed.  Thus, in the numerical results presented in this paper,
the SUSY Yukawa vertex corrections to both $h$ and $H$ couplings 
to up-type quark pairs have a negligible effect in the parameter regions of
interest and can be neglected.

By considering certain ratios of Higgs-fermion couplings, 
one can begin to isolate
various combinations of SUSY Yukawa vertex corrections.
We introduce the notation $\hat g_{\phi ff} \equiv
g_{\phi ff}/g_{h_{SM}ff}$ [$\phi\equiv h$, $H$]
for the Higgs-fermion couplings normalized to
their Standard Model values.
From Eqs.~\ref{eq:couplingsDeltab} and \ref{eq:couplingsDeltat}, we obtain
\begin{equation}
	\frac{\hat g_{hbb} - \hat g_{h\tau \tau}}
	{\hat g_{htt} - \hat g_{hbb}}=
	\frac{\hat g_{Hbb} - \hat g_{H\tau \tau}}
	{\hat g_{Htt} - \hat g_{hbb}}\simeq
        {\displaystyle
        \frac{\Delta_b - \Delta_{\tau}}{1 + \Delta_{\tau}}
	-\displaystyle\frac{\delta h_b}{h_b}+
         \left(\displaystyle\frac{1+\Delta_b}{1+\Delta_\tau}\right)
         \frac{\delta h_\tau}{h_\tau}\over
         1-\left(\displaystyle\frac{1+\Delta_b}{1+\Delta_t}\right)
        \displaystyle\frac{\Delta
	h_t}{h_t}\cot\beta+\displaystyle\frac{\delta h_b}{h_b}}\,.  
	\label{eq:ccratio}
\end{equation}
Note that dependence on the CP-even Higgs mixing angle has
conveniently canceled out.
At large $\tan\beta$, one can to first approximation keep only those
terms that are $\tan\beta$-enhanced at one loop.  We then obtain:
\begin{equation}
	\frac{\hat g_{hbb} - \hat g_{h\tau \tau}}
	{\hat g_{htt} - \hat g_{hbb}}=
        \frac{\hat g_{Hbb} - \hat g_{H\tau \tau}}
	{\hat g_{Htt} - \hat g_{hbb}}\simeq        
        \frac{\Delta_b - \Delta_{\tau}}{1 + \Delta_{\tau}}
	\simeq \Delta_b \,,
	\label{eq:ghat}
\end{equation}
where the last step follows if $|\Delta_{\tau}| \ll 1$
(we have already noted that $|\Delta_{\tau}| \ll |\Delta_b|$).  
Eq.~\ref{eq:ghat}
and the analogous result in which the Higgs couplings to top quarks are
replaced by the corresponding couplings to charm quarks
will be used in Sec.~\ref{sec:Deltab} when we discuss the extraction
of $\Delta_b$ from Higgs coupling measurements. 

%-----------------------------------------------------------------
\section{Behavior of Higgs decay observables}
\label{sec:partialwidths}

In this section, we examine in detail the behavior of the MSSM Higgs
boson partial widths.
In order to present quantitative results,
we consider three ``benchmark'' scenarios for the MSSM parameters
that lead to very
different behaviors of the SM-like Higgs boson of the MSSM.
Our three benchmark scenarios, summarized in Table~\ref{tab:scenarios},
correspond approximately to those discussed in Ref.~\cite{CHWW}.
All MSSM parameters are specified at the electroweak scale.
The three benchmark scenarios have the following properties:

\begin{description}
\item[No-mixing scenario:]  The top squark mixing angle $\theta_{\tilde t}$
is zero.
This scenario yields the lowest
value of $m_{h}^{\rm max}(\tan\beta)$
for given values of $\tan\beta$ and $M_S$.
For simplicity, we define the scenarios in terms of
$\msusy \equiv M_{\tilde Q} = M_{\tilde U} = M_{\tilde D}$,
where the latter are third generation squark mass parameters.
For $\msusy\gg m_t$, as is true in the scenarios considered here,
$M_{\rm SUSY}\simeq M_S$ [where
$M^2_S \equiv\half(M^2_{\tilde t_1} + M^2_{\tilde t_2})$].
Here we have chosen a large value for $\msusy = 1.5$ TeV
in order to obtain
a sufficiently large value of $m_{h}^{\rm max}(\tan\beta)$,
comparable to that obtained in the other
two scenarios (the case of $\msusy=1$ TeV is at the edge of the region
excluded by LEP2).
\item[Maximal-mixing scenario:]  The top squark
mixing is chosen to
give the maximal value of $m_{h}^{\rm max}(\tan\beta)$
for given values of $\tan\beta$ and $M_S$.
\begin{table}[t!]
    \begin{center}\begin{tabular}{c c c c c c l} \hline
        & \multicolumn{5}{c}{Mass parameters [TeV]} & \\
        Benchmark & $\mu$ & $X_t \equiv A_t - \mu \cot\beta$ & $A_b$
                & $M_{\rm SUSY}$ & $M_{\tilde g}$ & $m_h^{\rm max}$ [GeV]\\
        \hline
        No-Mixing & $-0.2$ & 0 & $A_t$ & 1.5 & 1 & 118 \\
        Maximal-Mixing & $-0.2$ & $\sqrt{6}$ & $A_t$ & 1 & 1 & 129\\
        Large $\mu$ and $A_t$ & $\pm 1.2 $ & $\mp 1.2(1 + \cot\beta)$
                & 0 & 1 & 0.5 & 119 \\
        \hline
        \end{tabular} \end{center}
        \caption{MSSM parameters for our benchmark scenarios,
and the derived maximal mass for the SM-like Higgs boson.}
        \label{tab:scenarios}
\end{table}
\item[Large $\mu$ and $A_t$ scenario:]
Large radiative corrections occur to both $\alpha$ and
$\Delta_b$.  In particular,
$\mathcal{M}_{12}^2$ can exhibit extreme variations in magnitude
depending on the sign of $A_t\mu$ and the magnitude of $A_t$.
The two possible sign combinations for $A_t$ and $\mu$ 
(for a fixed sign of $A_t\mu$)
yield small differences in
$\mathcal{M}_{12}^2$ through the dependence of $h_t$ and $h_b$
on $\Delta_t$ and $\Delta_b$, respectively.
The vertex correction $\Delta_b$ is dominated by the
bottom squark-gluino contribution, which can enhance or suppress
the Yukawa coupling $h_b$ for negative or positive $\mu$, respectively.
In the following we choose $A_t\mu< 0$ and consider the two possible sign
combinations for $A_t$ and $\mu$.\footnote{If the charged 
Higgs boson is light, then our choice of the sign of $A_t\mu$ is
favored by the experimental constraints on
$b\to s \gamma$ \cite{Degrassi:2000qf,Carena:2001uj}.}
\end{description}

To be conservative, we have chosen relatively large values for the SUSY 
breaking parameters, on the order of 1 TeV, so that some supersymmetric
particles may not be kinematically accessible at the LC.\footnote{Note 
that the $m_A$-independent decoupling in the large $\mu$ and $A_t$ 
scenario depends on  
$\mu/M_S$ and $A_t/M_S$.  Lower values of $M_S$ would allow for 
correspondingly lower values of $\mu$ and $A_t$.}
However, for simultaneously large $\mu$ and $M_{\tilde g}$, the size of the 
$\Delta_b$ corrections may drive the bottom Yukawa coupling out of the 
perturbative region.  
Thus the gluino mass is taken as $M_{\tilde g} = 0.5$ TeV for large $\mu$ and
$M_{\tilde g} = 1$ TeV for moderate $\mu$.  The other gaugino 
mass parameters are $M_2=2 M_1=200$ GeV 
($M_2$ is relevant for the one-loop $h \to \gamma \gamma$ amplitude).
%(these are relevant for
%the one-loop $h g g$ and $h \gamma \gamma$ amplitudes).
Finally, the masses of the remaining squarks and sleptons 
are set to 1 TeV.

We calculate the properties of the MSSM Higgs bosons for each of the
benchmark scenarios using the program {\sc Hdecay} \cite{HDECAY},
to which we have added the $\Delta_b$ and $\Delta_t$ Yukawa vertex
corrections in the two-loop Higgs boson squared-mass matrix \cite{CHHHWW}
and the $\Delta_b$ Yukawa vertex corrections in the Higgs couplings
to $b \bar b$.
By comparing the partial widths, total width
and branching ratios of the SM-like Higgs boson of the MSSM and the SM Higgs
boson of the same mass, we can evaluate the sensitivity of the various
observables for distinguishing the MSSM from the SM.
We define the fractional deviations of the MSSM Higgs 
partial widths from those of the SM Higgs boson of the same mass as follows:
\begin{equation}
        \delta \Gamma = \frac{|\Gamma_{\rm MSSM} - \Gamma_{\rm SM}|}
	{\Gamma_{\rm SM}}\,,
        \label{eq:deltaBR}
\end{equation}
and analogously for the branching ratios,
$\delta \BR = |\BR_{\rm MSSM} - \BR_{\rm SM}|/\BR_{\rm SM}$.
This allows us to demonstrate which Higgs decay quantities are the
most sensitive to the non-standard nature of the Higgs boson.
Later, we will combine these individual deviations into a $\chi^2$
variable to improve the experimental sensitivity.

In the next section, we discuss the expected behavior of
the Higgs BRs and partial widths in the MSSM, with particular emphasis on the
approach to the decoupling limit \cite{decoupling}.
In general, the couplings of the
SM-like Higgs boson of the MSSM deviate from those of the SM Higgs boson
of the same mass, except in the decoupling limit.

\subsection{Theoretical Expectations for Direct Higgs Couplings}
\label{sec:directcoup}

\begin{figure}
%       $\delta \Gamma(h \to WW^*)$, $\delta \Gamma(h \to c \bar c)$,
%       $\delta \Gamma(h \to b \bar b)$, $\delta \Gamma(h \to \tau^+ \tau^-)$.
\begin{center}
\resizebox{\textwidth}{!}{
\includegraphics*[19,142][529,682]{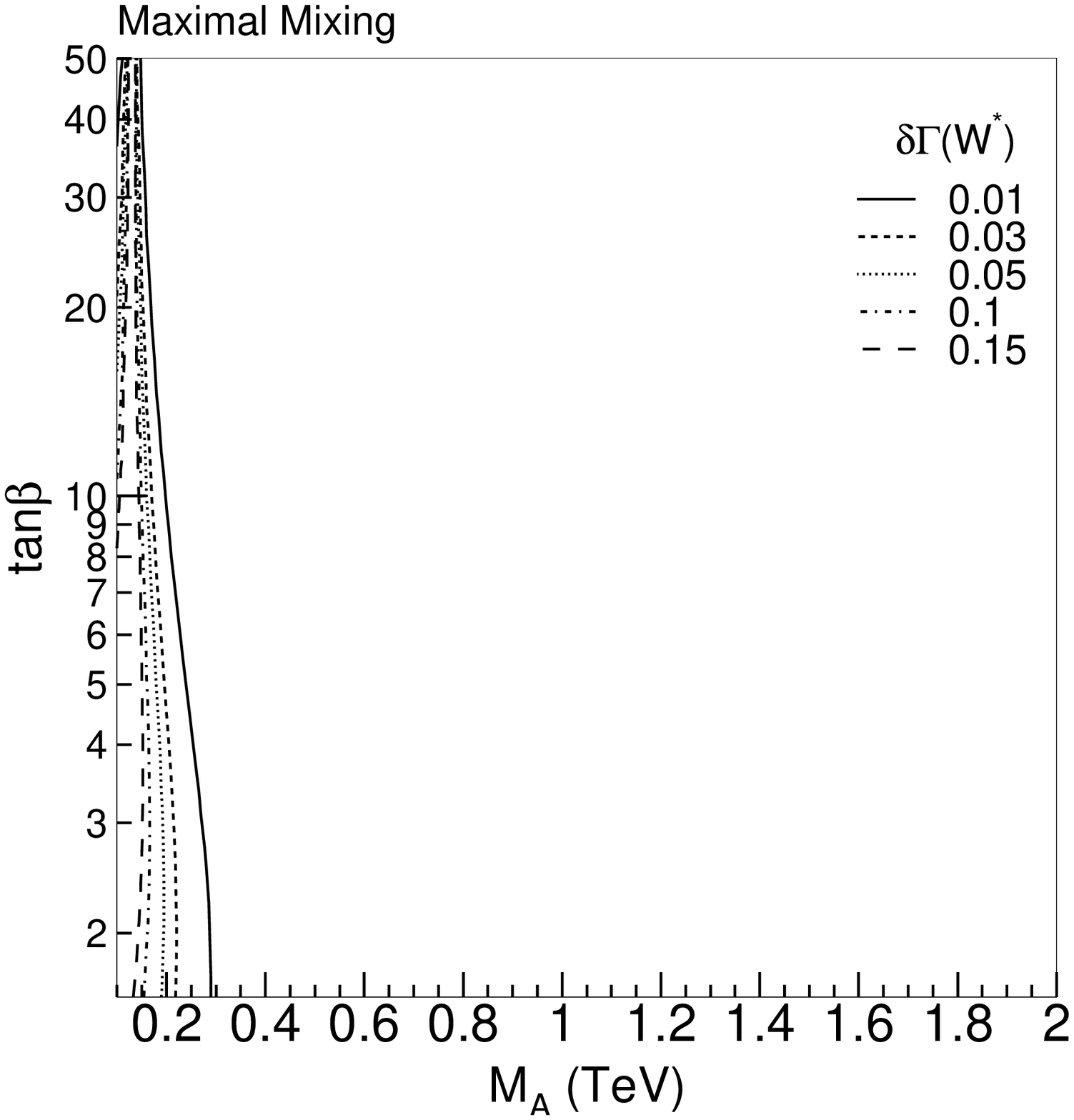}
\includegraphics*[19,142][529,682]{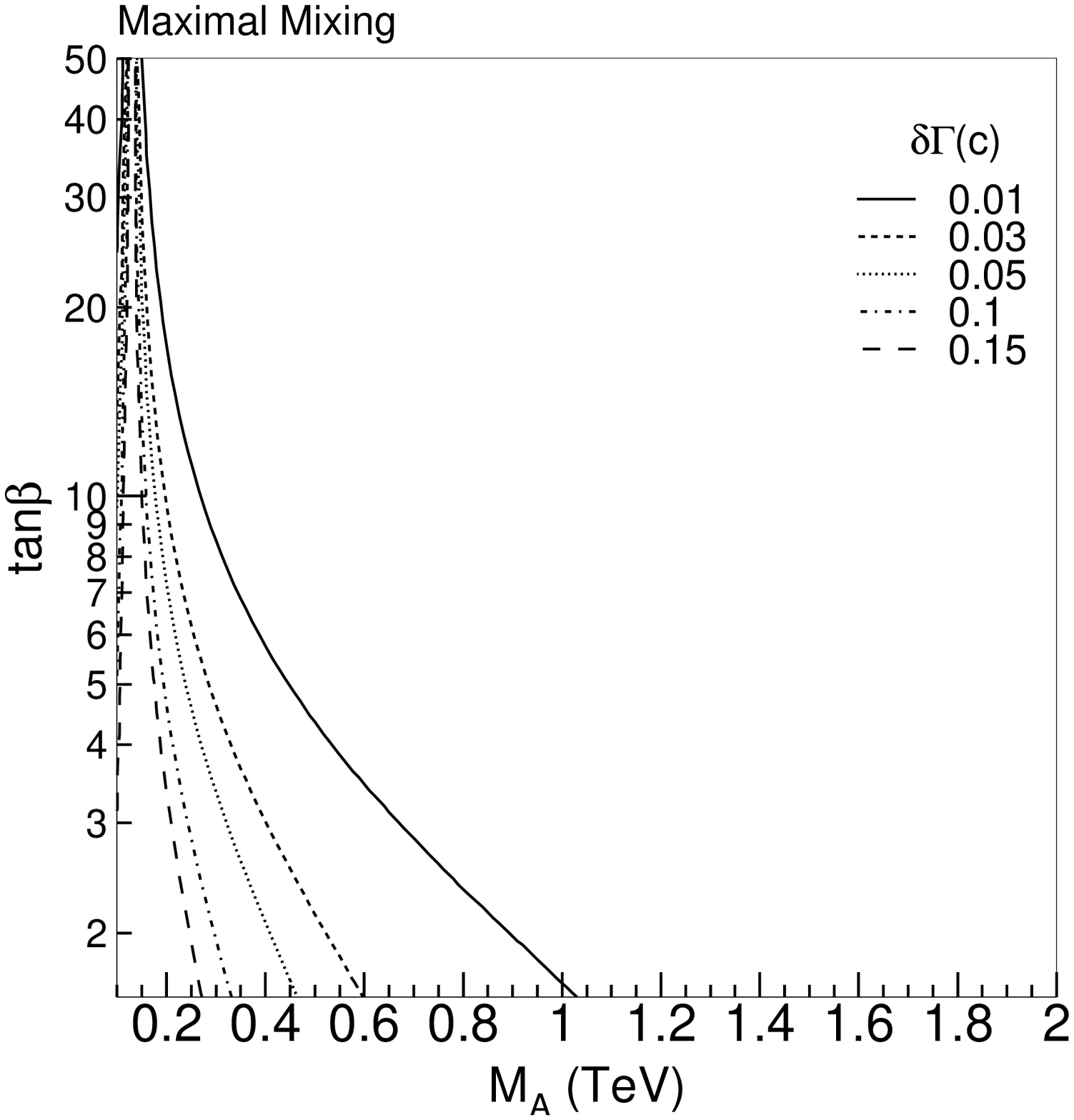}
}
\resizebox{\textwidth}{!}{
\includegraphics*[19,142][529,682]{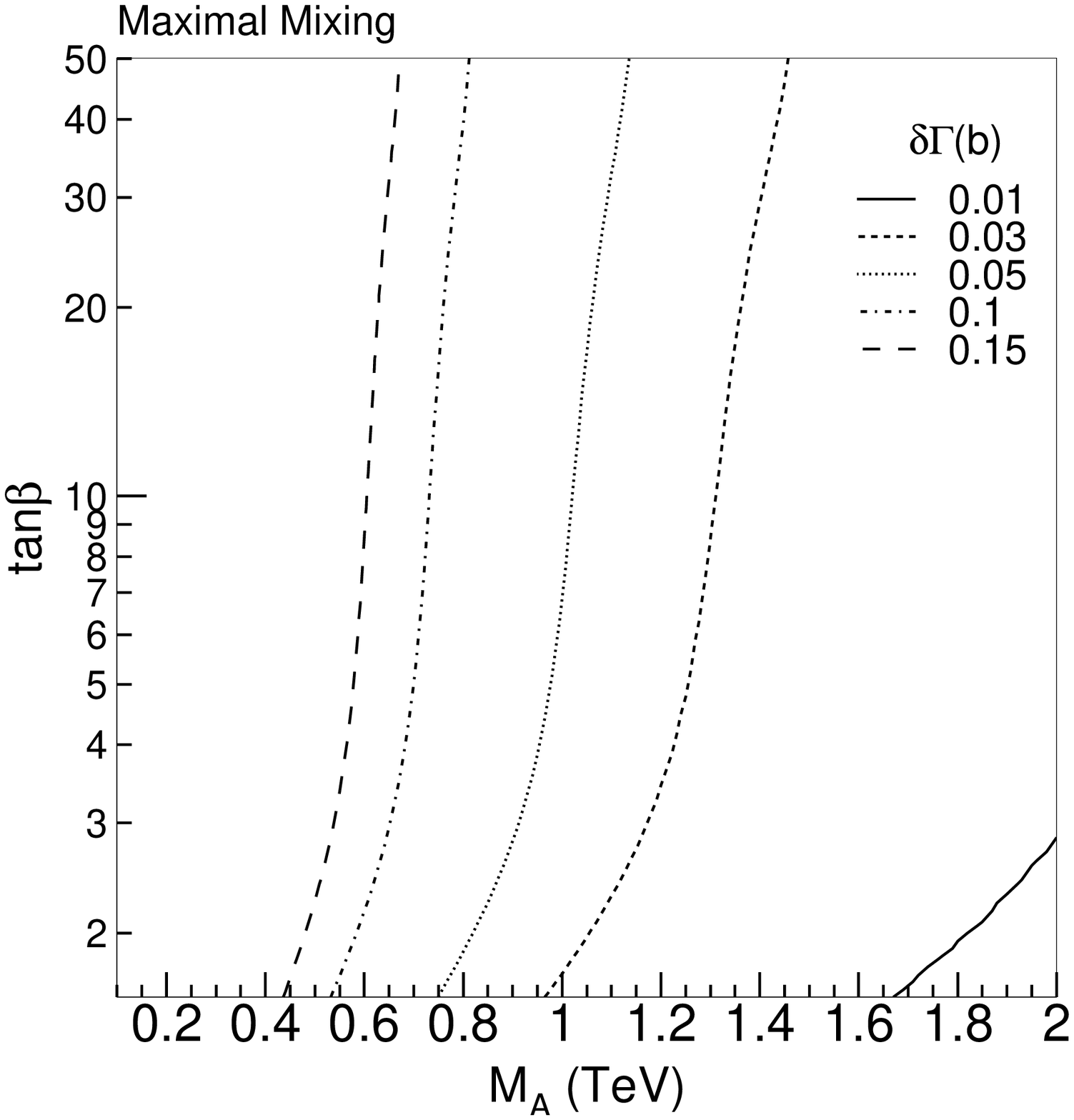}
\includegraphics*[19,142][529,682]{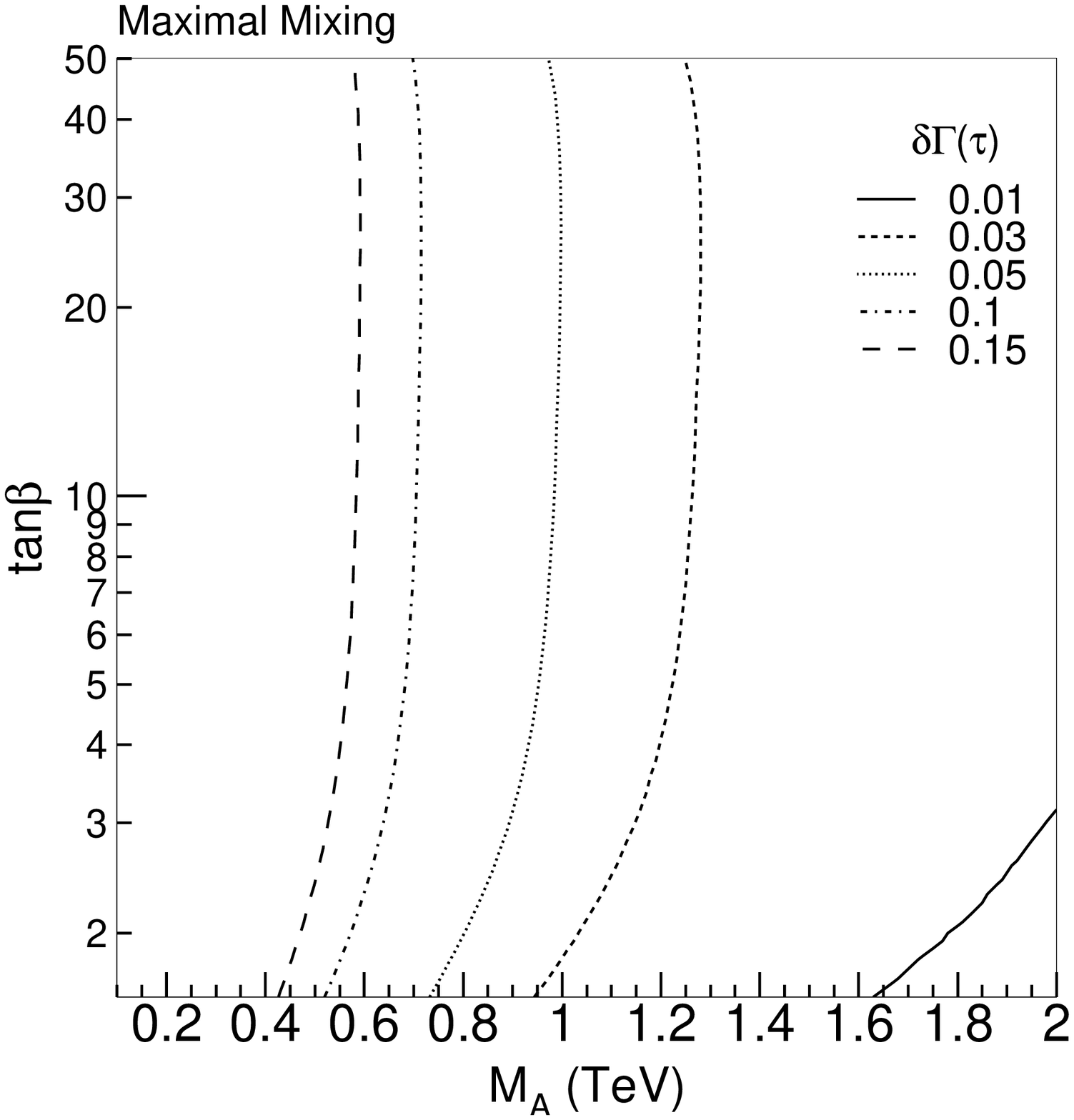}
}
\end{center}
        \caption{Deviations of Higgs partial widths from their SM values in
        the maximal-mixing scenario.}
        \label{fig:widths1}
\end{figure}

\begin{figure}
\begin{center}
\resizebox{\textwidth}{!}{
\includegraphics*[19,142][529,682]{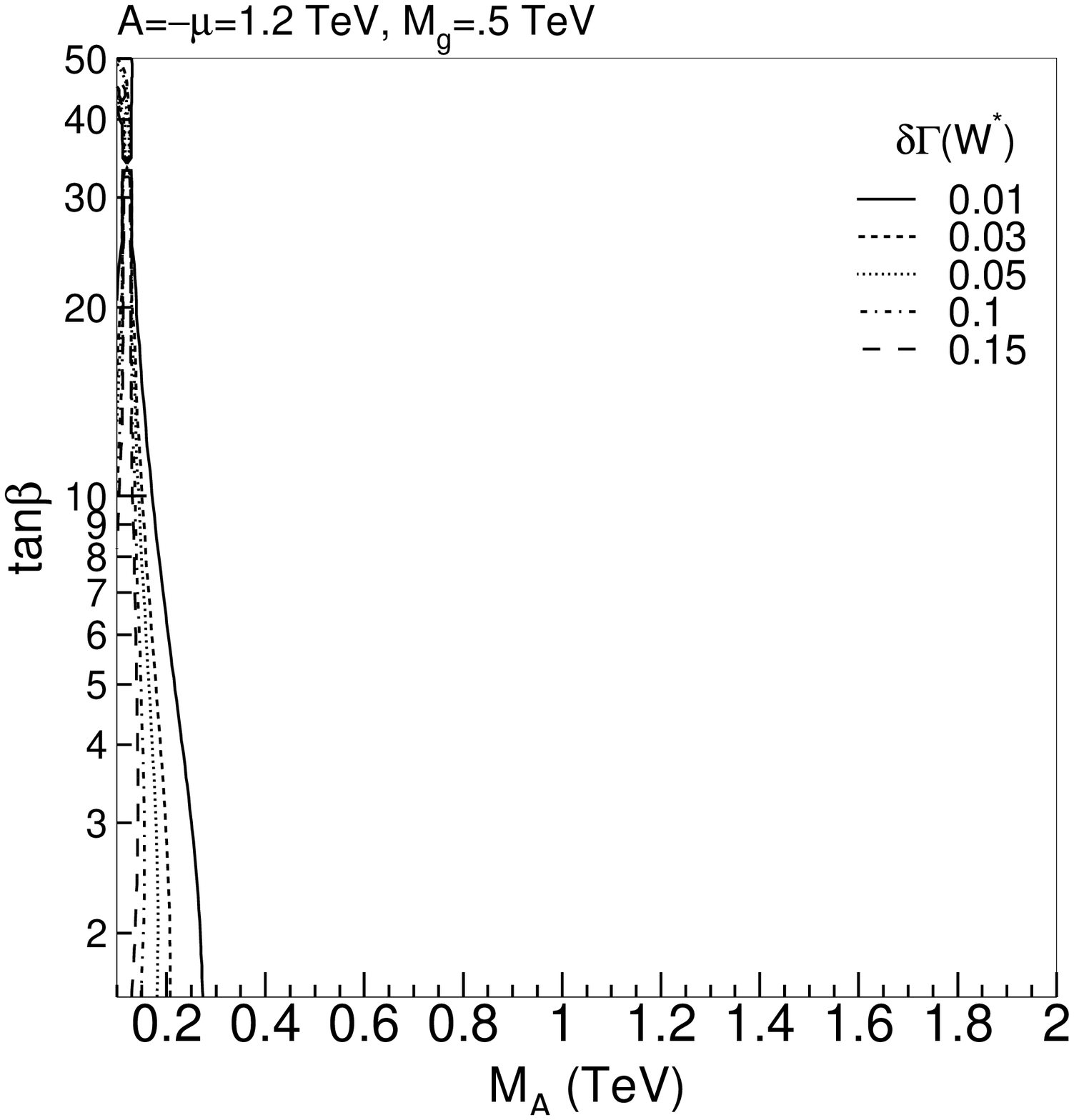}
\includegraphics*[19,142][529,682]{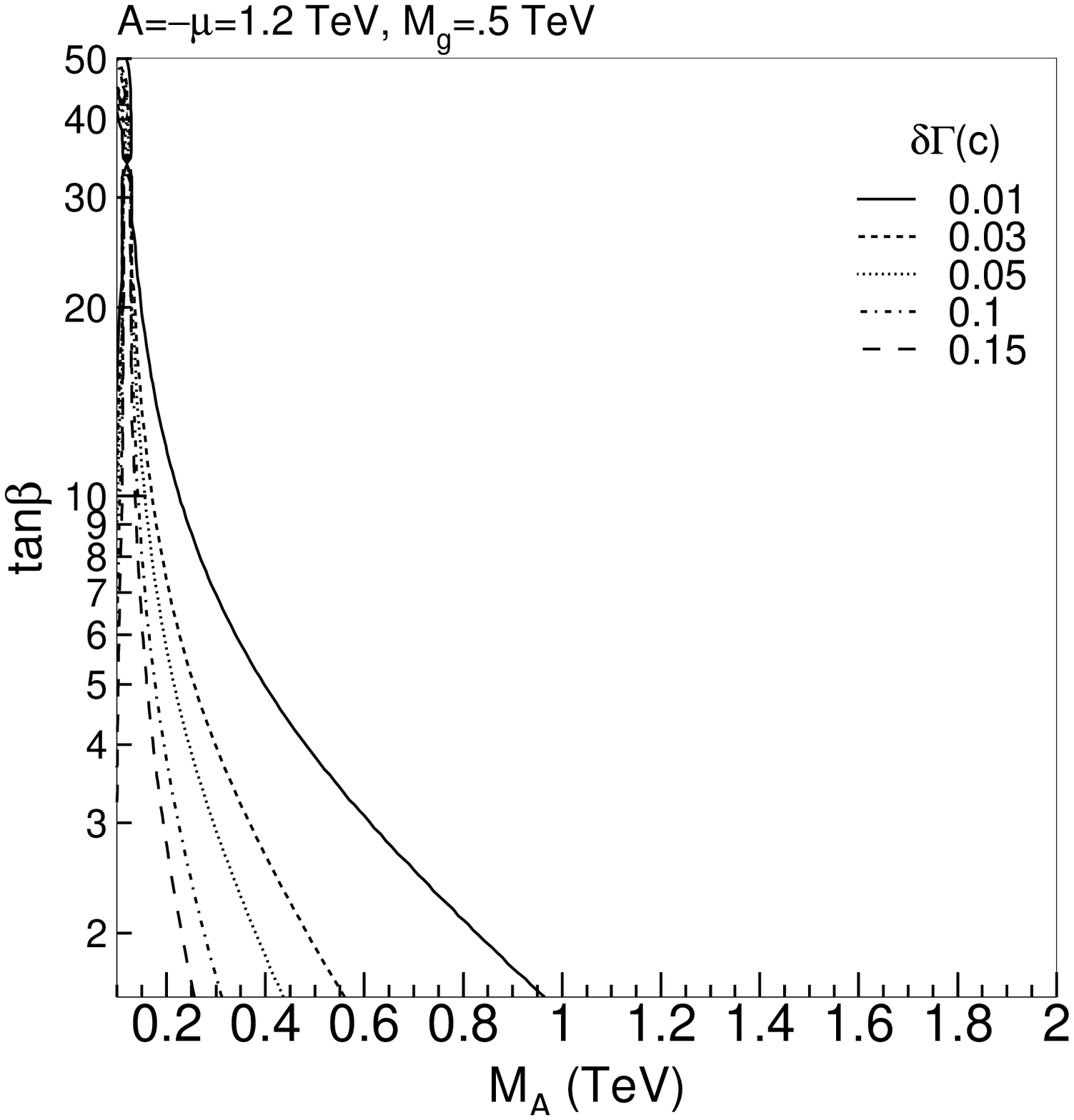}
}
\resizebox{\textwidth}{!}{
\includegraphics*[19,142][529,682]{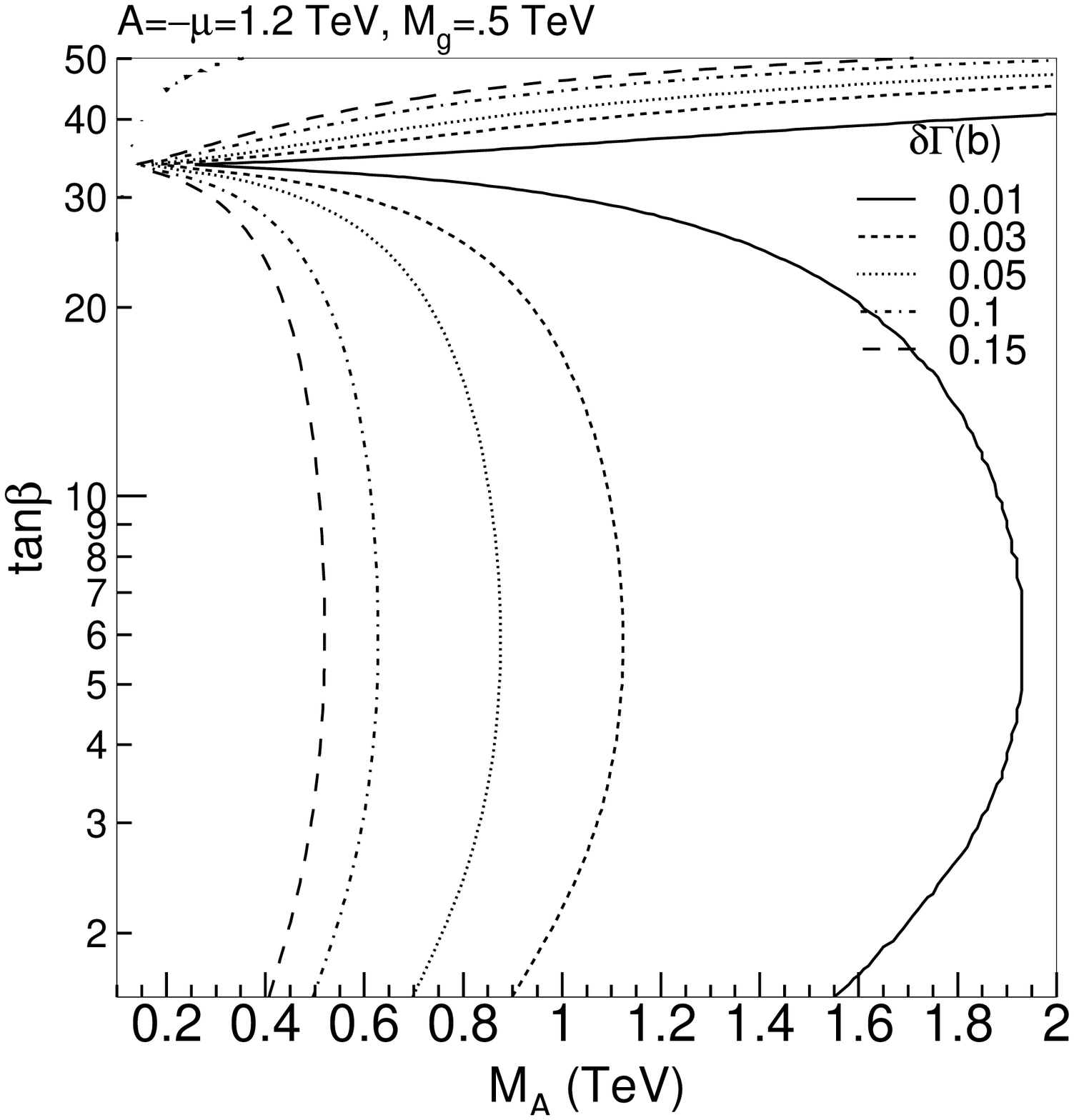}
\includegraphics*[19,142][529,682]{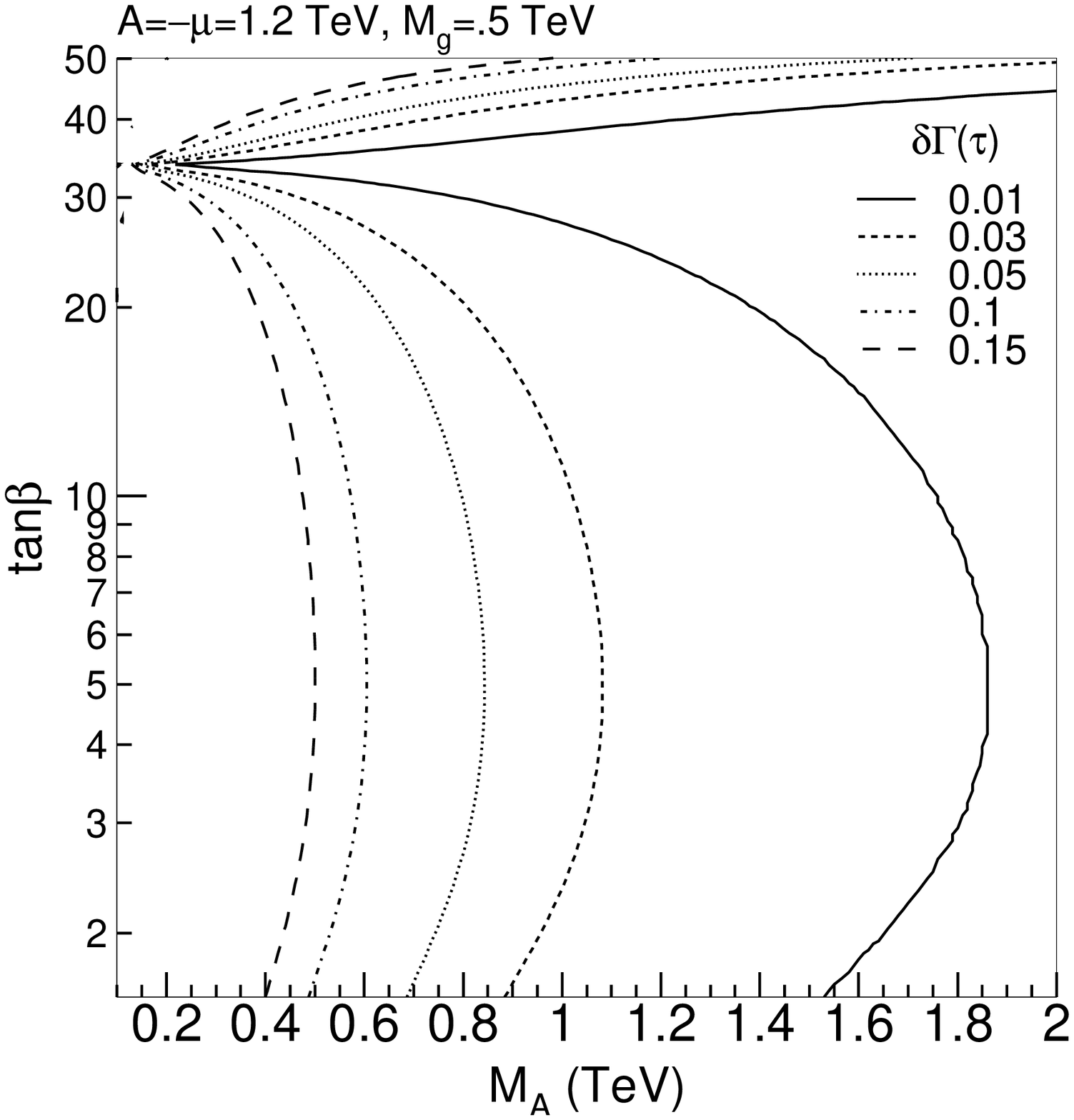}
}
\end{center}
        \caption{Deviations of Higgs partial widths from their SM values in
        the large $\mu$ and $A_t$ scenario, with $A_t = -\mu = 1.2$ TeV.}
        \label{fig:widthsmuAt}
\end{figure}

Consider first the couplings of the lightest
CP-even Higgs boson $h$ to vector bosons ($V=W$ or $Z$).  The
corresponding tree-level squared-coupling normalized to the SM value is:
\begin{equation}
{g^2_{hVV} \over g^{2}_{h_{\rm SM}VV} }= \sin^2(\beta-\alpha) \,.
\end{equation}
To a good approximation, the most important radiative correction to
this result can be incorporated by replacing the tree-level value of
$\alpha$ by its radiatively-corrected value [obtained in
Eq.~\ref{eq:alpha}].  In the decoupling limit, $\cos(\beta-\alpha)$ is
given by Eq.~\ref{cosbmadecoupling}.  It then follows that 
\begin{equation}
   {g^2_{hVV}\over g^2_{h_{\rm SM}VV}}\simeq 1-{c^2 m_Z^4\sin^2 4\beta\over
4m_A^4}\,.        \label{eq:g2wwdecoupling}
\end{equation}
At large $\tan\beta$, the approach to decoupling is even faster, since
$\sin 4\beta\simeq -4\cot\beta$ further suppresses the
deviation of the partial width
$\delta\Gamma(W) = |g^2_{hWW}/g^2_{h_{\rm SM}WW} -1|$.
Contours of $\delta\Gamma(W)$ for the maximal-mixing and large $A_t$ and $\mu$
scenarios are shown in the upper left panels of
Fig.~\ref{fig:widths1} and Fig.~\ref{fig:widthsmuAt}.
The behavior in the no-mixing scenario is quite similar, and is
therefore not shown here explicitly.

We next consider the couplings of $h$ to up-type fermions.  At
tree-level, we may write (using third-family notation):
\begin{equation}
{g^2_{htt}\over  g^{2}_{h_{\rm SM}tt}  }
        = \left[ \sin(\beta - \alpha) + \cot\beta \cos(\beta - \alpha)
        \right]^2,
%       \sin^2(\beta-\alpha) + \cot^2\beta\cos^2(\beta-\alpha)
%       +\cot\beta\sin 2(\beta-\alpha),
\label{eq:g2behavior}
\end{equation}
where the couplings are expressed in terms of $\sin(\beta - \alpha)$
and $\cos(\beta - \alpha)$ in order to better illustrate the decoupling
behavior.  Based on the discussion following Eq.~\ref{eq:couplingsDeltat},
we may neglect the effects of the SUSY vertex corrections.  Then, 
the dominant radiative corrections can be
incorporated simply by employing the radiatively-corrected value of
$\alpha$ in the tree-level coupling.  
Note that the approach to decoupling is significantly
slower [by a factor of $m_A^2/m_Z^2$] than in the case of the $hVV$ coupling
[Eq.~\ref{eq:g2wwdecoupling}].  For example, applying
Eq.~\ref{eq:g2behavior} to the Higgs coupling to charmed
quarks, one obtains
\begin{equation}
   {g^2_{hcc}\over g^2_{h_{\rm SM}cc}}\simeq 1+{c m_Z^2\sin 4\beta
\cot\beta\over m_A^2}\,.        \label{eq:g2ccdecoupling}
\end{equation}
At large $\tan\beta$, the approach to decoupling is faster due to the
additional suppression 
factor of $\cot^2\beta$ as in the case of the $hVV$ coupling.
The decoupling of $\delta\Gamma(c)$ as $m_A$ increases is exhibited in
the upper right panel of Figs.~\ref{fig:widths1} and \ref{fig:widthsmuAt}.
The behavior in the no-mixing scenario is again similar and is not shown.
In the three benchmarks considered, the $W$ and $c$ 
couplings to the light Higgs boson decouple quickly for increasing
$m_A$, with a somewhat slower decoupling for the $c$ quark in the low
$\tan \beta$ regime.  

Finally, we turn to the coupling of $h$ to down-type fermions,
and focus on the third-generation $b\bar b$ and $\tau^+\tau^-$ decay
modes.  The approach to the decoupling limit was given for $g_{hbb}$
in Eq.~\ref{hbbcorrection} [with a similar result for $g_{h\tau\tau}$
easily obtained].  For $m_A\gg m_Z$ (and neglecting $\delta h_b/h_b$
which is not $\tan\beta$-enhanced), it follows that
\begin{equation} 
  {g^2_{hbb}\over g^2_{h_{\rm SM}bb}}\simeq 1-{4c m_Z^2\cos 2\beta
\over m_A^2}\left[\sin^2\beta-{\Delta_b\over 1+\Delta_b}\right]\,.
        \label{eq:g2bbdecoupling}
\end{equation}
The approach to decoupling is again slower as compared to $g_{hVV}$.
However, in contrast to the previous two cases, there is no
suppression at large $\tan\beta$.  In fact, since
$\Delta_b\propto\tan\beta$, the approach to decoupling is further
delayed, unless $c\simeq 0$.  Thus, we expect the greatest
deviation from the SM in $\delta\Gamma(b)$ and
$\delta\Gamma(\tau)$, and this is confirmed
in the lower panels of Figs.~\ref{fig:widths1} and \ref{fig:widthsmuAt}.
As before, the behavior in the no-mixing scenario is quite
similar to that of the maximal-mixing scenario and is not shown here.
There is a small difference in the behavior of
$\delta \Gamma(b)$ with respect to  $\delta \Gamma(\tau)$
at large $\tan\beta$ due to the effect
of $\Delta_b$, as discussed in Sec.~\ref{sec:theory}.
%Such effects will always be present, and have often been
%ignored in the past.
Note that in the large $A_t$ and $\mu$ scenario,
Fig.~\ref{fig:widthsmuAt} exhibits the
$m_A$-independent decoupling phenomenon, corresponding to the case
where $c\simeq 0$.\footnote{In principle, this phenomenon could also
occur in the case of no-mixing or maximal-mixing.  However, in these
cases, since $\delta\mathcal{M}^2_{12}$ is quite small, the value of
$\tan\beta$ one obtains from Eq.~\ref{earlydecoupling} would be so  large
(way beyond what is plotted in Fig.~\ref{fig:widths1} and 
Fig.~\ref{fig:widthsmuAt}), that the $m_A$-independent decoupling
takes place in a  $\tan\beta$ region
that is no longer theoretically meaningful.}  
If the MSSM parameters are such that the $m_A$-independent 
decoupling is realized, then the 
experimental sensitivity to $m_A$ is greatly compromised.
The value of $\tan\beta$ at which this decoupling occurs 
[Eq.~\ref{earlydecoupling}] depends slightly on the
two possible sign choices for $\mu$ and $A_t$ (for a fixed value of $\mu A_t$)
through the dependence of
$\delta \mathcal{M}_{ij}^2$  on $h_t$ and $h_b$ (which depend
on $\Delta_t$ and $\Delta_b$, respectively).

\begin{figure}
        \begin{center}
        \resizebox{\textwidth}{!}{
        \includegraphics*[19,162][529,682]{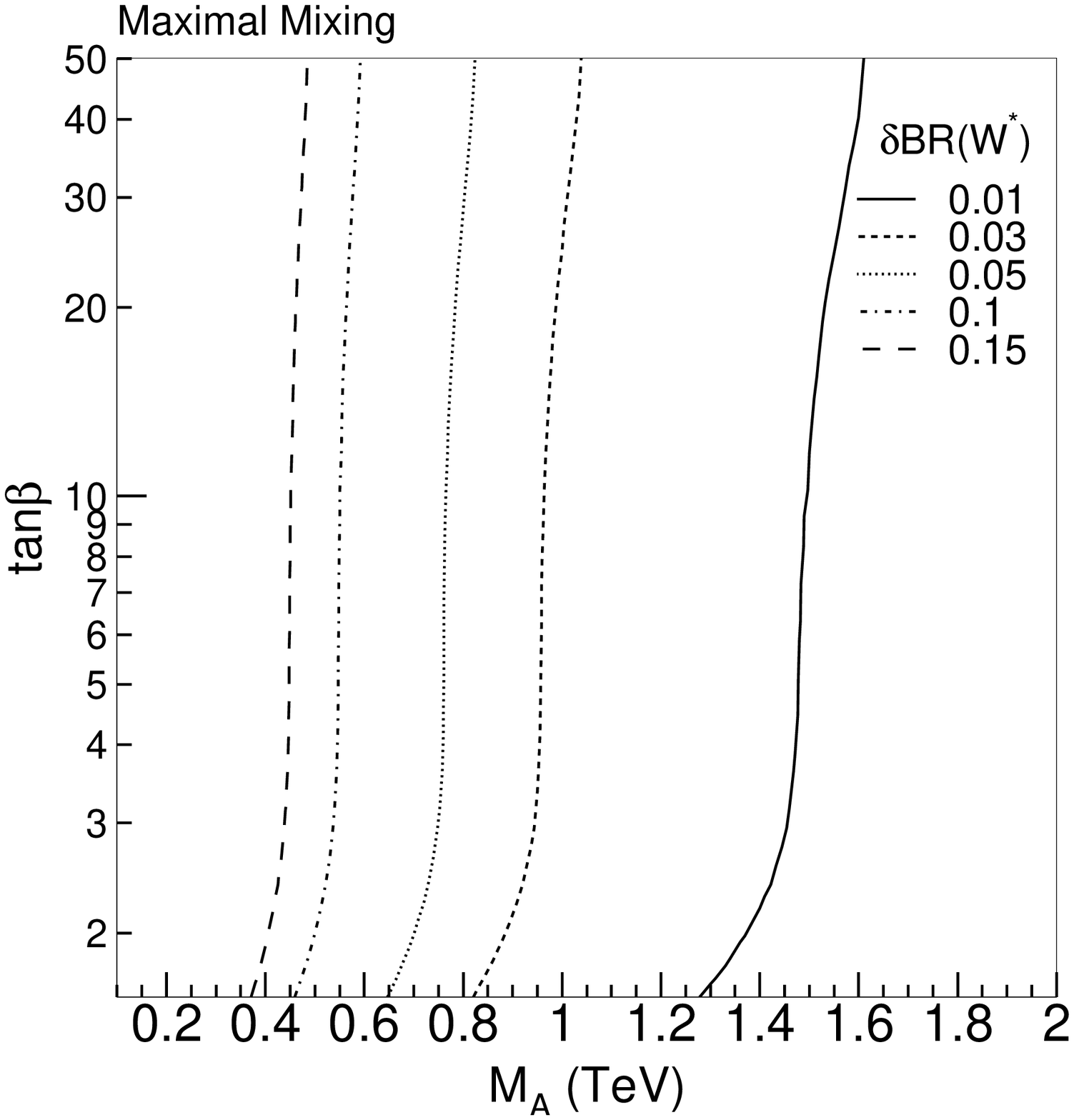}
        \includegraphics*[19,162][529,682]{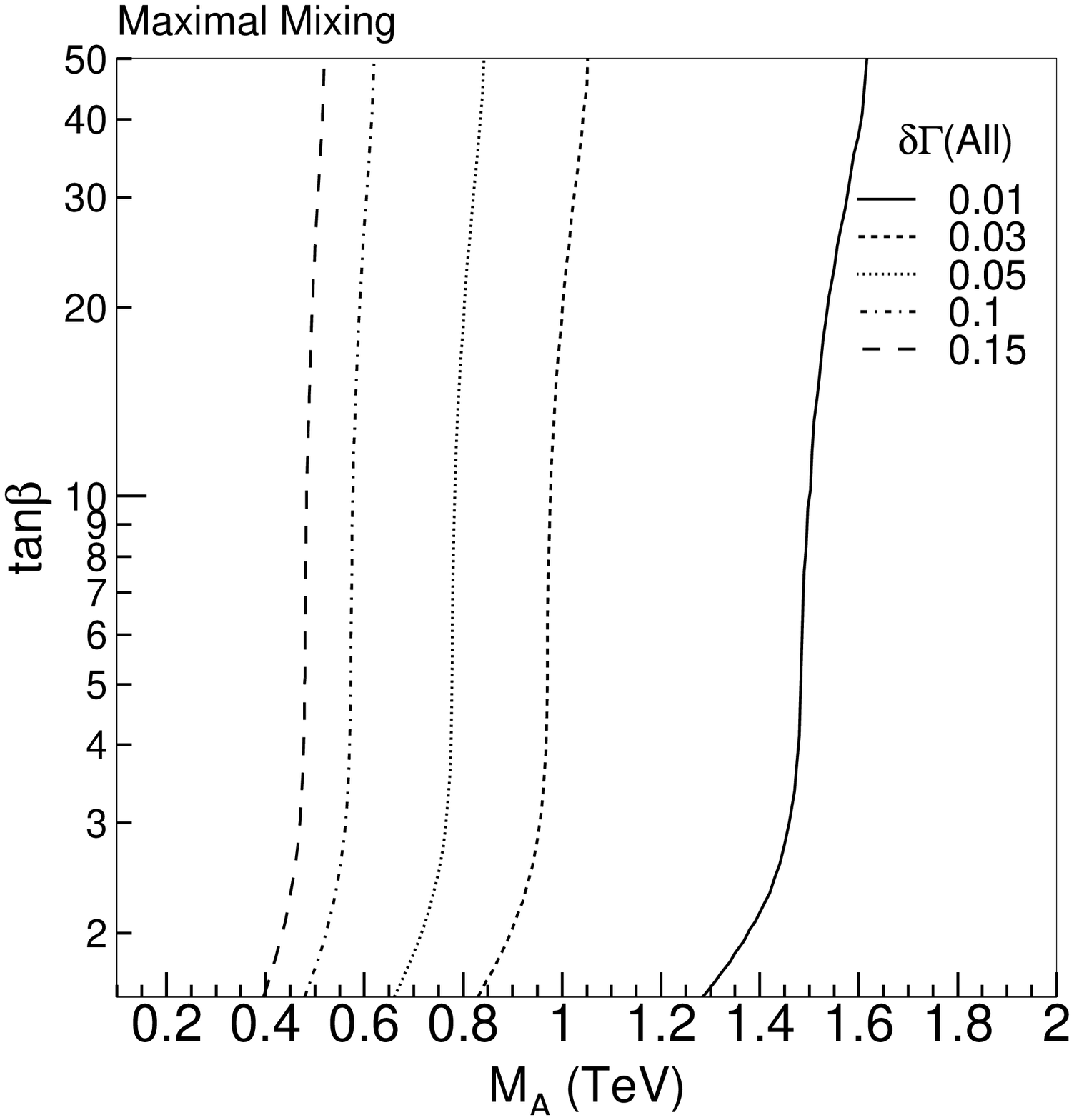}
}
        \end{center}
        \caption{Deviations of BR($W$) and $\Gamma_{\rm tot}$ from
        their SM values in the maximal-mixing scenario.}
        \label{fig:widths2}
\end{figure}

So far, we have discussed the sensitivity of the Higgs couplings to
the MSSM parameters.  However, experiments at the LC
will also measure the Higgs branching ratios.  Since
$h \to b\bar b$ is the dominant decay mode of a Higgs boson
lighter than about 135 GeV (unless the $h b \bar b$ coupling is anomalously
suppressed),
%because of the relative size of $m_b$ and the phase space suppression of
%$WW^*$ and $ZZ^*$ decays.
%Because
$\Gamma(b)$ dominates
the total Higgs width.  Thus, $\BR(b)$
is not as sensitive to deviations of $\Gamma(b)$ from its SM
value as the BRs for other decay modes.
In particular, since $\Gamma(W)$ quickly
approaches its SM value, $\delta \BR(W) \simeq \delta \Gamma_{\rm tot}$
almost independently of the value of $\tan\beta$.
This is illustrated in Fig.~\ref{fig:widths2} for the case of maximal-mixing,
and it is generically true in the other benchmark scenarios.

\subsection{Loop induced couplings}

The decay modes $h \to gg$ and $h \to \gamma \gamma$ proceed only at
the loop level.  In the MSSM, SUSY particles and additional Higgs
bosons also run in the loops.  Thus the deviations of $\Gamma(g)$ and
$\Gamma(\gamma)$ from their SM values depend not only on deviations in
the fermion and $W$ pair couplings to $h$ but also on
SUSY loop contributions.  $\Gamma(g)$ depends mainly
on the $t$ quark loop, which has a SM-like coupling for large
$m_A$ except at small $\tan\beta$.  $\Gamma(\gamma)$ depends mainly on
the $W$ boson loop, which has a very rapid decoupling behavior
with $m_A$.  Note that, in all the scenarios considered, deviations in the 
$t$ and $c$ couplings are significant at low $\tan\beta$, and deviations
in the $b$ coupling are significant for all $\tan\beta$ values
[see Figs.~\ref{fig:widths1} and~\ref{fig:widthsmuAt}].

For the loop-induced Higgs couplings, there are two separate decoupling
limits of relevance.  In the first decoupling limit, discussed often
in this paper, the non-zero tree-level Higgs couplings 
approach their SM values for $m_A\gg m_Z$.\footnote{In principle, 
deviations can arise from radiative corrections due to loops of SUSY
particles, but these corrections will be a small fraction of the
corresponding SM tree-level Higgs coupling.}  
The second decoupling
limit applies to loop-induced Higgs couplings in the limit of large
SUSY particle masses.  In this limit, the effects of the SUSY-loops
decouple, and the loop-induced Higgs couplings are determined by loops
of SM particles.  If we now take $m_A\gg m_Z$ so that the Higgs
couplings to SM particles approach their SM limits, then the resulting
loop-induced Higgs coupling should likewise approach its SM limit.
However, suppose that $m_A\gg m_Z$ but the masses of
some of the SUSY particles that contribute to the
loop-induced Higgs couplings are not too large.  In this case, the resulting
loop-induced Higgs couplings will deviate from the corresponding SM
values due to the SUSY loop 
contributions, which can be a sizable fraction of the SM loop contributions.

%In general, the behavior of the
%one-loop amplitudes is complicated and depends on several MSSM
%parameters.  
%Naively, one does not expect to observe large variations
%from the SM predictions for the partial widths, because of the
%decoupling properties discussed earlier.  

\begin{figure}[!ht]
\begin{center}
\resizebox{\textwidth}{!}{
\includegraphics*[19,142][529,682]{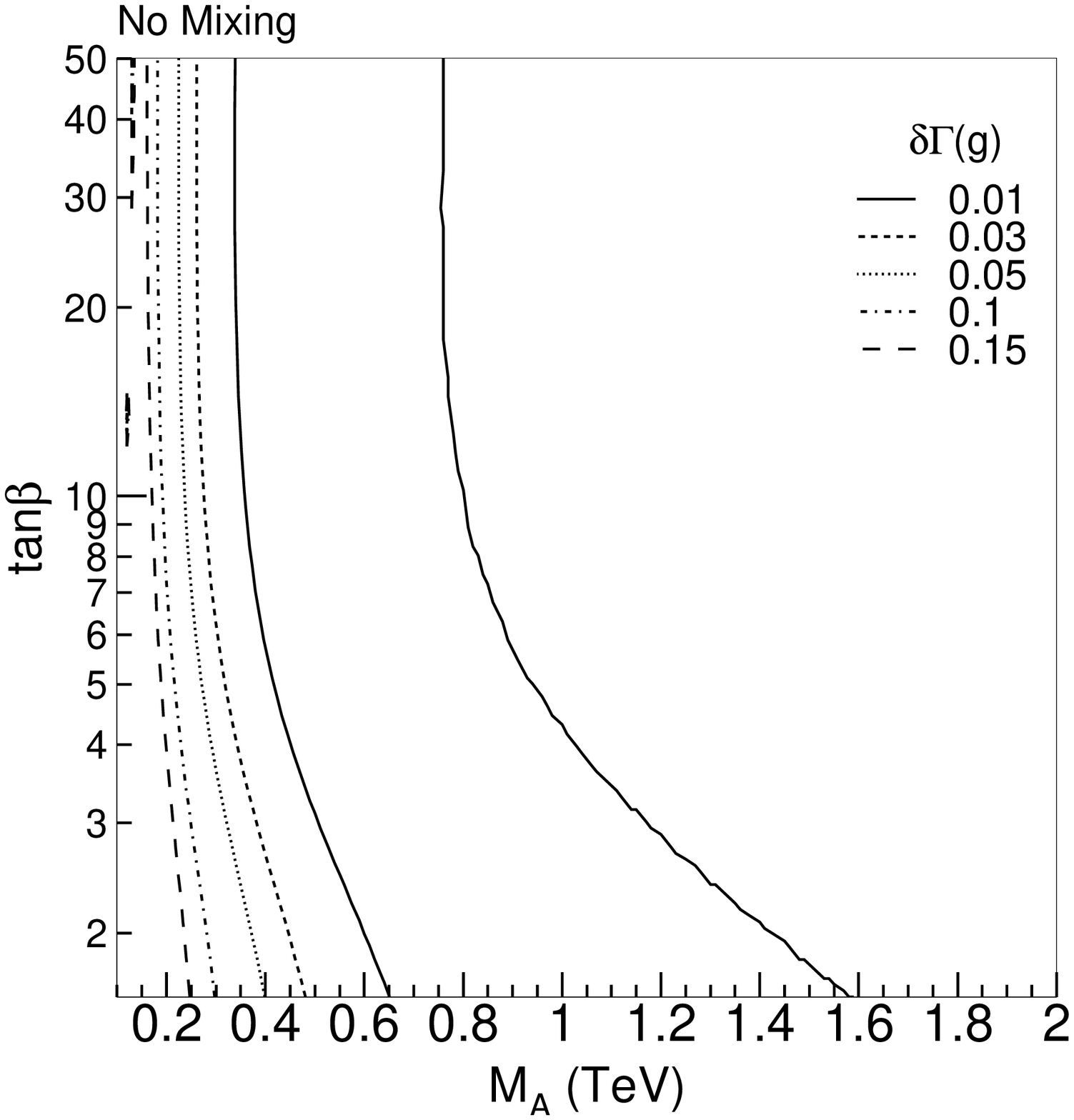}
\includegraphics*[19,142][529,682]{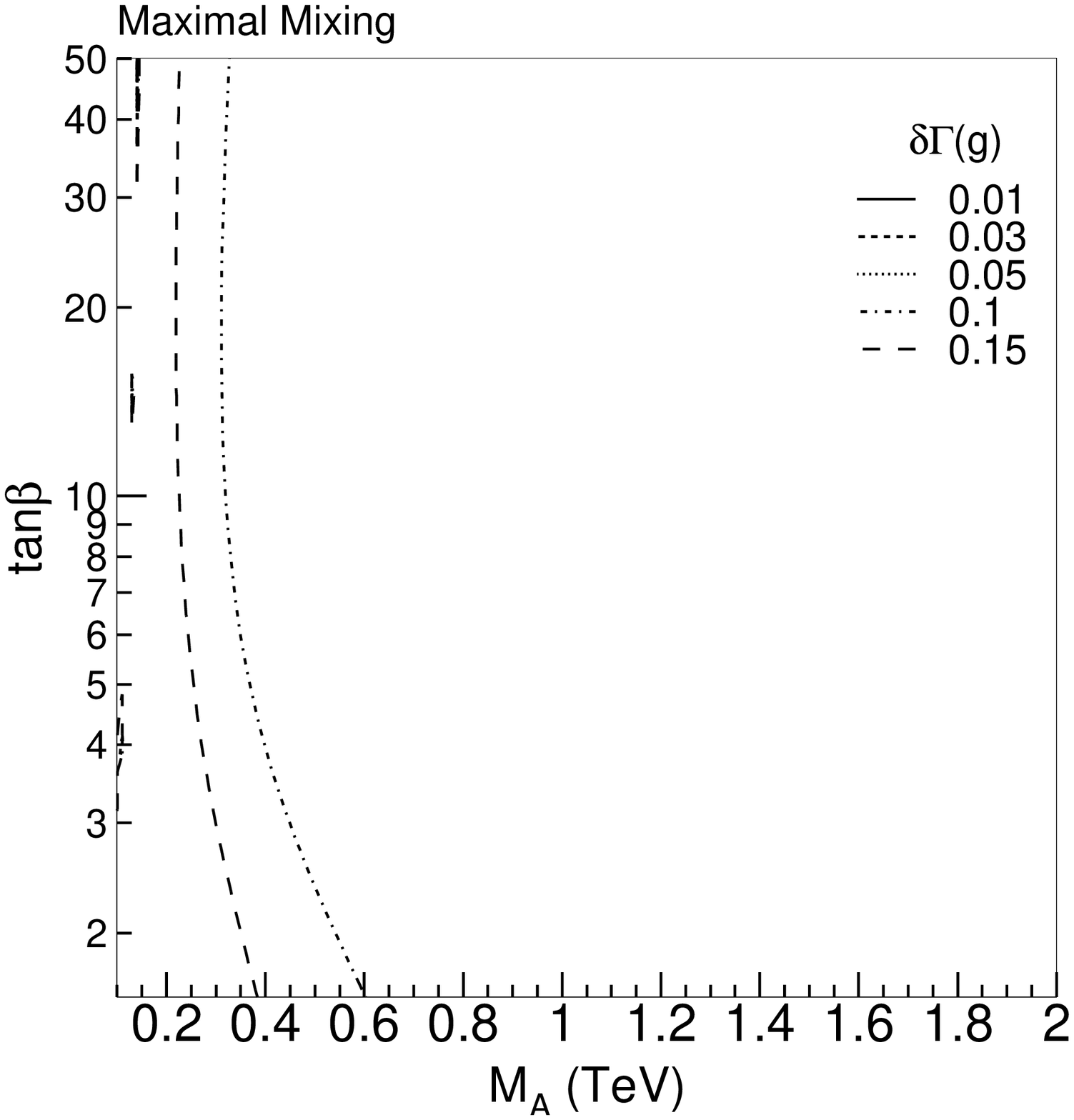}
}
\resizebox{\textwidth}{!}{
\includegraphics*[19,142][529,682]{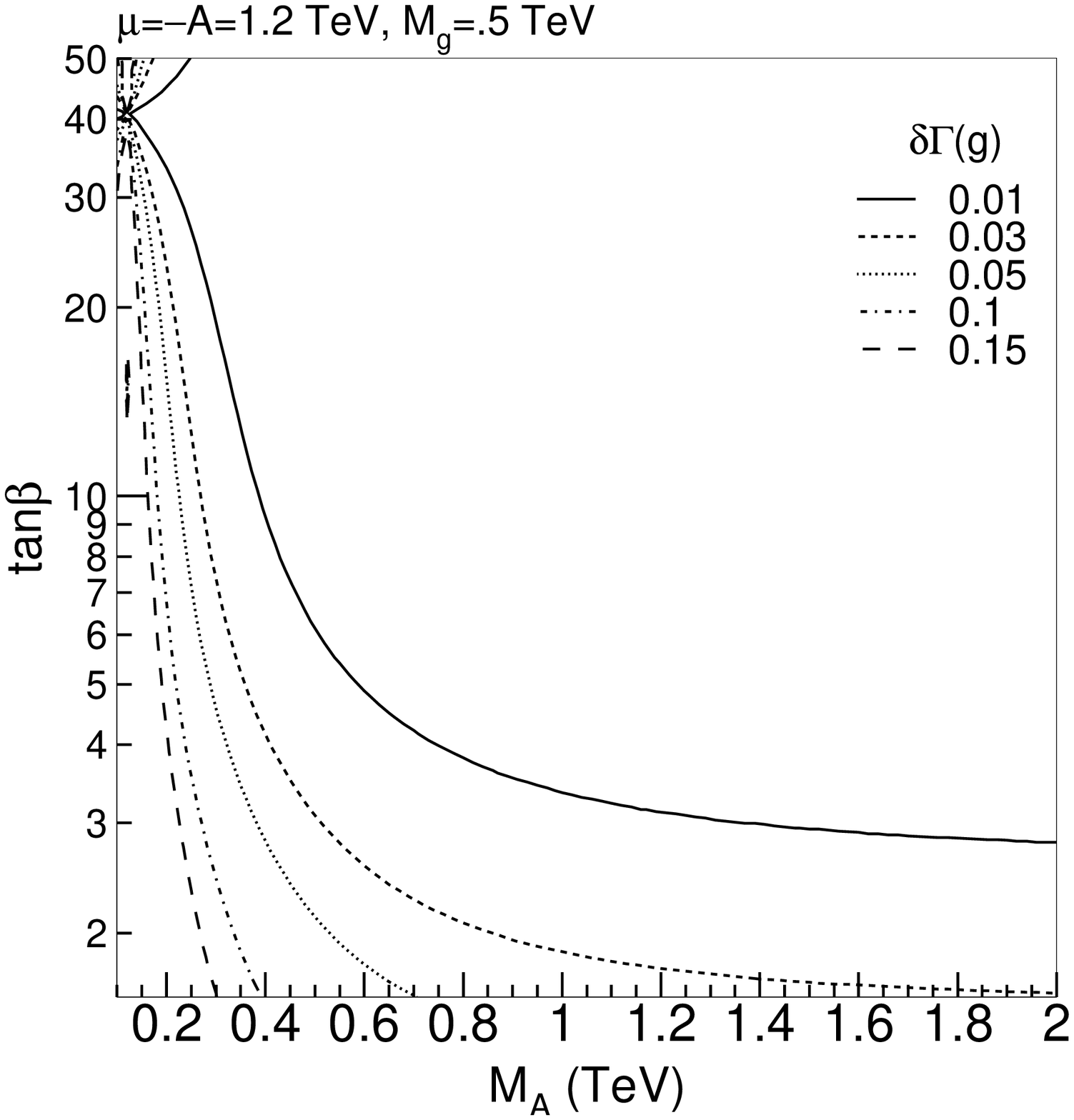}
\includegraphics*[19,142][529,682]{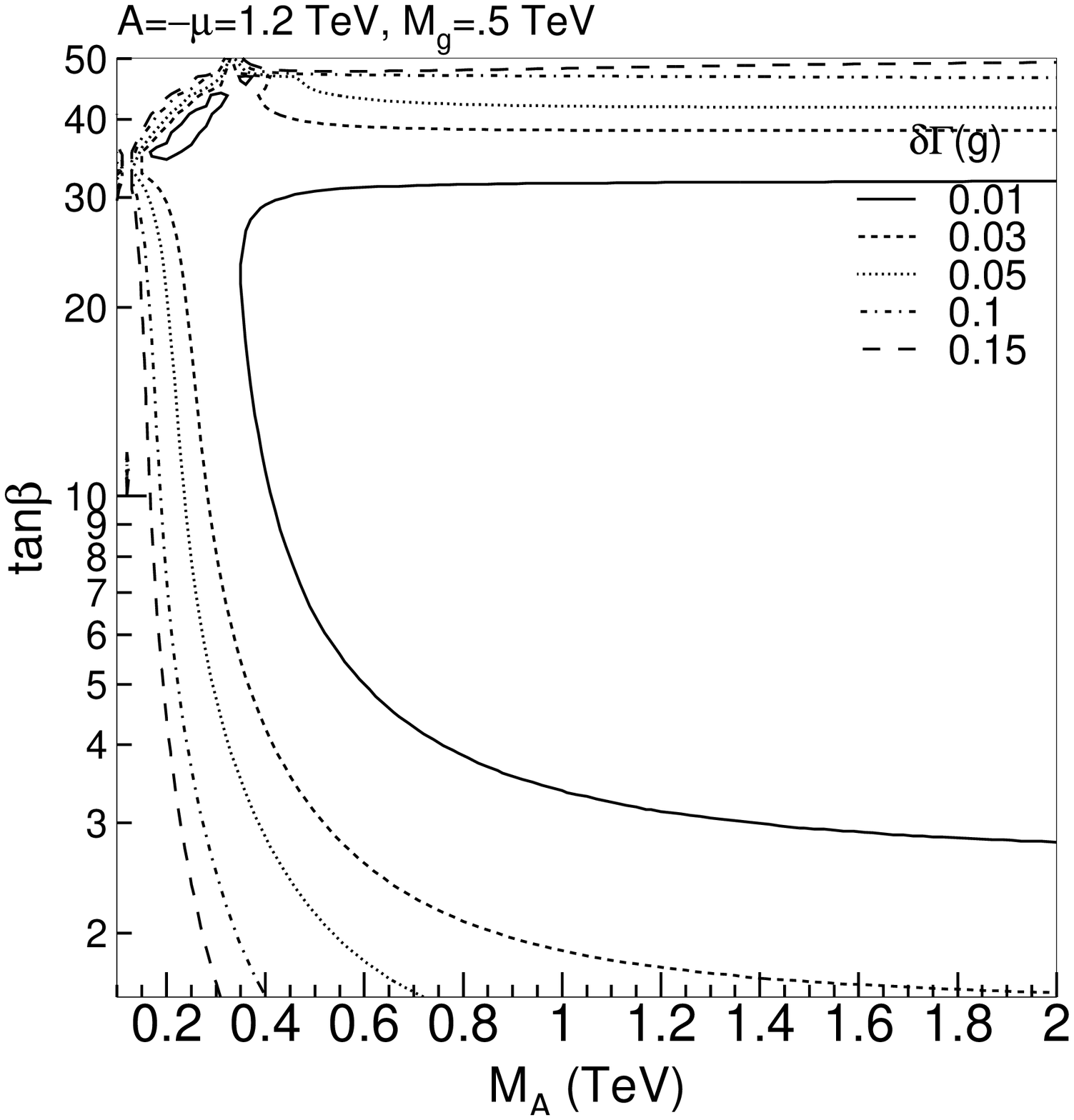}
}
\end{center}
        \caption{Deviations of the partial width $\Gamma(g)$ from its
        SM value in the 
no-mixing scenario (top left),
the maximal-mixing scenario (top right),
and the large $\mu$ and $A_t$ scenario with
$\mu = -A_t = 1.2$ TeV (bottom left) and
$\mu = -A_t = -1.2$ TeV (bottom right).
In the no-mixing scenario, $\Gamma(g)_{\rm MSSM}-\Gamma(g)_{\rm SM}$
changes sign and thus passes through zero [{\it i.e.}, $\delta\Gamma(g)$=0]  
along a contour between the two $\delta\Gamma(g)=0.01$ 
contours (solid lines) exhibited in the top left panel.
}  
        \label{fig:Gamma_gg}
\end{figure}

In the decay $h \to gg$, 
the bottom and charm quark
contributions  destructively interfere with
the top quark contribution, reducing the SM amplitude by several percent.
Additional deviations arise from squark contributions.
The top squark couplings
to $h$ are
\begin{equation}
        g_{h \tilde t_{1,2} \tilde t_{1,2}}=
          g_D- \frac{gm_t^2 \cos\alpha}{m_W \sin\beta} 
         \mp \frac{gm_t}{2m_W \sin\beta}
        \left( \mu \sin\alpha + A_t \cos\alpha \right)
        \sin 2 \theta_{\tilde t}\,,
        \label{eq:hstopstop}
\end{equation}
where $g_D$ arises from the so-called $D$-term contribution to the
scalar potential.\footnote{Explicitly, $g_D\equiv gm_Z
(1 \mp (\nicefrac{8}{3} \sin^2 \theta_W - 1) \cos 2 \theta_{\tilde t})
\sin(\alpha + \beta)/ 4 \cos\theta_W$.  This
term is independent of $m_t$ and gives rise to a
small coupling of $h$ to all squark species,
independent of the corresponding quark Yukawa coupling,
so that the squarks of
the first two generations also contribute to $h \to gg$.
Thus, in our numerical
calculations we must specify the masses of the squarks of the first two
generations; we set $M_{\tilde Q} = M_{\tilde U} = M_{\tilde D} = 1$ TeV
in all cases, as mentioned before.  With this choice, the contributions
of the first two generations of squarks are negligible.}
In addition, the minus (plus) sign in Eq.~\ref{eq:hstopstop} 
corresponds to $h \tilde t_1 \tilde t_1$
($h \tilde t_2 \tilde t_2$), and
$\sin 2 \theta_{\tilde t} = 2 m_t X_t /(M_{\tilde t_1}^2 - M_{\tilde
t_2}^2)$,
where $M_{\tilde t_1} < M_{\tilde t_2}$.
The top-squark contribution to the $h \to gg$ amplitude behaves as
$g_{h \tilde t_i \tilde t_i}/M^2_{\tilde t_i}$ for large top squark
mass, where the 
$1/M^2_{\tilde t_i}$ suppression is due to the loop integral.  This
suppression can be partially compensated by a large
$h \tilde t_i \tilde t_i$ coupling.
In the decoupling limit 
[where $\cos (\beta - \alpha) \simeq 0$], it follows
that 
$\mu \sin\alpha + A_t \cos\alpha = \sin\beta X_t$.  Thus if $X_t$
is large, the third term in Eq.~\ref{eq:hstopstop} gives a contribution
to the $h \to gg$ amplitude proportional to
$m_t X_t \sin 2 \theta_{\tilde t}/M_{\tilde t_i}^2$ [$i$ = 1,2].
%the $\sin 2 \theta_{\tilde t}$ factor indicates that this contribution
%which is large only if there is significant mixing between the top squarks.
Note that if $|X_t| \sim M_{\tilde t_i}$ and the diagonal elements of 
the top squark squared-mass matrix are of the same order, then 
$|\sin 2 \theta_{\tilde t}| \simeq 1$
and this contribution to the amplitude decouples like
$m_t/M_{\tilde t_i}$, {\it i.e.}, suppressed by one power of $M_{\tilde t_i}$.  
This should be contrasted with the case of small top-squark mixing
angle ({\it i.e.}, $|\sin 2\theta_{\tilde t}|\ll 1$), which arises
when $m_t X_t$ is small compared to the difference of the 
diagonal entries of the squark squared-mass matrix.  
In this case, the dominant
contribution to the $h \tilde t_i \tilde t_i$ couplings comes from the second
term in Eq.~\ref{eq:hstopstop}, and
the top squark loop contribution
to the $h \to g g$ amplitude decouples like $m_t^2/M^2_{\tilde t_i}$.
This behavior of the top squark contribution with $X_t$ explains the 
behavior of $\Gamma(g)$ in the no-mixing and maximal-mixing scenarios
(the top panels of Fig.~\ref{fig:Gamma_gg}).  
At low $m_A$, the $h b \bar b$ coupling is enhanced over its SM value so that 
$\Gamma(g)$ is suppressed due to the destructive interference between 
the bottom quark and top quark loops.
In the no-mixing scenario, the top squark contribution enters with the 
same sign as the dominant top quark loop; as $m_A$ increases, the $b$
quark contribution approaches its SM value and the top squark contribution
then leads to a small enhancement of $\Gamma(g)$ at large $m_A$.
Thus in the no-mixing scenario (the top left panel of Fig.~\ref{fig:Gamma_gg})
we see that $\Gamma_{\rm MSSM}(g)-\Gamma_{\rm SM}(g)$ 
passes through zero for $m_A \sim 0.6\,$--$\,1$~TeV
and reaches an asymptotic value of $1$--$2$\% in the large $m_A$ limit.
In the maximal-mixing scenario, the top squark loop contribution is enhanced
by the large value of $X_t$ and enters with the opposite sign as the 
dominant top quark loop; thus it leads to a further suppression of $\Gamma(g)$
in addition to the suppression due to the enhanced $hb\bar b$ coupling 
at low $m_A$.  As $m_A$ increases, then, the $b$ quark contribution
approaches its SM value and $\delta \Gamma(g)$ reaches an asymptotic value
of about 6\% in the large $m_A$ limit, 
without  $\Gamma_{\rm MSSM}(g)-\Gamma_{\rm SM}(g)$ 
passing through zero at any value of $m_A$.

In the large $\mu$ and $A_t$ scenarios, the behavior of $\Gamma(g)$ is more
complicated.  At low $\tan\beta$, 
the behavior of $\delta \Gamma(g)$ is dominated by the decoupling behavior
of the $h t \bar t$ coupling (see Fig.~\ref{fig:widthsmuAt}).
At large $\tan\beta$, $m_A$-independent decoupling effects and $\Delta_b$
play an important role.
For $\mu<0$ (shown in the bottom right panel of Fig.~\ref{fig:Gamma_gg}),
$\delta \Gamma(g)$ is quite significant at large $\tan\beta$, 
and the deviations from the SM in this region are independent of $m_A$
for $m_A \gsim 0.5$ TeV.  
For this sign of $\mu$, $\Delta_b$ is large and
negative (the two terms in Eq.~\ref{eq:Deltab} enter with the same sign),
%Because $\Delta_b$ enters non-linearly in the $b$ quark Yukawa coupling
%as $1/(1 + \Delta_b)$, in this scenario $\Delta_b$ leads to 
resulting in a significant
enhancement of $h_b$ in the large $\tan\beta$ region.  
Because of this enhancement of $h_b$, the bottom squark loop contributions 
are significant in this scenario at large $\tan\beta$; they modify the
$h \to gg$ amplitude by a few percent, leading to several percent deviations
in $\Gamma(g)$.
At lower values of $m_A$, the $b$ quark loop also contributes due to
the deviation in the $h b \bar b$ coupling (see Fig.~\ref{fig:widthsmuAt}).
For $\mu>0$ (the bottom left panel of Fig.~\ref{fig:Gamma_gg}),
$\delta \Gamma(g)$ exhibits $m_A$-independent decoupling for 
$\tan\beta \simeq 40$.  For this sign of $\mu$, the two terms contributing
to $\Delta_b$ in Eq.~\ref{eq:Deltab} enter with opposite signs, leading
to a partial cancellation, and the overall sign of $\Delta_b$ is positive.  
Thus $\Delta_b$ leads to a small suppression of $h_b$, and the bottom 
squark loop contribution is not significant.

\begin{figure}[!ht]
\begin{center}
\resizebox{\textwidth}{!}{
\includegraphics*[19,142][529,682]{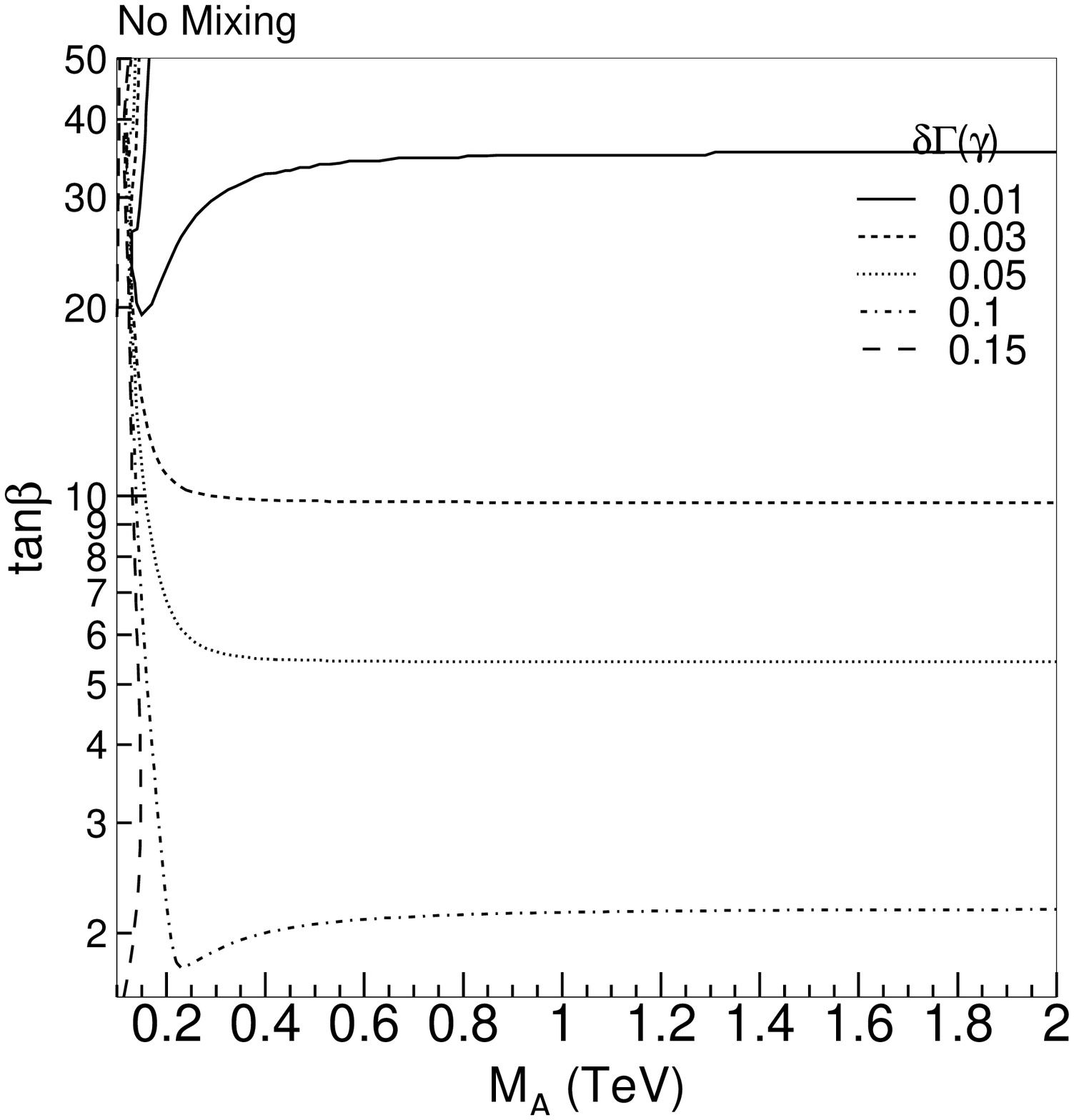}
\includegraphics*[19,142][529,682]{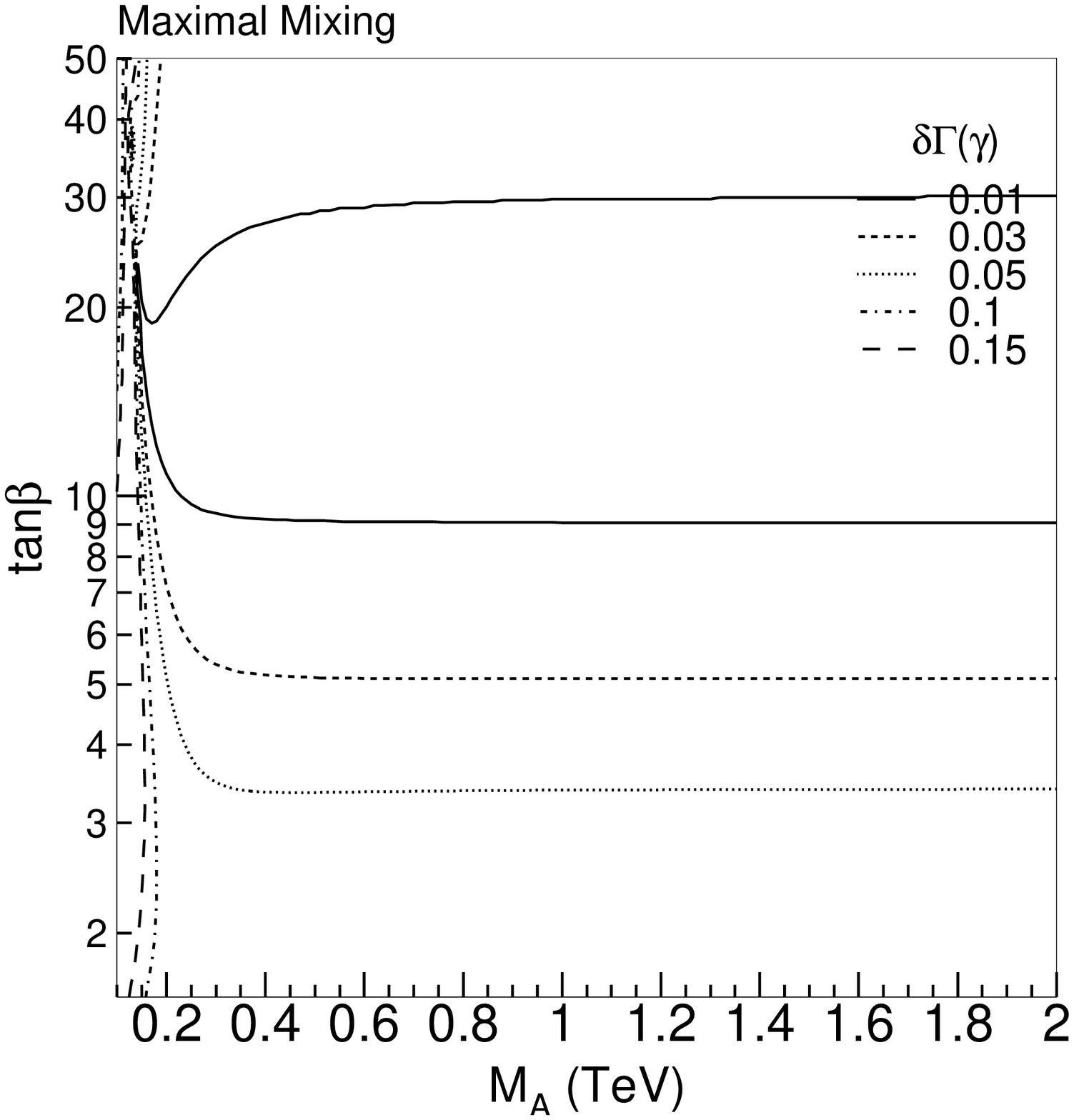}
}
\resizebox{\textwidth}{!}{
\includegraphics*[19,142][529,682]{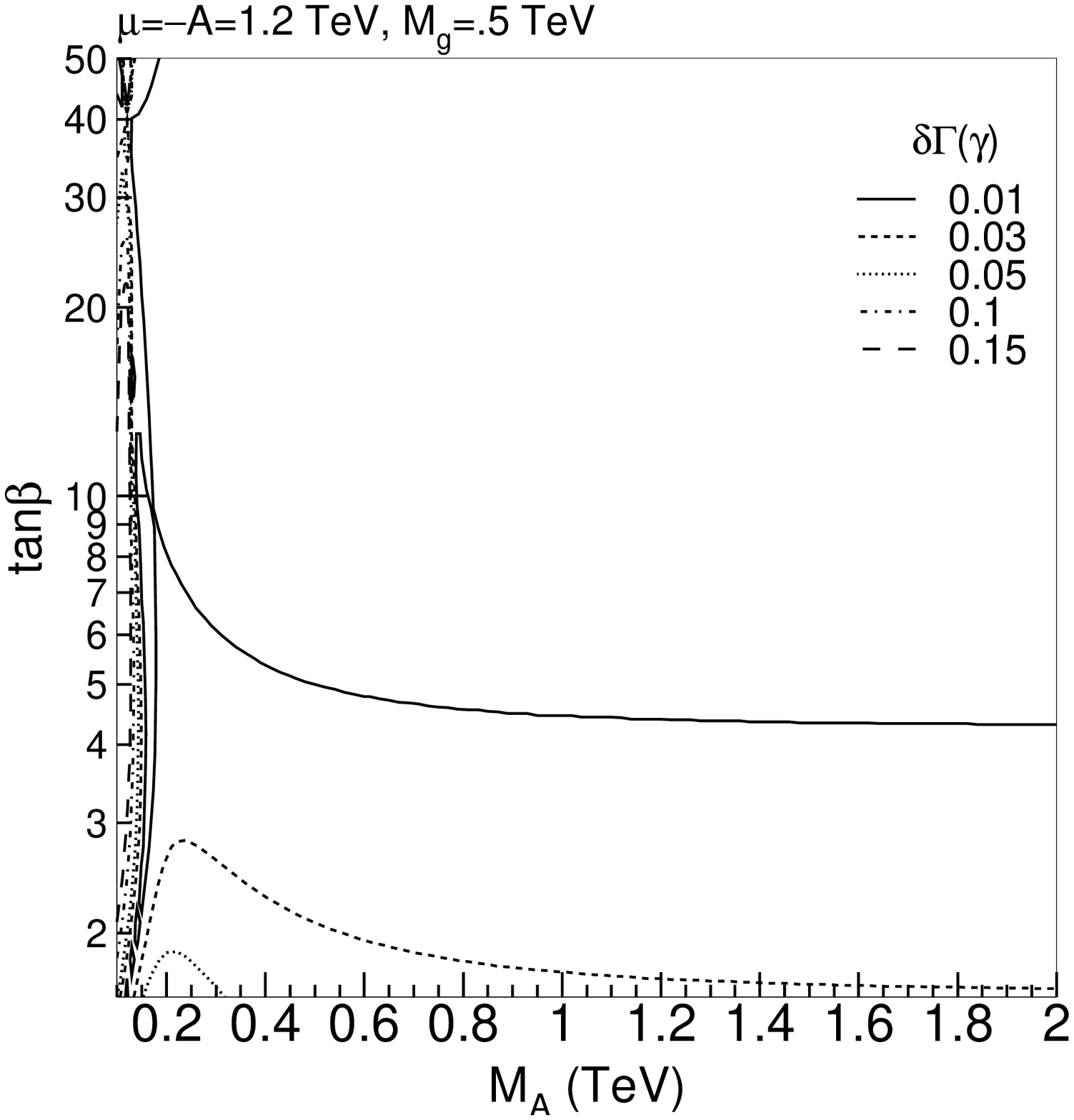}
\includegraphics*[19,142][529,682]{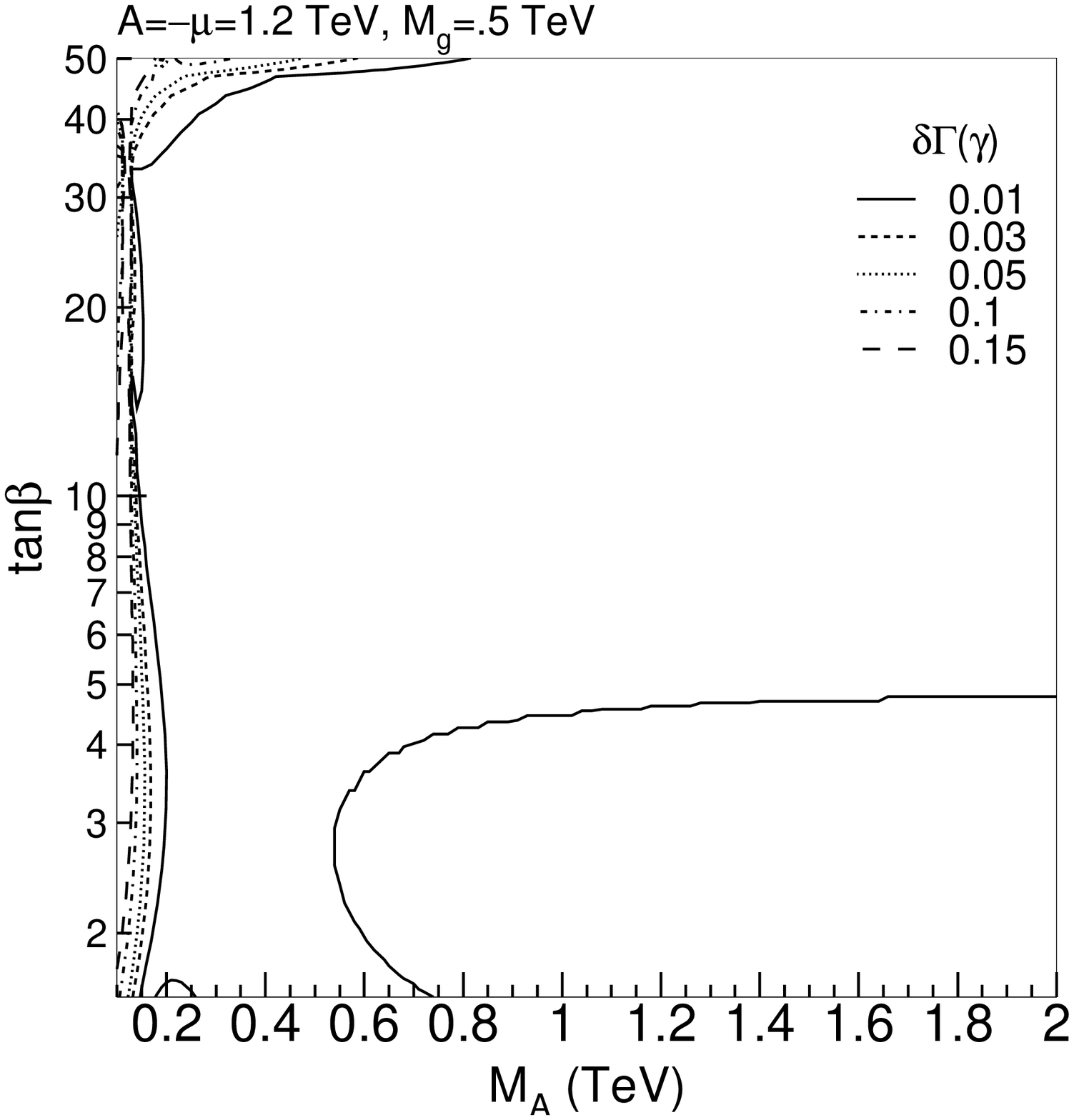}
}
\end{center}
\caption{Deviations of the partial width
$\Gamma(\gamma)$ from
its SM value in the three benchmark scenarios.}
\label{fig:Gamma_photons}
\end{figure}

Contours of $\delta\Gamma(\gamma)$ are shown in
Fig.~\ref{fig:Gamma_photons}.
In the SM, $\Gamma(\gamma)$ is dominated by the $W$ boson contribution;
at low $m_A \lsim 0.2$ TeV, the deviation of $\Gamma(\gamma)$ from 
its SM value is dominated by the deviation of the $hWW$ coupling 
(see Figs.~\ref{fig:widths1} and~\ref{fig:widthsmuAt}).
Because this one-loop decay gets contributions
from all charged particles that couple to $h$,
it depends not only on the parameters of the squark and slepton
sectors but also on the charginos.  It was shown in Ref.~\cite{Djouadi}
that near the decoupling regime, only top and bottom squark 
and chargino contributions
can generate a sizable deviation from the SM partial width.
%\footnote{Note
%however that in Ref.~\cite{Djouadi} bottom squark mixing effects were 
%neglected.  Thus these conclusions may be modified at large $\tan\beta$,
%where bottom squark mixing can be sizable, especially if $h_b$ is enhanced
%by $\Delta_b$ effects.}
%
In the no-mixing and maximal-mixing scenarios, we choose $M_2=-\mu=200$ GeV.
This choice leads to large Higgsino-gaugino mixing, so that the couplings
of $h$ to chargino pairs are large.  Thus at low $\tan\beta$ and 
$m_A \gsim 0.2$ TeV, the chargino contribution to $\delta \Gamma(\gamma)$ 
is responsible for the bulk of the deviation from the SM.
As $\tan\beta$ increases, the Higgsino-gaugino mixing
decreases and the chargino contribution becomes smaller (see the top
two panels of Fig.~\ref{fig:Gamma_photons}).  
As in the case
of $h \to gg$, there are also top squark contributions to $h \to \gamma\gamma$;
these are very small in the no-mixing scenario but somewhat larger in the
maximal-mixing scenario due to the large $X_t$ enhancement of the 
$h \tilde t_i \tilde t_i$ couplings, as discussed before.  In the 
maximal-mixing scenario, 
the top squark and chargino contributions to the amplitude
enter with opposite signs.  Then as $\tan\beta$ increases, the chargino 
contribution shrinks until it is the same size as the top squark contribution;
this occurs for $\tan\beta \simeq 15-20$, where $\delta \Gamma(\gamma)$ 
goes to zero in this scenario.  
At larger values of $\tan\beta$, the top squark contribution
is responsible for the bulk of the deviation from the SM.
%
%Note that the no- and maximal mixing scenarios exhibit non-zero
%deviations even for large $m_A$.
%For the no mixing case, the main effect is from chargino loops.
%For maximal mixing, charginos are most important at small $\tan\beta$, but
%are eventually surpassed by the top squark.  Since the chargino and top squark
%contributions destructively interfere, the $\delta\Gamma$ variation decreases
%with increasing $\tan\beta$ until the top squark becomes most important.
%For these two benchmarks, $M_2=-\mu=200$ GeV is chosen to increase
%the mass of the SM-like Higgs boson.  However, this choice induces a large
%chargino coupling from substantial Wino--Higgsino mixing in the chargino
%wave functions.  Since this mixing decreases with increasing $\tan\beta$,
%the chargino contribution becomes less important.
%
In the large $\mu$ and $A_t$ scenarios (the bottom two panels of 
Fig.~\ref{fig:Gamma_photons}), we have $M_2 \ll |\mu|$, so that
the Higgsino-gaugino mixing is very small and the couplings of $h$ to 
chargino pairs are suppressed.  Thus in these scenarios 
the chargino contribution is very small, and appears only at low
$\tan\beta$.
For $\mu < 0$, the deviation at very large $\tan\beta$ and low $m_A$ is 
due to the deviation in the $h b \bar b$ coupling in the $b$ quark loop,
as in the case of $h \to gg$.
%
%For $M_2$ small and $|\mu|$ large, the chargino
%contributions will be suppressed.  The small deviations from the SM value
%at large $m_A$ for the large $A_t$ and $\mu$ scenario arises mainly
%from modifications in the fermion Yukawa couplings.

%------------------------------------------------------
\section{Higgs boson measurements at future colliders}
\label{sec:brmeas}

Figs.~\ref{fig:widths1}--\ref{fig:Gamma_photons}
demonstrate our theoretical expectations for
the behavior of partial widths and branching ratios in the MSSM with respect
to the SM.  In this section, we fold those results with the expected
experimental resolution from
the next generation of experiments at linear colliders and compare it
to what will be known from hadron collider experiments.

\subsection{Anticipated experimental uncertainties in Higgs branching ratios 
and couplings}
\label{expunc}

We expect
quite sensitive measurements at the LC of both the Higgs production cross
sections and the BRs for the most important Higgs boson decay modes.
Combining these measurements, it is possible to extract the total
Higgs boson width and the partial widths for the various decay modes.
The measurement of the total cross section for $e^+e^- \to Zh$
yields a measurement of $g^2_{hZZ}$.
Branching ratios are determined
by selecting Higgs events in $Zh$ production, where
$Z\to\ell^+\ell^-$,  using the recoil mass
technique and counting the different types of events in the invariant
mass region of the Higgs boson signal.
These branching ratio determinations are
model independent, even if invisible Higgs decays are present.
Finally, by including other $Z$ decay modes in $hZ$ production as well
as Higgs production via vector boson fusion, one can further improve
the statistical precision of the Higgs branching ratios and Higgs couplings.

The expected experimental uncertainties in the measurement of BRs at the 
LC for a 120 GeV SM-like Higgs boson are summarized in Table~\ref{tab:BRmeas}.
The first row shows the results assuming
500 fb$^{-1}$ of integrated luminosity at 
$\sqrt{s}=350$ GeV \cite{BattagliaDesch}.
\begin{table}[t]
        \begin{center} \begin{tabular}{l c c c c c c}
        Decay mode: & $b \bar b$ & $W W^*$ & $\tau^+ \tau^-$ & $c \bar c$
                & $gg$ & $\gamma \gamma$ \\
        \hline
        Ref.~\cite{BattagliaDesch} & 2.4\% & 5.1\% & 5.0\% & 8.5\%
                & 5.5\% & 19\% \\
%        Ref.~\cite{BattagliaDesch} (scaled) & 3.6\% & 7.7\% & 7.5\% & 12.8\%
%                & 8.3\% & 29\% \\
        Ref.~\cite{Brau}           & 2.9\% & 9.3\% & 7.9\% & 39\%
                & 18\% &  \\
        Ref.~\cite{Boos} (scaled)  &       &       &       &
                &      & 14\% \\
        \hline
        theory uncertainty  &  1.4\% & 2.3\% & 2.3\% & 23\%  & 5.7\%
                & 2.3\% \\
        \hline
        \end{tabular} \end{center}
        \caption{Expected fractional uncertainty of BR measurements
at an $e^+e^-$ LC for a 120 GeV SM-like Higgs boson.
Results are shown from Ref.~\cite{BattagliaDesch} (500 fb$^{-1}$ at
$\sqrt{s} = 350$ GeV) (first row); 
%Ref.~\cite{BattagliaDesch} naively
%scaled by $(\sigma \times \int \mathcal{L} dt)^{1/2} \simeq 1.5$ to
%approximate the expectation for 500 fb$^{-1}$ at $\sqrt{s} = 500$ GeV
%(second row); 
Ref.~\cite{Brau} (500 fb$^{-1}$ at $\sqrt{s} = 500$ GeV)
(second row); and Ref.~\cite{Boos} (1 ab$^{-1}$ at $\sqrt{s} = 500$ GeV,
scaled to 500 fb$^{-1}$) (third row).  The theoretical uncertainty of
the predicted Standard Model branching ratios is given in the fourth
row (see Sec.~\ref{thunc}).
}
        \label{tab:BRmeas}
\end{table}
At $\sqrt{s}=500$ GeV, the $e^+e^- \to Zh$ cross section for $m_h=120$ GeV
is about a factor of two smaller than at 350 GeV \cite{tesla_tdr};
thus roughly twice as much integrated luminosity ({\it i.e.}, 
about 1 ab$^{-1}$) would be needed at 500 GeV to
obtain the same statistical precision on Higgs BRs.  To estimate the precision
on Higgs BRs that can be obtained with 500 fb$^{-1}$ at $\sqrt{s} = 500$ GeV,
the results for $\sqrt{s}=350$ GeV shown on the first line of
Table~\ref{tab:BRmeas}
should be reduced by a factor of 1.5 (corresponding to the square root of
the ratio of the corresponding $Zh$ cross sections for
$m_h=120$~GeV).%
%$(\sigma \times \int \mathcal{L} dt)^{1/2} \simeq 1.5$.
\footnote{This
scaling is only approximate since it does not take into account the 
Higgs production by vector boson fusion employed in the analysis of
Ref.~\cite{BattagliaDesch}.}
%The BR precisions of Ref.~\cite{BattagliaDesch} scaled to $\sqrt{s} = 500$ GeV
%are shown in the second row of Table~\ref{tab:BRmeas}.
The second row of Table~\ref{tab:BRmeas}
shows the results of a similar study \cite{Brau} for the branching
ratios of a 120 GeV SM-like Higgs boson with
500 fb$^{-1}$ at 
$\sqrt{s} = 500$ GeV.\footnote{Note the very different predictions 
in Table~\ref{tab:BRmeas} for the
precisions of $\BR(c)$ and $\BR(g)$, which depend on very good charm
and light quark separation.
%These discrepancies are not entirely understood and
The authors of Refs.~\cite{BattagliaDesch} and \cite{Brau} are working
together to resolve these discrepancies \cite{Deschcomm}.} 
Finally, we consider the results of a dedicated study of the
$\BR(\gamma)$ measurement \cite{Boos}
for $\sqrt{s} = 350$ and 500 GeV, both without and with beam polarization
(80\% left-handed electron polarization and 40 or 60\% right-handed positron
polarization) chosen to enhance the Higgsstrahlung and $WW$ fusion cross
sections.
At $\sqrt{s} = 500$ GeV and the highest polarizations,
a measurement of $\BR(\gamma)$ with an experimental uncertainty of
9.6\% is possible with
1 ab$^{-1}$.  Scaling this to 500 fb$^{-1}$ to compare with the other
studies yields a precision of about 14\%, as shown 
in the third row of Table~\ref{tab:BRmeas}.
Without beam polarization, this deteriorates
to 16\% (23\%) with 1 ab$^{-1}$ (500 fb$^{-1}$).

The $\delta\Gamma(\gamma)$ deviations shown in
Fig.~\ref{fig:Gamma_photons} are typically
too small to be observed at an $e^+e^-$ linear collider.
However, the operation
of the LC as a photon-photon collider
offers the possibility of a direct measurement of
$\Gamma(\gamma)$ from the $s$-channel Higgs production
cross section $\sigma(\gamma \gamma \to h)$.
Estimates for the precision obtainable on
$\sigma(\gamma \gamma \to h) \times \BR(h \to b \bar b)$ range from
$2$--$10$\% for a light Higgs boson with mass between
120 and 160 GeV \footnote{The variation in precision is due to the
decline of $\BR(h \to b \bar b)$ as $m_h$ increases.}
from a $\gamma\gamma$ collider running at
$\sqrt{s_{ee}} \simeq m_h/0.8$
(giving peak $\gamma\gamma$ luminosity at the Higgs mass)
and an integrated luminosity corresponding
to 400 fb$^{-1}$ of $e^-e^-$ luminosity \cite{SoldnerJikia}.
Combining a 2\% measurement of
$\sigma(\gamma \gamma \to h) \times \BR(h \to b \bar b)$
at a $\gamma\gamma$ collider for a 120 GeV Higgs boson with a
3\% measurement of $\BR(h \to b \bar b)$ at the $e^+e^-$ LC,
%(see Table~\ref{tab:BRmeas}), 
we find that $\Gamma(\gamma)$
could be extracted with an uncertainty of about $3.6$\%.
As Fig.~\ref{fig:Gamma_photons} shows, such a measurement would not exhibit
a significant deviation from the SM prediction
in our benchmark scenarios unless $m_A \lsim 200$ GeV or
$\tan\beta$ is very small; however, a more detailed analysis of the MSSM
parameter space should be performed in light of this expected precision.

From the measurement of $g^2_{hWW}$ based on the production cross section,
the partial width $\Gamma(W)$ can be calculated and combined
with the measurement of BR$(W)$ to determine the total Higgs width 
$\Gamma_{\rm tot}$.  The expected resolution on $g^2_{hZZ}$
of about 3\% yields a measurement of the total width to roughly 6\%
accuracy \cite{BattagliaDesch,tesla_tdr}.
This method for extracting the total Higgs width is more accurate
than using the photon collider mode.\footnote{As noted earlier,
$\Gamma(\gamma)$ can be measured to about 3--4\% accuracy
in $\gamma\gamma$ collisions.  The total Higgs width is
then extracted by combining this measurement with BR$(\gamma)$.
%The uncertainty in the total Higgs width extracted by this method
%is dominated by the large uncertainty in BR($\gamma$);
%using the numbers from Table~\ref{tab:BRmeas} we find a 14--29\% uncertainty
%on the total Higgs width.  
Using the range of uncertainties in BR$(\gamma)$ given in
Table~\ref{tab:BRmeas}, we see that the uncertainty in the total Higgs
width extracted by this method is dominated by the large uncertainty in
BR($\gamma$).  This can be improved somewhat by combining the LC measurements
and the LHC data on $\Gamma(\tau)/\Gamma(\gamma)$ (see Table~\ref{tab:LHC}),
as discussed at the end of Sec.~4.5.  Even in this case, the
uncertainty in BR($\gamma$) still dominates the total Higgs width
as extracted from the $\gamma\gamma$ collider measurements.}
However, the
photon collider mode is still quite useful for the
high precision measurement that it provides of the partial width
$\Gamma(\gamma)$.  In other models, such as the (non-SUSY)
two Higgs doublet model, this measurement can be essential for distinguishing
the extended model from the SM in some regions of parameter space
\cite{MariaKrawczyk}.

Combining the measurements of Higgs boson BRs and production cross
sections that can be obtained at the LC, the Higgs couplings
to SM particles can be extracted \cite{tesla_tdr}.  The results 
of a $\chi^2$ minimization 
%constrained fit 
using {\sc HFitter} \cite{BattagliaDesch,HFITTER} 
are summarized in Table~\ref{tab:g^2}.
\begin{table}
\begin{center} 
\begin{tabular}{l c c c c c c c}
       & $g^2_{hbb}$ &$g^2_{hWW}$  & 
 $g^2_{hZZ}$ & $g^2_{hcc}$ &  $g^2_{h\tau\tau}$ & $g^2_{hgg}$ & $g^2_{htt}$  \\
        \hline
experimental uncertainty & 
4.4\% & 2.4\% &  2.4\% &  7.4\% & 6.6\% & 7.4\% & 10\% \\
\hline
theory uncertainty & 3.5\% & -- & -- & 24\% & -- & 3.9\% & 2.5\% \\
\hline
\end{tabular}
\end{center}
\caption{Expected uncertainty of measurements of squared couplings
(equivalently partial widths) for a 120 GeV SM-like Higgs boson from 
%.  Results are taken from Ref.~\cite{tesla_tdr} using
{\sc HFitter} \cite{BattagliaDesch,HFITTER}, 
assuming 500 fb$^{-1}$ at $\sqrt{s} = 500$ GeV, except for the
measurement of $g^2_{htt}$ which assumes  1000 fb$^{-1}$ at
$\sqrt{s} = 800$ GeV.  The second line shows the theoretical uncertainty 
(see Sec.~\ref{thunc}).}
\label{tab:g^2}
\end{table}
The first six couplings listed in Table~\ref{tab:g^2} are extracted from 
$h$ BRs into $b\bar b$, $WW^*$, $c\bar c$, $\tau\tau$ and $gg$ and 
the production cross sections in the Higgsstrahlung and $WW$ fusion modes,
assuming 500 fb$^{-1}$ at $\sqrt{s} = 500$ GeV [the first five of these
are given in Ref.~\cite{tesla_tdr}].
The $ht\bar t$ coupling can be measured indirectly from the LC measurements
of $h \to gg$ and $h \to \gamma\gamma$ if one assumes that SUSY loop 
contributions are negligible; however, this is a model-dependent assumption
that we wish to avoid.
A direct measurement of $g^2_{htt}$ can be obtained from the 
$e^+e^- \to t\bar t h$ cross section \cite{tesla_tdr,JusteMerino}.
Such a measurement requires running at higher $\sqrt{s} = 800\,$--$\,1000$~GeV
in order to avoid kinematic suppression of the cross section;
the result in Table~\ref{tab:g^2} assumes 1000 fb$^{-1}$ at 
$\sqrt{s} = 800$ GeV.~\footnote{The experimental uncertainty in $g^2_{htt}$
can be reduced by combining the $e^+e^- \to t\bar t h$ cross section 
measurement with the measurements of $h \to gg$ and $h \to \gamma\gamma$
if one assumes that SUSY loop contributions to the latter are negligible;
the resulting uncertainty in $g^2_{htt}$ is 6.0\% \cite{tesla_tdr}.}
%We include also the experimental uncertainty in
%$g^2_{hgg}$ that we obtained from {\sc HFitter}.
%but that is not quoted in Ref.~\cite{tesla_tdr}.
%
The studies summarized in Tables~\ref{tab:BRmeas} and~\ref{tab:g^2}
were conducted
for the SM Higgs boson of mass 120 GeV, and thus are directly
applicable to the
study of a SM-like Higgs boson of the MSSM with a mass
near 120 GeV, especially near the
decoupling limit.  In our benchmark scenarios, the value of $m_h$
varies between 118 and 129~GeV.  
We shall assume that the mass dependence of
the results quoted in Tables~\ref{tab:BRmeas} and~\ref{tab:g^2} 
is minimal in this mass range, but this assumption should be tested
with detailed simulations.

\subsection{Theoretical uncertainties in Higgs branching ratios and couplings}
\label{thunc}

In order to gauge the significance of an observed deviation of Higgs
boson properties at the LC from the Standard Model expectation, one
must take into account both the experimental uncertainties
(statistical and systematic) described in Sec.~\ref{expunc} and the
corresponding theoretical uncertainties for the Standard Model Higgs
boson.  Sources of theoretical uncertainty include higher order loop
corrections to Higgs decay rates not yet computed and parametric
uncertainties due to the choice of input parameters.  The largest
sources of uncertainty arise from the choice of $\alpha_s$, $m_c$
and $m_b$.\footnote{The observed uncertainty in $m_t$ has only a small
effect on the predictions for the $h\to gg$ and $h \to \gamma\gamma$
decay rates.  This is not surprising given that the top quark mass is
the most accurately known of all the quark masses!}  To determine the
current theoretical uncertainty in Higgs branching ratios and
couplings, we choose $\alpha_s= 0.1185\pm 0.0020$ \cite{pdg},
$m_c(m_c)= 1.23\pm 0.09$~GeV \cite {ej} and $m_b(m_b)=4.17\pm
0.05$~GeV \cite{hoang} (see Ref.~\cite{mbmc} for an alternative
evaluation of $\overline{\rm MS}$ quark masses).  By varying these
input parameters in the program {\sc Hdecay} \cite{HDECAY}, we
obtain the theoretical fractional uncertainties 
for the Higgs branching ratios quoted in 
Table~\ref{tab:BRmeas}.  For the Higgs squared-couplings listed in 
Table~\ref{tab:g^2}, the only significant theoretical uncertainties
reside in $g_{hbb}^2$ and $g_{hcc}^2$, due to the uncertainties in
the $b$ and $c$ quark masses and in $\alpha_s$ (which governs the
running of the quark masses from the quark mass to the Higgs mass).
The resulting theoretical uncertainties for $g_{hbb}^2$ and $g_{hcc}^2$ 
(for a SM Higgs boson of mass 120 GeV) are 3.5\% and 24\%, respectively.
In addition, we find a theoretical uncertainty in $g_{hgg}^2$ of 
3.9\% due to the uncertainty in $\alpha_s$.

For a SM Higgs boson with $m_h=120$~GeV, about 2/3 of the width is
due to $h\to b\bar b$.  The theoretical 
fractional uncertainties for the Higgs
branching ratios to $WW^*$, $\tau^+\tau^-$ and $\gamma\gamma$ 
listed in Table~\ref{tab:BRmeas} are
due primarily to the fractional uncertainty of the total width,
which for a SM Higgs boson with $m_h=120$~GeV is mainly governed
by the corresponding uncertainty in the $h\to b\bar b$ width.%
\footnote{For larger values of the Higgs mass, the $h\to b\bar b$
branching ratio is smaller and the uncertainty in the total width, 
which is now dominated by $h\to WW^{(*)}$, is
correspondingly reduced.}
The theoretical uncertainty in the $h\to c\bar c$ decay rate is 
particularly large
due to the relatively large uncertainty in the charmed quark mass.  
It is hoped that further theoretical work and numerical improvements in
lattice calculations \cite{charmlattice} will reduce this theoretical
uncertainty by the time that LC data is available.
%is ready to take data.

Finally, a scan of the $t\bar t$ threshold at the
LC will yield a value of $m_t$ with an uncertainty of
about 100~MeV \cite{Hoang:2000yr}.  Thus, the theoretical error
expected for the Standard Model Higgs coupling to $t\bar t$ due to the
top-quark mass uncertainty will be negligible.  The remaining uncertainty
in $g^2_{htt}$ is due to uncalculated higher order QCD corrections
to the $e^+e^- \to t \bar t h$ cross section.  We estimate this uncertainty
to be about 2.5\%
based on the renormalization scale dependence in the next-to-leading order
(NLO) QCD result for $m_h = 120$ GeV and $\sqrt{s} = 1$ TeV \cite{DawsonReina}.

\subsection{Branching ratio analysis}
\label{sec:BRanalysis}
\begin{figure}[!ht]
\resizebox{\textwidth}{!}{
\includegraphics*[0,142][472,682]{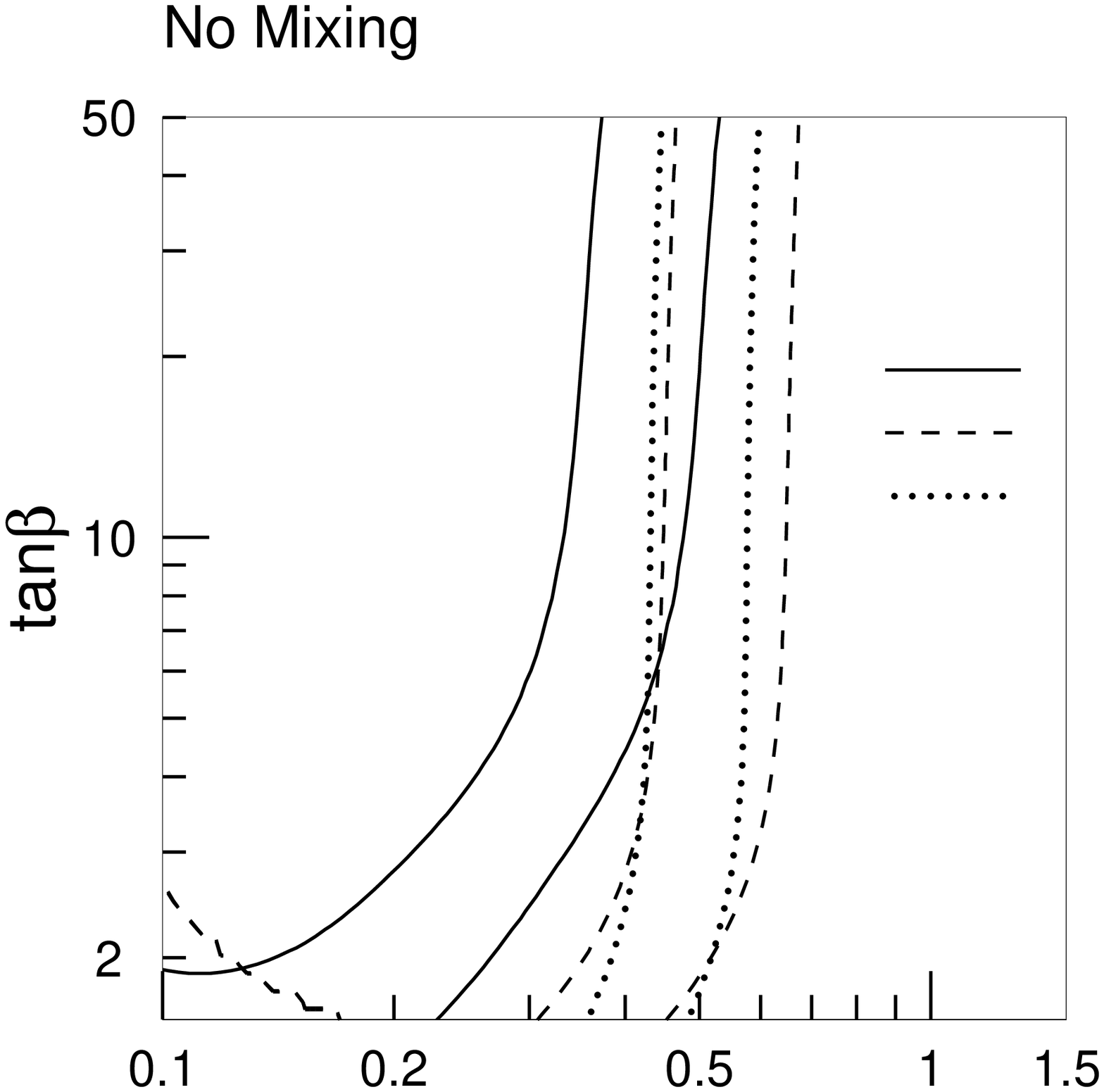}
\includegraphics*[57,142][529,682]{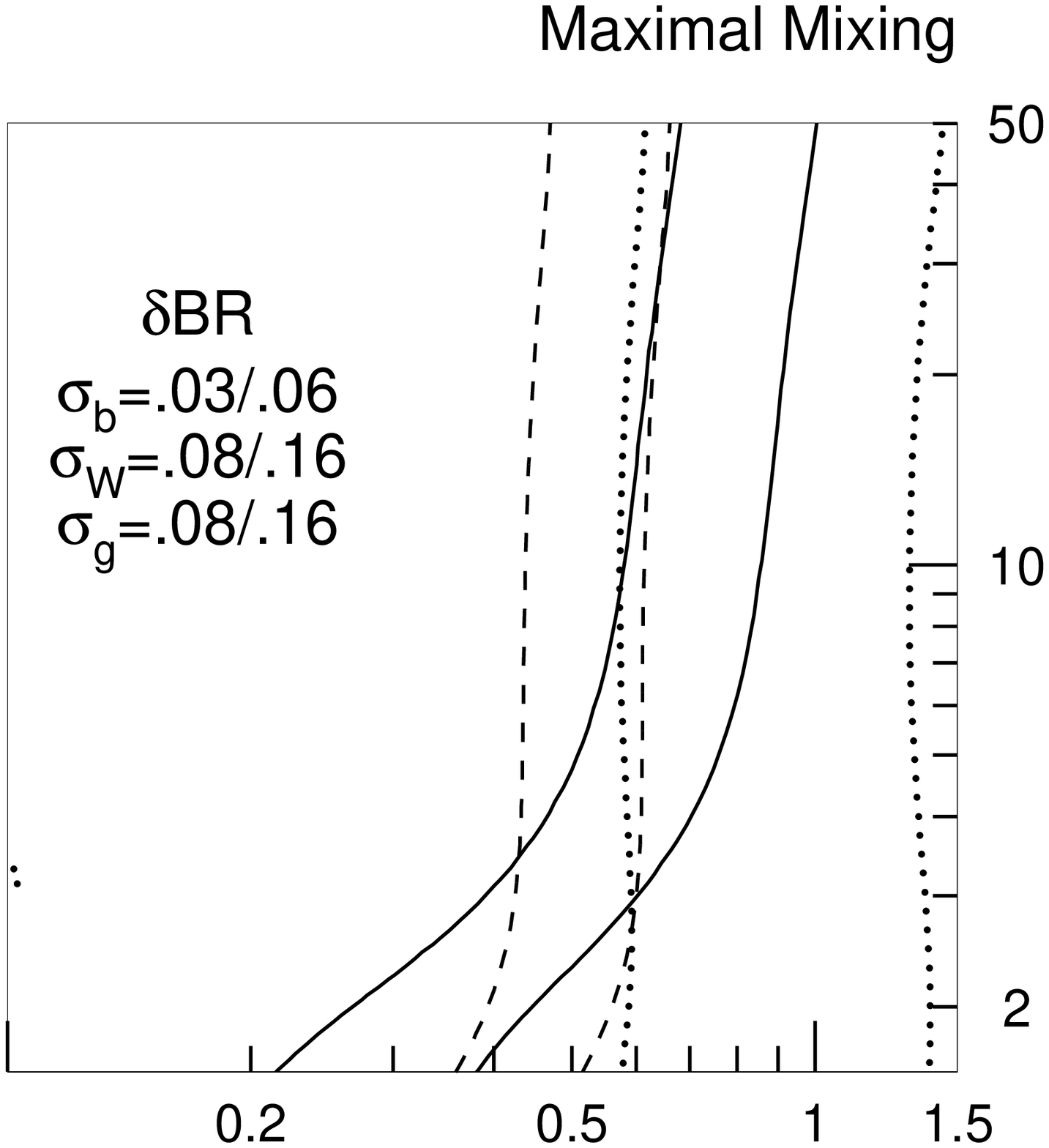}             
}
\resizebox{\textwidth}{!}{
\includegraphics*[0,142][472,682]{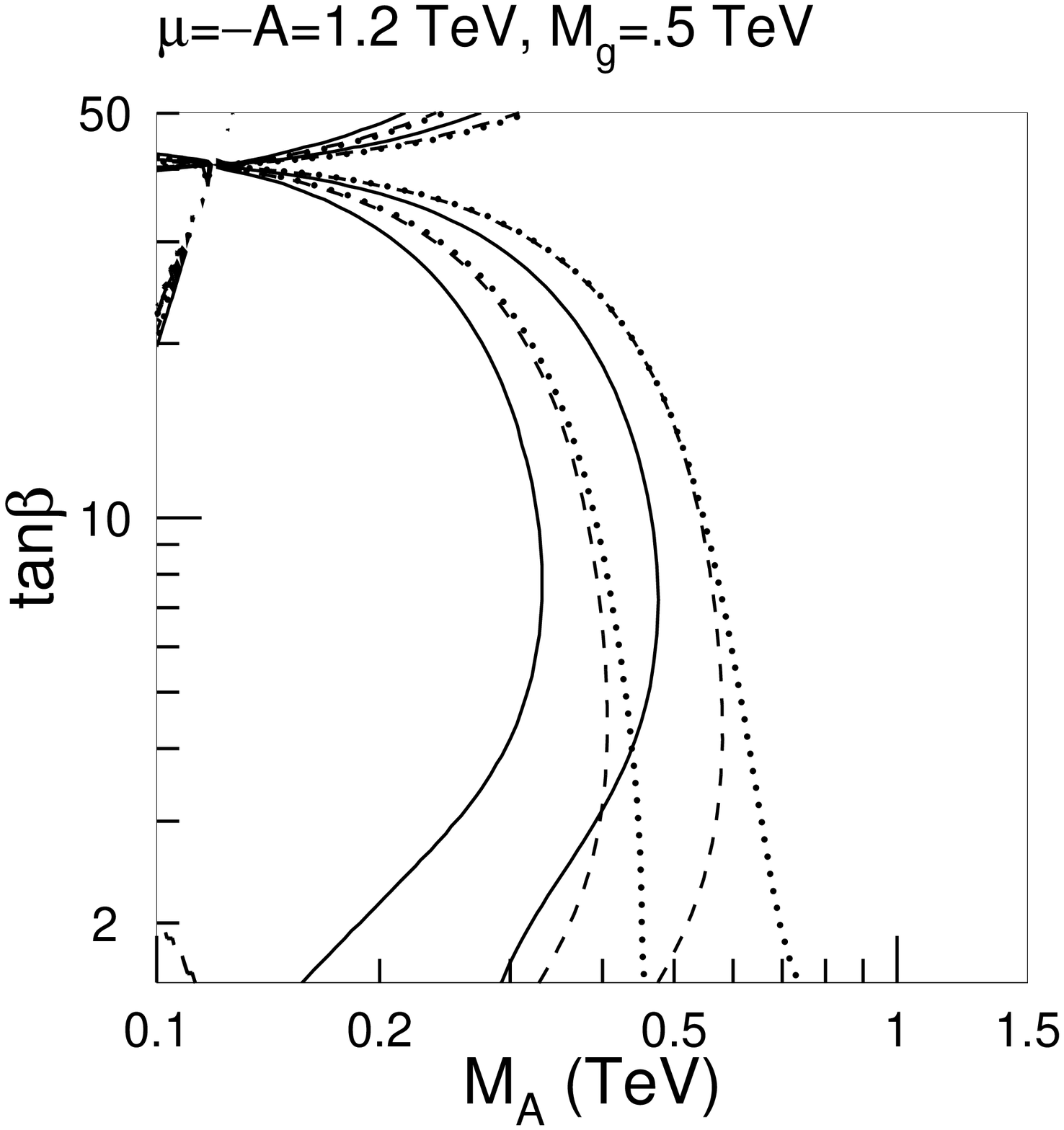}
\includegraphics*[57,142][529,682]{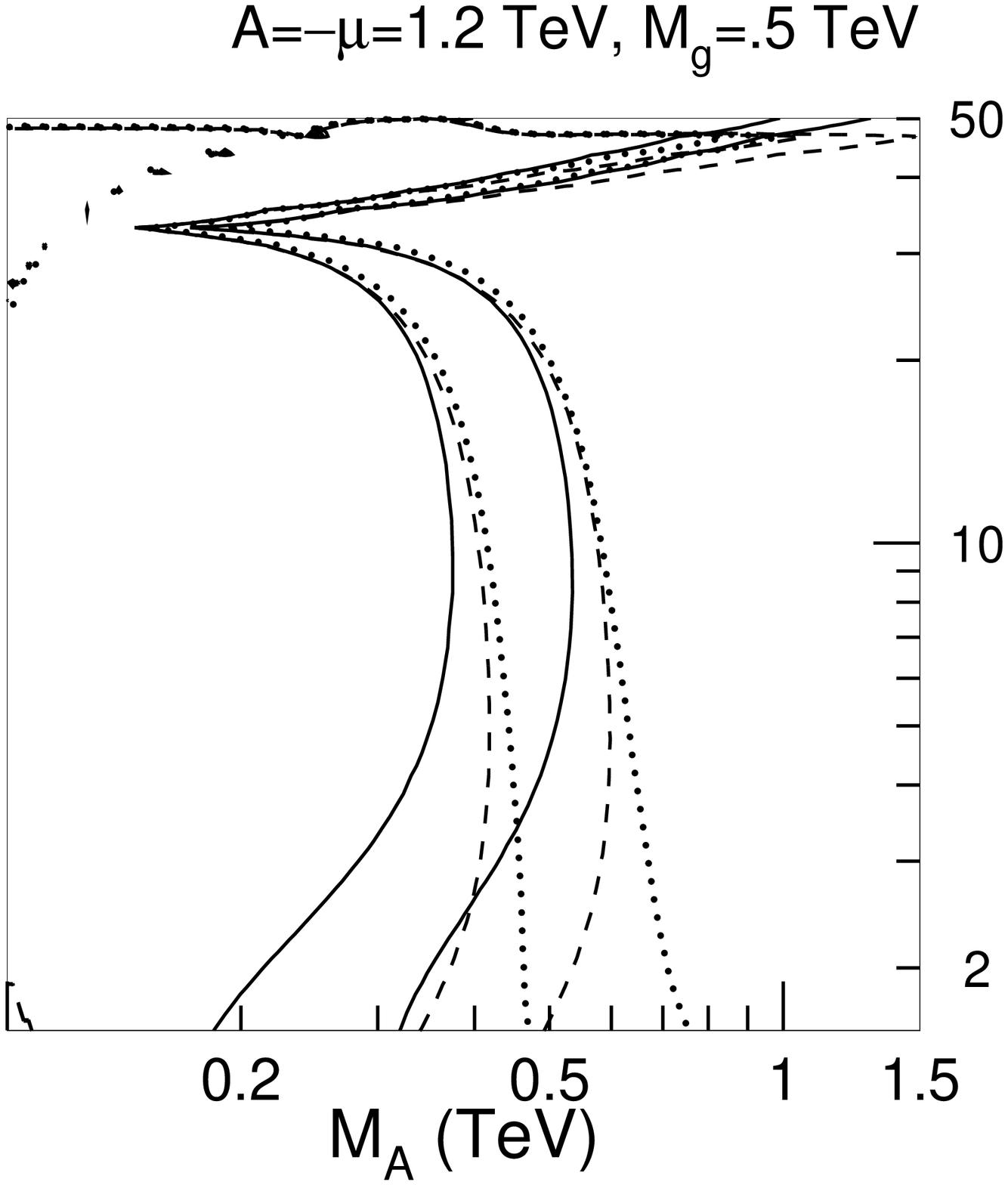}
}
\caption{Contours of
$\delta \BR(b) = 3$ and 6\% (solid),
$\delta \BR(W) = 8$ and 16\%
(long-dashed) and
$\delta \BR(g) = 8$ and 16\% (short-dashed)
in the three benchmark scenarios.
%In the maximal-mixing scenario (top right) we plot $m_A$ between 0.1 and
%1.5 TeV.
}
\label{figure1}
\end{figure}

A number of the Higgs BRs can be measured to higher accuracy than
the total Higgs width.
Thus BR measurements alone are valuable for distinguishing
the SM Higgs boson from a MSSM Higgs boson.
To illustrate the potential of the LC, contours of $\delta\BR$
(Eq.~\ref{eq:deltaBR}) are shown in Fig.~\ref{figure1}
over the $m_A$---$\tan\beta$ plane for the benchmark scenarios.
Contours of $\delta \BR(b) = 3$ and 6\% and
$\delta \BR(W)$, $\delta \BR(g) = 8$ and 16\% were chosen,
corresponding roughly
to one and two times
the expected experimental uncertainties
quoted in Table~\ref{tab:BRmeas}, or approximately one and two
sigma deviations from the SM.\footnote{We choose typical
values rather than using the precisions quoted in the experimental studies
\cite{BattagliaDesch,Brau} because we expect the exact results of these
studies to change as the LC detector design evolves and experimental
techniques improve.}
As shown in Table~\ref{tab:BRmeas},
the theoretical uncertainties in these three BRs for the SM 
Higgs boson are smaller than the expected experimental 
uncertainties.
%~\footnote{The dominant theoretical
%uncertainties in the BRs of the SM Higgs boson are due to the uncertainties
%in the bottom and charm quark masses~\cite{mbmc} and $\alpha_s$.}

In the four panels of Fig.~\ref{figure1}, the solid, long-dashed, 
and short-dashed lines are
contours of $\delta\BR(b)$, $\delta\BR(W)$ and $\delta\BR(g)$, respectively. 
Although $\delta \Gamma(b)$ is quite large
over much of the parameter space, $\delta \BR(b)$ is smaller because the
increase in $\Gamma(b)$ also significantly increases $\Gamma_{\rm tot}$.
Because $\delta \Gamma(W)$ quickly approaches zero for increasing $m_A$,
$\delta\BR(W)$ indicates variation
in the total Higgs width, and is more sensitive than $\delta\BR(b)$,
except for the case of maximal mixing.
In regions of parameter space where $\delta \Gamma(g)$ approaches zero
(see Fig.~\ref{fig:Gamma_gg}), $\delta\BR(g)$, like $\delta\BR(W)$,
is sensitive to variations in the total width.
%Except for low $\tan\beta$ or for large values of $X_t$ as in the
%maximal mixing case, $\delta\Gamma(g)$ also approaches zero with
%increased $m_A$, and thus $\delta\BR(g)$, like $\delta\BR(W)$,
%is sensitive to variations in the total width.  
%For the maximal
%mixing scenario, however, the behavior of $\delta\BR(g)$ should
%indicate deviations from the decoupling limit for $m_A$ as
%large as $1.4-1.5$ TeV.

For the maximal-mixing scenario,
the mass of the SM-like Higgs boson near the decoupling limit is
roughly 10 GeV heavier than in the other benchmarks
(see Table~\ref{tab:scenarios}), so that
the relative contribution of $\Gamma(b)$ to $\Gamma_{\rm tot}$ is
decreased.  Therefore, deviations in $g_{hbb}$ are not as
diluted in the BR measurement as in the other scenarios,
and  the measurement of $\delta\BR(b)$ yields superior
sensitivity at large $\tan\beta$, around $m_A \lsim 600$--700 GeV at
$2\sigma$.  
One should interpret
this result with caution, however, since the accuracies for
BR measurements are based on the simulation of a 120 GeV
SM Higgs boson.
In the maximal-mixing scenario,
$\BR(g)$ deviates by more than 8\% from its SM value for
$m_A \lsim 1.4$ TeV.
At $2 \sigma$
the reach in $\delta \BR(g)$ is roughly $m_A \lsim 600$ GeV.
In the no-mixing scenario,
$\delta\BR(g)$ and $\delta\BR(W)$
give comparable
reach in $m_A$; at $2 \sigma$ the reach is $m_A \lsim 425$ GeV.
For comparison, in the no-mixing scenario 
deviations in $\BR(b)$ yield sensitivity at
$2 \sigma$ for $m_A \lsim 300$ GeV for $\tan\beta \gsim 5$.

The large $\mu$ and $A_t$ scenario demonstrates the 
complementarity of the LC and the hadron colliders.
For $\tan\beta \lsim 20$, where the heavy MSSM Higgs bosons
can be missed at the LHC, $\BR(g)$ gives the
greatest reach in $m_A$, allowing one to distinguish the MSSM from the
SM Higgs boson at $2 \sigma$ for
$m_A \lsim 350$--450 GeV, depending
on the value of $\tan\beta$.
At larger values of $\tan\beta$, the large $\mu$ and $A_t$ scenarios
have regions of $m_A$-independent decoupling where the SM-like MSSM Higgs
boson cannot be distinguished from the SM Higgs boson even for very low
values of $m_A$.  
In fact, in these scenarios it is possible for $h$ to be indistinguishable
from the SM Higgs boson at the LC, while at the same time $m_A < 250$ GeV
so that the heavy Higgs bosons will be directly observed at a 500 GeV
LC through $e^+e^- \to HA$, $H^+H^-$. 
Moreover, in our scenarios with $A_t\mu<0$, 
for $\mu > 0$ [$\mu < 0$] the $m_A$-independent decoupling occurs
for $\tan\beta \simeq 40$ [$\tan\beta \simeq 33$].\footnote{As explained in 
Sec.~\ref{sec:directcoup},
the value of $\tan\beta$ at which this $m_A$-independent decoupling occurs
has a small dependence on the signs of $\mu$ and $A_t$.}
For such large values of $\tan\beta$, the heavy MSSM Higgs bosons 
would be discovered at the LHC even for $m_A$ above 500 GeV \cite{atlas_tdr}.
Note also that for large $\mu$ and $A_t$ and large $\tan\beta$,
the correction $\Delta_b$ is quite large, and modifies the $b$ quark 
Yukawa coupling from its SM value.  The effect of $\Delta_b$ on the 
couplings of the heavy MSSM Higgs bosons does {\it not} decouple 
for $m_A\gg m_Z$ ({\it i.e.}, for 
$\tan\alpha\tan\beta = -1$).  Thus $\Delta_b$ could have a significant 
effect on the discovery of the heavy Higgs bosons at the Tevatron and 
LHC in this region of parameter space by modifying their production cross
sections and decay branching ratios.  In particular, the value of $\tan\beta$
extracted from the $A$ and $H$ production rates at the 
LHC \cite{atlas_tdr}
could not be unambiguously determined without knowledge of the value 
of $\Delta_b$.  LC data would then be of great 
value for disentangling the $\Delta_b$ dependence 
(see Sec.~\ref{sec:Deltab}) so that the value
of $\tan\beta$ extracted from heavy Higgs boson measurements can be compared
to the value obtained from other sectors of the theory.

Clearly, from Fig.~\ref{figure1}, the regions of the $m_A$---$\tan\beta$ plane
in which the MSSM and SM Higgs bosons can be distinguished from one another
depend strongly on the supersymmetric parameters, and the sensitivity
comes from different measurements for different sets of MSSM parameters.
In isolation, measurements of one or two sigma deviations 
in individual BRs would not be a significant probe of the MSSM.
%Contours of
%one and two $\sigma$ sensitivity for measurements of
%individual decay modes, however, do not give the full picture,
%and in isolation would not
%be a significant probe of MSSM parameter space.
In combination, however, the
measurements are much more powerful, as indicated by the $\chi^2$ analysis
presented in the next section.

%------------------------------------------
\subsection{$\chi^2$ analysis of couplings}

In order to make a quantitative assessment of the ability of the LC to
discriminate between the SM-like Higgs boson of the MSSM and
the SM Higgs boson, we 
combine several observations and compute the compatibility
with the SM using a $\chi^2$ test.  In particular,
\(\chi^2 = \sum_{i}^{}(X_i^{\rm MSSM}-X_i^{\rm SM})^2/ \sigma_i^2\),
where $i$ is a decay product and $X$ is a BR, $\Gamma$, a
Higgs squared-coupling or a ratio
of these quantities.
The $\sigma_i$ values include the experimental resolution and any
theoretical uncertainty on the quantity $X_i$.
Motivated by the {\sc HFitter} results \cite{tesla_tdr}, we choose the 
$X_i$ to be the squared couplings of Higgs bosons to various final states.
The significance of a particular value for $\chi^2$ depends on the number of
observables that have been combined.
Our results indicate that the addition of several variables
does not necessarily improve the significance.
For example, the $h\gamma\gamma$ coupling is not measured very well, and
the relative theoretical error in the predicted $h c\bar c$ coupling
is too large.  Moreover, 
the $h WW$ and $hc\bar c$ couplings quickly approach their 
SM values for increasing
$m_A$.  Thus measurements of these three couplings do not add
to the significance of our results.
Therefore, we compute the $\chi^2$ combining $b\bar b$, $\tau^+\tau^-$
and $gg$ squared-coupling measurements, adding in quadrature the
experimental and theoretical uncertainties in each squared-coupling
[see Table~\ref{tab:g^2}].

\begin{figure}[!ht]
\begin{center}
\resizebox{\textwidth}{!}{
\includegraphics*[19,142][529,682]{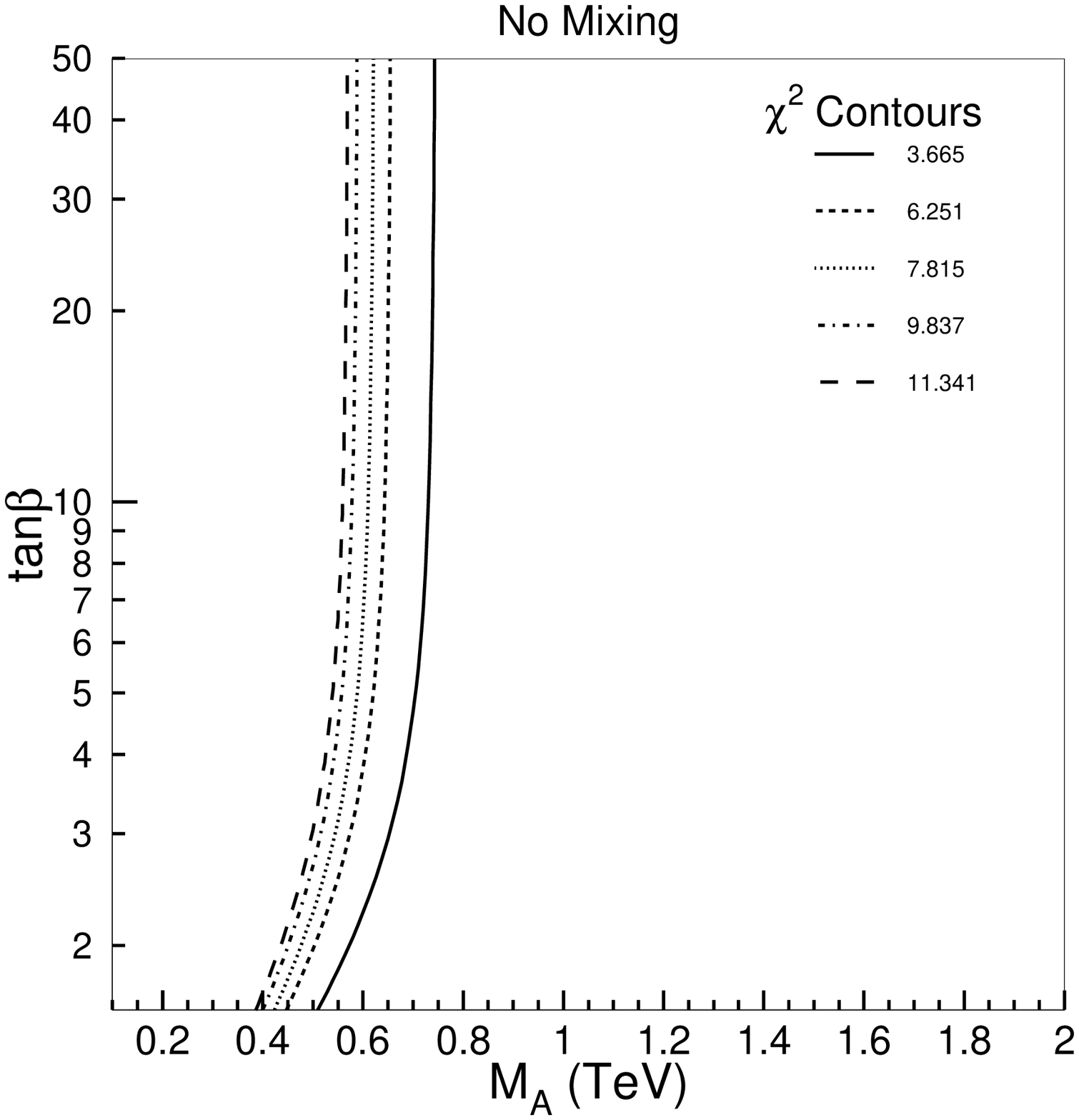}
\includegraphics*[19,142][529,682]{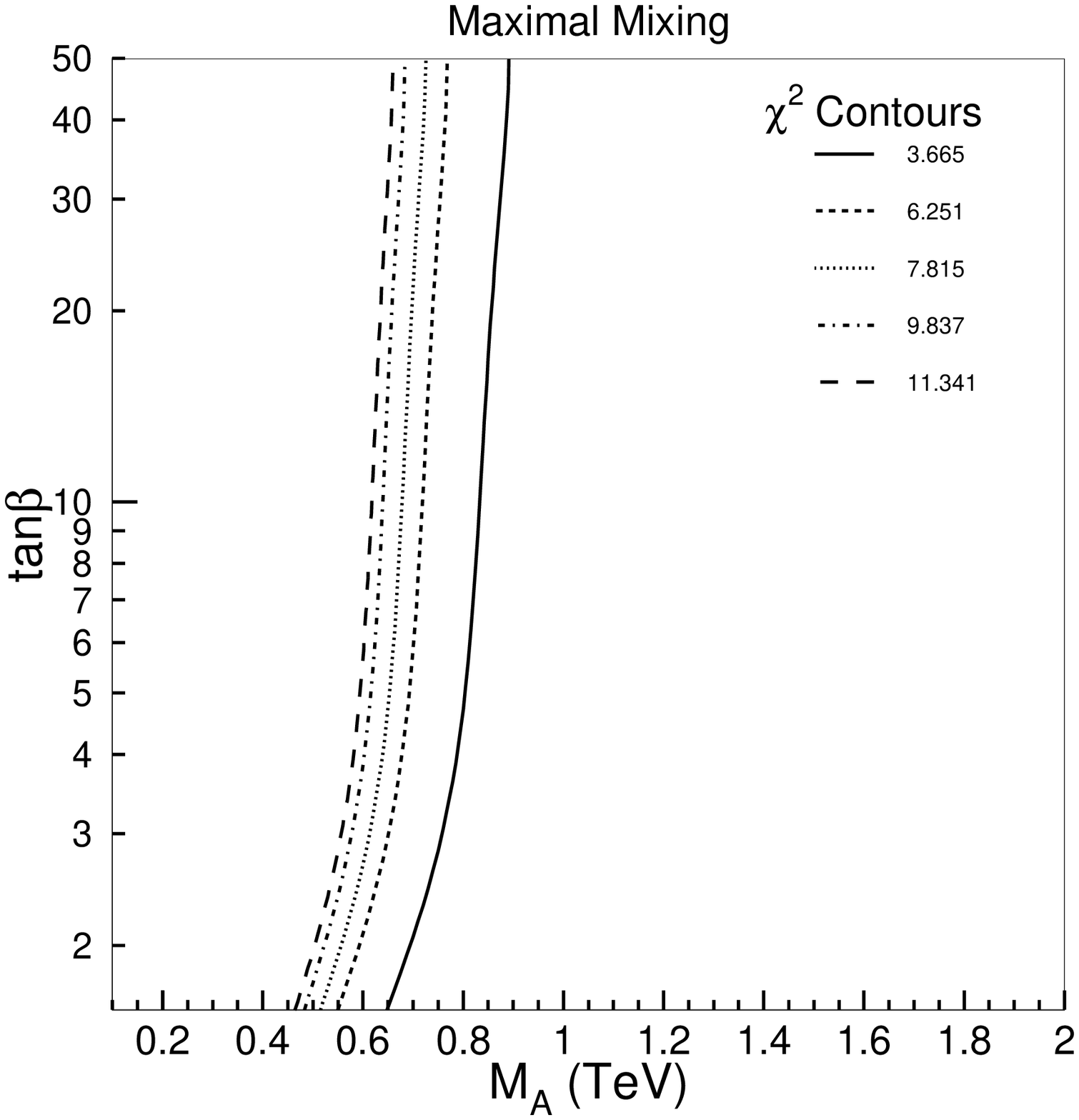}
}
\resizebox{\textwidth}{!}{
\includegraphics*[19,142][529,682]{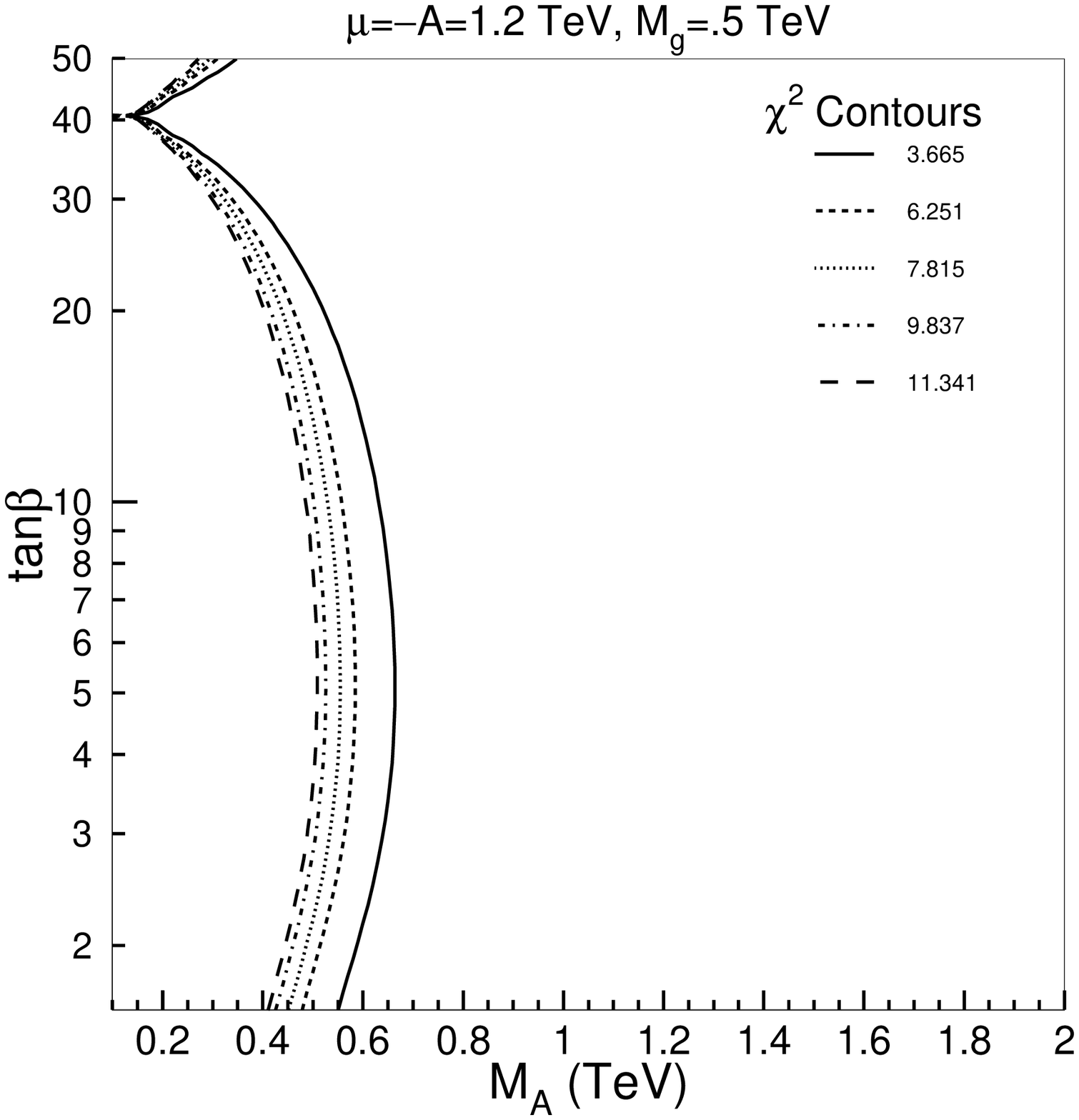}
\includegraphics*[19,142][529,682]{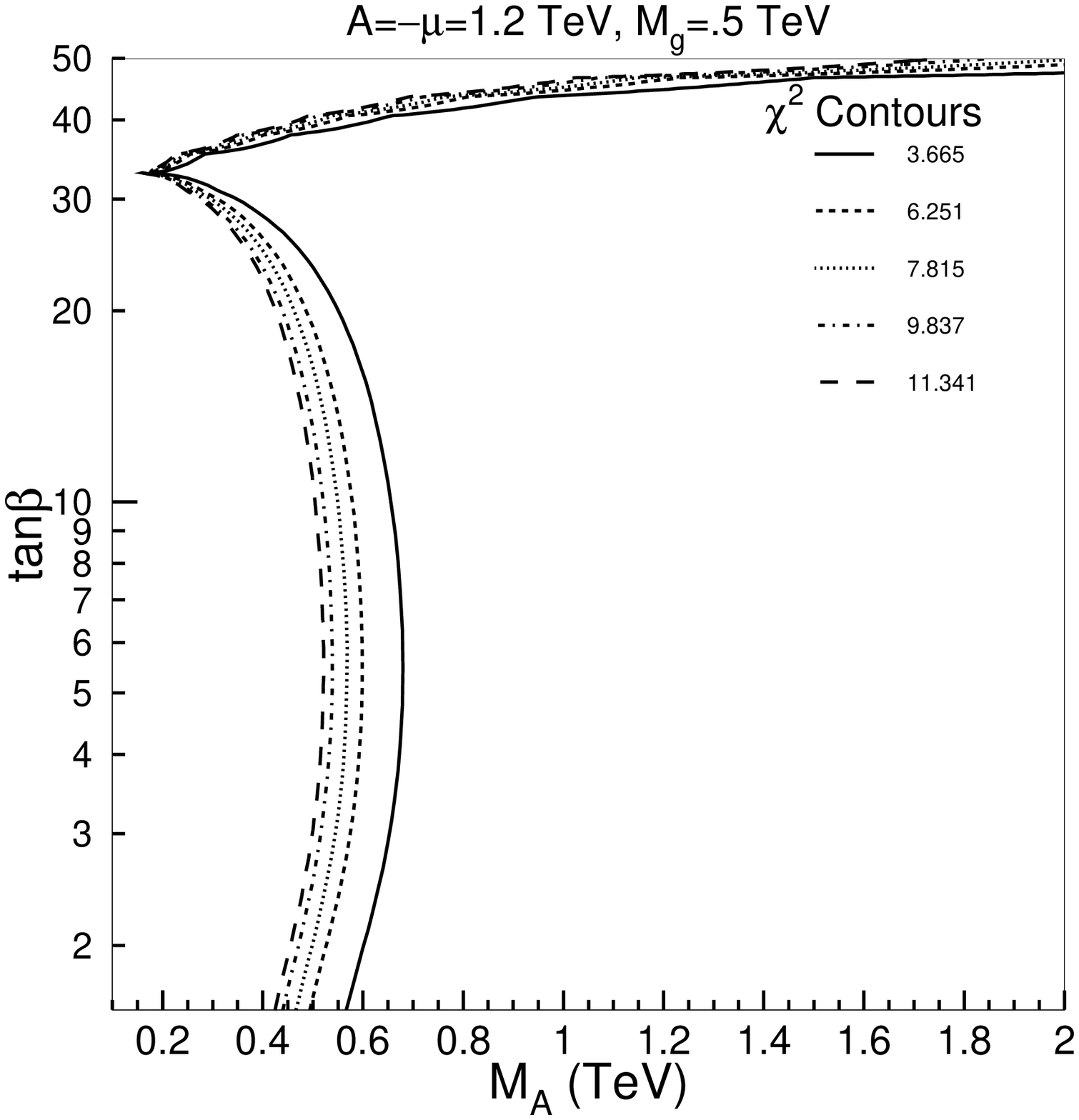}
}
\end{center}
\caption{Contours of $\chi^2$ for Higgs
boson decay observables in the benchmark scenarios.
The contours correspond to
68, 90, 95, 98 and 99\% confidence levels (right to left) for the three
observables $g^2_{hbb}$, $g^2_{h\tau\tau}$, and $g^2_{hgg}$.  }
\label{fig:chisquare}
\end{figure}

The $\chi^2$  results are shown in Fig.~\ref{fig:chisquare} for 
the benchmark scenarios
with contours corresponding to $68, 90, 95, 98$ and $99\%$ confidence levels.
For $\tan\beta \gsim 5$, the no-mixing and maximal-mixing scenarios 
can be distinguished from the SM at
95\% [99\%] confidence level for $m_A \lsim 600$ [500] GeV
and 650 [600] GeV, respectively.
The large $A_t$ and $\mu$ scenarios have regions at some large $\tan\beta$ 
values that are indistinguishable from the SM for any value of $m_A$.
For $\tan\beta \lsim 20$, however, the large $\mu$ and $A_t$ scenarios 
can be distinguished from the SM at
95\% [99\%] confidence level for $m_A \lsim 450$ [425] GeV.
Note also that for $\mu < 0$, the large $\Delta_b$ effects lead to
significant deviations from the SM even for $m_A \lsim 2$
TeV at the largest values of $\tan\beta$.

%-----------------------------------------------------------
\subsection{Complementarity to hadron collider measurements}
Measurements of Higgs boson properties
will be available from the Tevatron
and the LHC.  This data will most likely be available before the LC is
operational.  Based on our current understanding of the MSSM Higgs boson
properties and experimental capabilities, these hadron colliders will
observe a light, SM-like Higgs boson and, perhaps, other non-SM-like
Higgs bosons if their couplings to heavy flavor are enhanced over the
SM ({\it e.g.}, $H$ and $A$ will be observed in $b \bar b H / A$
if $\tan\beta$ is sufficiently large \cite{HiggsWGrep,atlas_tdr}).

If a light, SM-like Higgs boson is discovered, various combinations of
production cross sections times branching ratios will be measured at the
LHC to about $10$--$20\%$ \cite{Zeppenfeld} assuming an integrated luminosity
of 100 fb$^{-1}$ at each of the two detectors.
The uncertainties on most of these measurements
are dominated by statistical error.
From these measurements, various ratios of the partial widths
of a 120 GeV SM-like Higgs boson
to $ZZ^*$, $WW^*$, $\gamma\gamma$, $\tau^+\tau^-$, and $gg$
can be extracted with uncertainties between 15 and 30\%.  
Expected uncertainties from Ref.~\cite{Zeppenfeld}
are summarized in Table~\ref{tab:LHC},\footnote{
The decay modes to $\gamma\gamma$, $ZZ^*$ and $WW^*$ were considered for
inclusive Higgs production (dominated by gluon fusion) and the decay
modes $\gamma\gamma$, $\tau^+\tau^-$ and $WW^*$ were considered for
Higgs production through vector boson fusion.}
\begin{table}
\begin{center}
\begin{tabular}{lcccccc}
        &
        $\Gamma(Z)/\Gamma(W)$ &
        $\Gamma(\gamma)/\Gamma(W)$ &
        $\Gamma(\tau)/\Gamma(W)$ &
        $\Gamma(\tau)/\Gamma(\gamma)$ &
        $\Gamma(g)/\Gamma(W)$ &
        $\Gamma(b)/\Gamma(W)$ \\
\hline
        LHC \cite{Zeppenfeld} &
        29\% &
        16\% &
        15\% &
        15\% &
        15\% &
        -- \\
\hline
        LC \cite{BattagliaDesch} &
        -- &
        20\% &
        7.1\% &
        20\% &
        7.5\% & 5.6\% \\
        {\sc HFitter} \cite{tesla_tdr} &
        3.4\% &
        -- &
        6.6\% &
        -- &
        -- &
        2.4\% \\
\hline
\end{tabular}
\end{center}
\caption{Expected uncertainties for the ratios of partial widths of a
120 GeV SM-like
Higgs boson from the LHC \cite{Zeppenfeld} (100 fb$^{-1}$ at each of the two
detectors), the $e^+e^-$ LC \cite{BattagliaDesch} (ratios of BRs;
500 fb$^{-1}$ at $\sqrt{s} = 350$ GeV), and the $e^+e^-$ LC using
{\sc HFitter} \cite{BattagliaDesch,HFITTER,tesla_tdr} 
(500 fb$^{-1}$ at $\sqrt{s} = 500$ GeV).}
\label{tab:LHC}
\end{table}
along with the corresponding uncertainties from
LC measurements with
500 fb$^{-1}$ at $\sqrt{s} = 350$ GeV \cite{BattagliaDesch} 
and from {\sc HFitter} \cite{BattagliaDesch,HFITTER} 
at the LC with 500 fb$^{-1}$ at $\sqrt{s} = 500$ GeV.

Assuming that
the $hWW$ and $hZZ$ couplings are related by the usual
SU(2) relation, that $\Gamma(b)$ is related to
$\Gamma(\tau)$ by the SM relation, and that
only the decay modes $b \bar b$, $ZZ^*$, $WW^*$, $\gamma\gamma$, 
$\tau^+\tau^-$,
and $gg$ are needed to estimate the total Higgs width, then
$\Gamma(W)$ and $\Gamma_{\rm tot}$ can also be extracted from LHC measurements
with uncertainties of 10 and 20\%, respectively \cite{Zeppenfeld}.
(The second assumption is violated by the $\Delta_b$ Yukawa vertex corrections;
this affects the determination of both $\Gamma(W)$ and $\Gamma_{\rm tot}$.)
The branching ratio for the decay to $b \bar b$ is more difficult to measure
at the LHC due to QCD backgrounds.
$\Gamma(b)$ can be extracted to about 50\% from
vector boson fusion $Whjj$ events \cite{Davehbb}.

From Table~\ref{tab:LHC}, it is clear that LC measurements would give
a significant improvement over the hadron collider measurements.
The LC also offers the advantage of
model-independent measurements of the Higgs boson branching ratios,
even if invisible Higgs boson decays are present, and a model-independent
determination of the Higgs total width.
Generally speaking, except for the rare $\gamma\gamma$ decay, the LC has
measurement uncertainties that are smaller by at least a factor of 2
compared to the LHC.
The precision of
$b\bar b$ measurements are improved by an order of magnitude at the LC.

Some information may also be
available from the hadron colliders
regarding $m_A$ or $m_{H^\pm}$ and $\tan\beta$ \cite{Gianottiplot}.
For example, if the SUSY spectrum is such that $\Delta_b$ 
and $\Delta_{\tau}$ are negligible, then 
$\tan\beta$ can be measured at the LHC from $b \bar b H/A$, 
$H/A \to \tau^+\tau^-$
event rates with a $5$--$25$\% statistical uncertainty
and about a 10\% luminosity uncertainty \cite{atlas_tdr,Mitsou}.
However, as indicated earlier, the extraction of $\tan\beta$ from LHC 
Higgs boson measurements will require knowledge of $\Delta_b$ and 
$\Delta_{\tau}$.
Moreover, several interpretations of high-mass excesses of $\tau$'s may exist,
especially within the MSSM at large $\tan\beta$.
As we have discussed already, in most regions of parameter space 
the LC can provide an indirect measurement
of $m_A$ for values of $m_A$ significantly beyond the $e^+e^- \to AH$
kinematic limit.  
%This indirect measurement of $m_A$ would
%be strong support of the MSSM hypothesis.

From Tables~\ref{tab:BRmeas} and~\ref{tab:LHC}, we see that
the combination of LHC and LC measurements will significantly
improve our knowledge of
BR($\gamma$) in the absence of
a $\gamma\gamma$ collider.
In particular, 
the ratio of Higgs partial widths $\Gamma(\tau)/\Gamma(\gamma)$
can be determined to 15\% accuracy using ratios of production cross
sections times branching ratios in Higgs production through vector
boson fusion.  Combining this with the LC measurement of BR($\tau$) to
5--8\%,
we find that BR($\gamma$) can be extracted with a precision of 16--17\%.
If the BR($\gamma$) measurement at the LC of 14--19\%
is combined with this BR($\gamma$)
extraction from LHC data, a ``world
average'' precision of 11--13\% can be obtained.

%----------------------------------------------------------------------
\section{Extracting SUSY parameters:  $\Delta_b$}
\label{sec:Deltab}

In this section 
we examine the possibility of extracting MSSM parameters from
measurements of the properties of the SM-like Higgs boson.  We 
concentrate on the SUSY vertex corrections to the $h b
\bar b$ coupling, $\Delta_b$ (see Sec.~\ref{sec:theory}).  As noted in
Sec.~\ref{sec:BRanalysis}, when $\Delta_b$ is large it can have a
significant effect on the phenomenology of the heavy MSSM Higgs
bosons.  In particular, at large $\tan\beta$, knowledge of $\Delta_b$
will be needed to extract the value of $\tan\beta$ from measurements
of the $A$ and $H$ production rates at the LHC~\cite{atlas_tdr}.
Because $\Delta_b$ depends on a combination of MSSM parameters
(Eq.~\ref{eq:Deltab}), a measurement of this quantity provides a
constraint on MSSM parameter space.  However, additional information
is needed to fully disentangle these parameters; {\it e.g.}, the
gluino and bottom squark masses can be obtained from measurements at
hadron colliders.  To extract $\Delta_b$ directly from the data in the
large $\tan\beta$ regime, we shall consider the following ratio of
Higgs couplings [see Eq.~\ref{eq:ghat}] 
\begin{equation}
	\frac{\hat g_{hbb} - \hat g_{h\tau \tau}}
	{\hat g_{htt} - \hat g_{hbb}}\simeq
	\frac{\hat g_{hbb} - \hat g_{h\tau \tau}}
	{\hat g_{hcc} - \hat g_{hbb}}
        \simeq\Delta_b\,,
	\label{eq:ghatreprise}
\end{equation}
where we have employed 
the notation introduced at the end of Sec.~\ref{sec:theory} for 
Higgs-fermion couplings normalized to
their Standard Model values.

At the LC the $hb \bar b$ and $h\tau \tau$ couplings can be extracted from 
measurements of $\BR(b)$, $\BR(\tau)$ and the total width of $h$,
as discussed in Sec.~\ref{expunc}.
In the up-type quark sector, the measurements of $h$ couplings to 
charm- and top-quark pairs are possible at the LC.
In $h$ decays, we have access to the $hc\bar c$ coupling through 
$\BR(c)$.  This measurement, like those of the $hb \bar b$ and $h \tau \tau$ 
couplings, can be made in the initial stage of the LC
running at $\sqrt{s} = 350$ or 500 GeV.
However, as discussed in Sec.~\ref{thunc}, the theoretical 
uncertainty in $g_{h_{SM}cc}$ is very large due to the 
large uncertainty in the 
charm quark mass.  If the uncertainty in the charm quark mass is not 
significantly reduced by the time that LC data is available, then the 
resulting theoretical uncertainty in $\hat g_{hcc}$ 
would dominate the uncertainty
in the value of $\Delta_b$ extracted from Eq.~\ref{eq:ghatreprise}.
In contrast, the theoretical uncertainty in the $ht\bar t$ coupling is very
small due to the precision in the top quark mass measurement.
As discussed in Sec.~\ref{expunc}, the coupling
$g_{htt}$ can be measured directly from the 
$e^+e^- \to t \bar t h$ cross section; however,
this requires LC running at a higher energy, {\it e.g.}, $\sqrt{s}=800$ GeV.

The LC measurements of $h$ decay properties are sensitive to the squares
of the Higgs couplings $g_{hff}^2$
through the $h$ partial widths; therefore the signs of the normalized couplings
$\hat g_{hff}$ in Eq.~\ref{eq:ghatreprise} are not measured directly in the 
corresponding BRs. This leads to a potential four-fold ambiguity 
in the extraction of~$|\Delta_b|$.
%$(\Delta_b - \Delta_{\tau})/(1 + \Delta_{\tau})$.
However, LC measurements of $h \to gg$ and $h \to \gamma \gamma$ are 
sensitive to the relative signs of the Higgs 
couplings to the particles running 
in the loops; for example, flipping the sign of $g_{htt}$ while leaving
that of $g_{hWW}$ fixed leads to a large deviation in the $h\gamma \gamma$
coupling from its SM value, 
while the $hgg$ coupling is sensitive to the relative sign of 
$g_{htt}$ and $g_{hbb}$ (assuming that SUSY loops are not important).
Near the decoupling limit, the radiative corrections
to the CP-even Higgs mixing angle $\alpha$ are not so large 
as to change the sign of $\tan\alpha$
from its tree-level negative value, and $|\Delta_{\tau}| < |\Delta_b| < 1$.
In this case the signs of $g_{hff}$ are the same as in the SM, so all
the $\hat g_{hff}$ in Eq.~\ref{eq:ghatreprise} are positive.

In the discussion above, we have assumed that $h$ is SM-like, which is
appropriate once $m_A$ is sufficiently above $m_h^{\rm max}$ (in practice,
$m_A\gsim 150$~GeV is usually sufficient).  For values of $m_A\lsim
m_h^{\rm max}$ and $\tan\beta\gg 1$, it is $H$ which has SM-like
couplings to vector boson pairs ({\it i.e.}, the roles of $h$ and $H$  
are interchanged).  However, a separate analysis is not
required since we also have [see Eq.~\ref{eq:ghat}] 
\begin{equation}
	\frac{\hat g_{Hbb} - \hat g_{H\tau \tau}}
	{\hat g_{Htt} - \hat g_{Hbb}}\simeq
	\frac{\hat g_{Hbb} - \hat g_{H\tau \tau}}
	{\hat g_{Hcc} - \hat g_{Hbb}}
        \simeq\Delta_b\,.
	\label{eq:gHatreprise}
\end{equation}
The extraction of the normalized couplings $\hat g_{Hff}$ follows the
same procedure as outlined above.

\begin{figure}
        \begin{center}
\resizebox{\textwidth}{!}{
\includegraphics*[19,142][529,692]{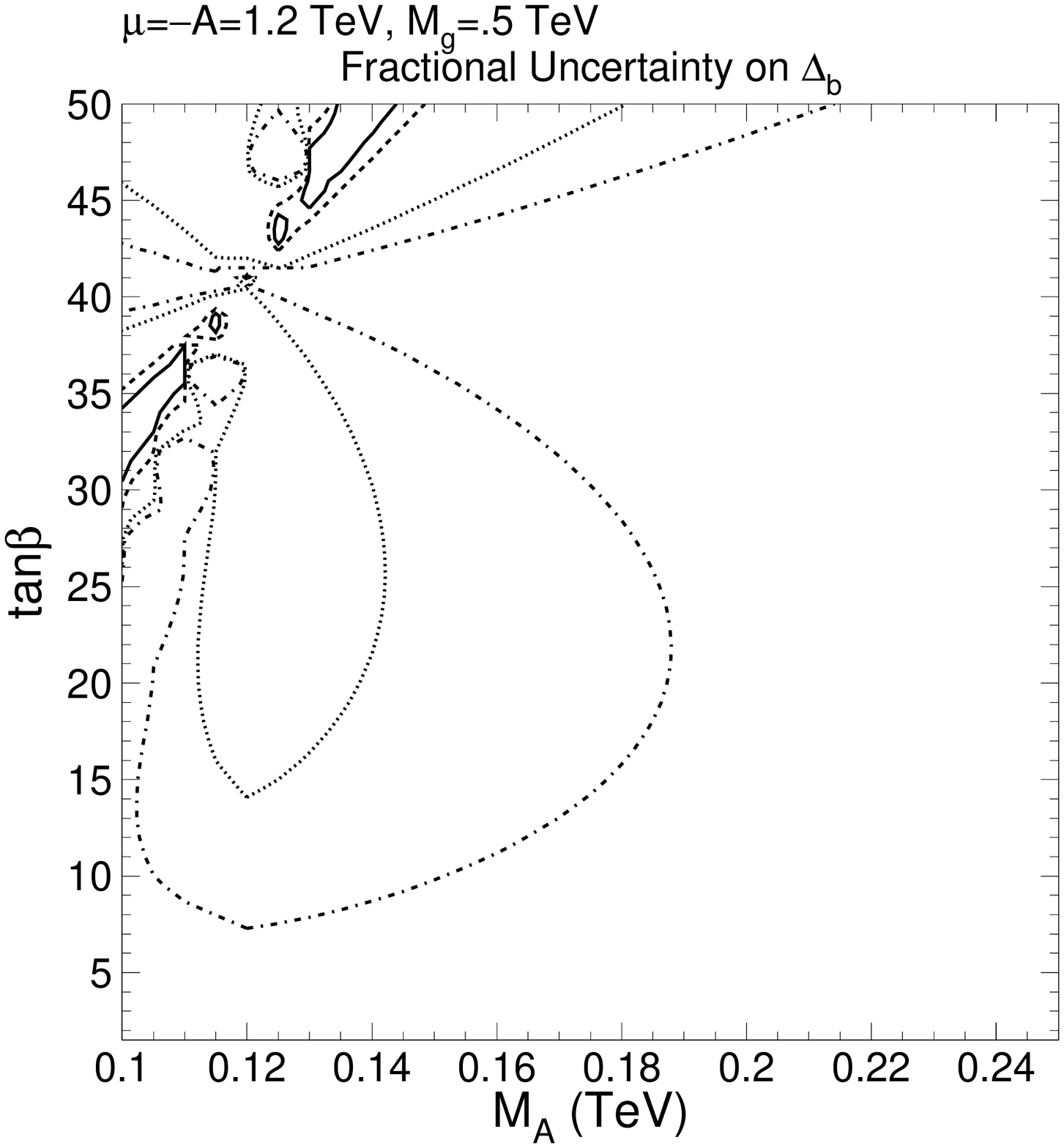}
\includegraphics*[19,142][529,692]{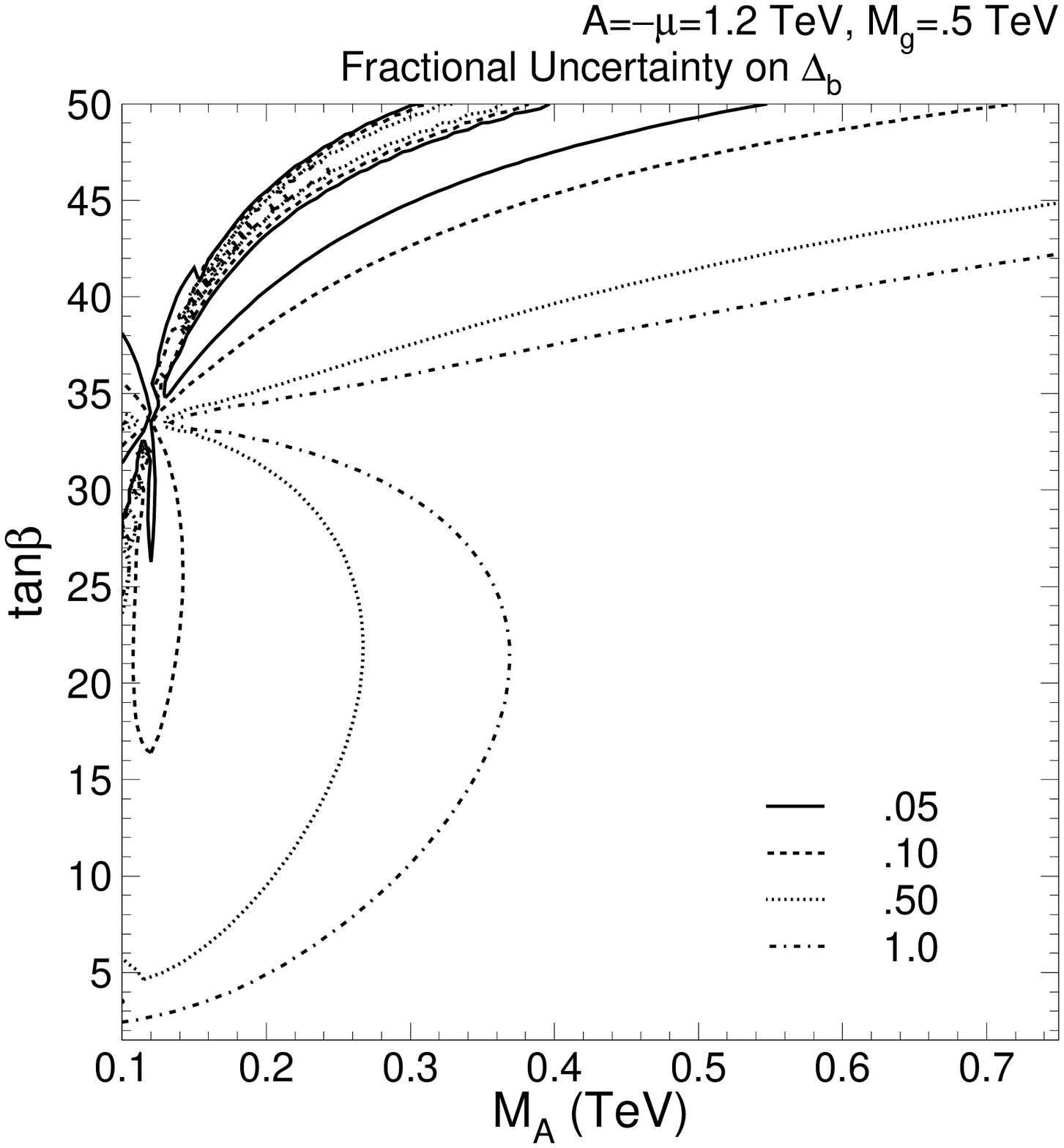}
}
\resizebox{\textwidth}{!}{
\includegraphics*[19,142][529,692]{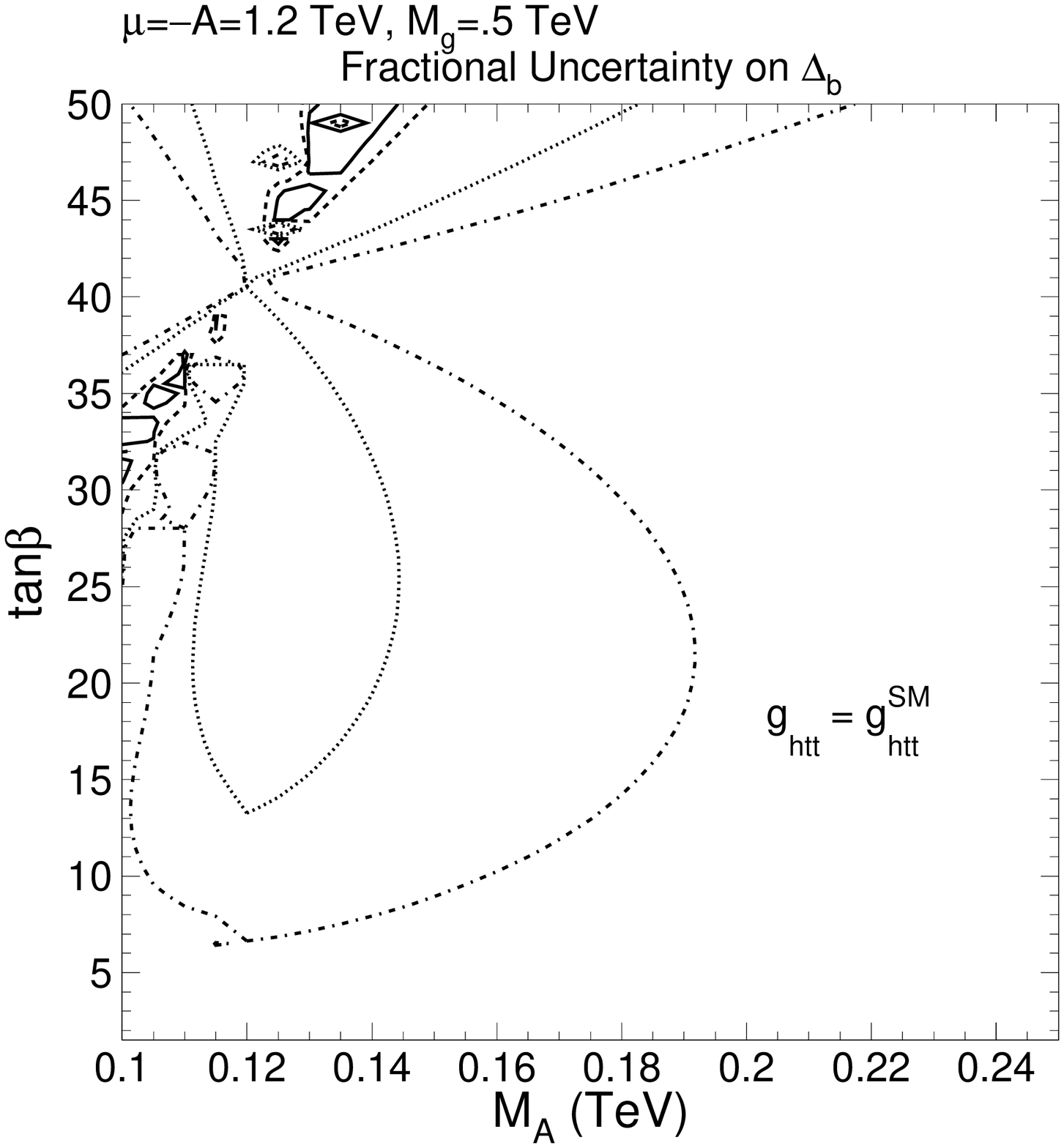}
\includegraphics*[19,142][529,692]{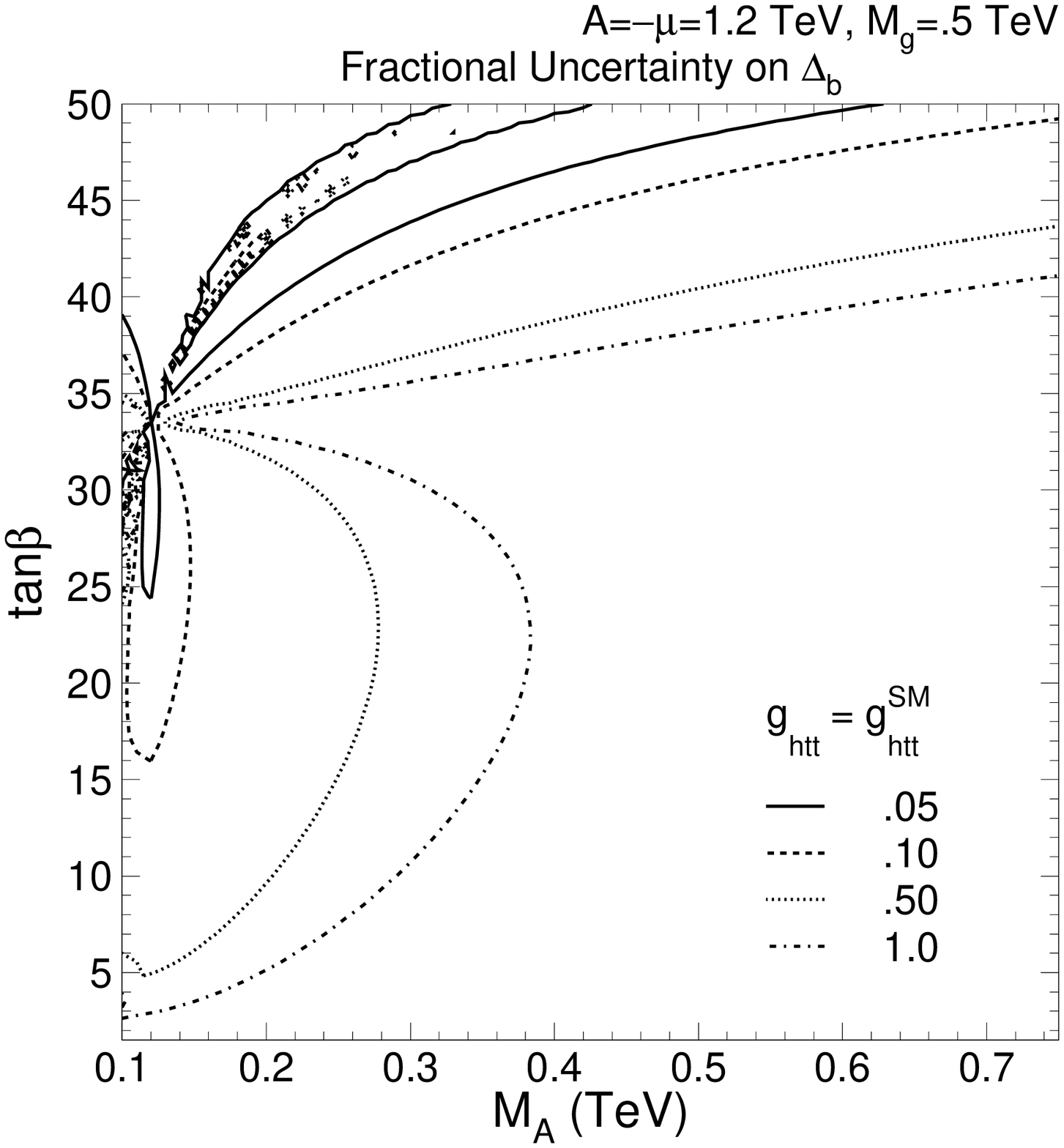}
}
        \end{center}
        \caption{Contours of the fractional uncertainty in the
determination of $\Delta_b$ 
%from $g^2_{hbb}$, $g^2_{h\tau\tau}$ and $g^2_{htt}$
in the large $\mu$ and $A_t$ scenario with
$\mu = -A_t = 1.2$ TeV ($\Delta_b > 0$) (left)
and $\mu = -A_t = -1.2$ TeV ($\Delta_b < 0$) (right).
The upper two plots use the Higgs couplings to $b\bar b$,
$\tau^+\tau^-$ and $t\bar t$ 
as inputs while the lower two plots 
use only the first two couplings as input while taking the
Higgs coupling to $t\bar t$ equal to its SM value (see 
Eqs.~\ref{hatone} and \ref{Hatone}).
In the $\mu = -A_t = 1.2$ TeV scenario (left), we plot 
$0.1\, {\rm TeV} < m_A < 0.25\, {\rm TeV}$,
while in the $\mu = -A_t = -1.2$ TeV
scenario (right), we plot $0.1\, {\rm TeV} < m_A < 0.75\, {\rm TeV}$.
}
        \label{fig:dmb}
\end{figure}

The upper two panels of Fig.~\ref{fig:dmb} show the fractional error 
in the determination of $\Delta_b$
%of $(\Delta_b - \Delta_\tau)/(1 + \Delta_\tau)$ 
from measurements of the Higgs couplings to $b\bar b$,
$\tau^+ \tau^-$ and $t\bar t$,
for the benchmark large
$\mu$ and $A_t$ scenario (see Table~\ref{tab:scenarios}),
in which $\Delta_b$ is quite sizable.
%\footnote{In this case
%$\Delta_b$ is much larger than $\Delta_\tau$, because $\Delta_\tau$ is 
%proportional to the electroweak gauge couplings [Eq.~\ref{eq:deltatau}],
%which justifies the approximation of \Eq{eq:ghat}.}
% $\Delta_\tau$ can be neglected in Eq.~\ref{eq:ghat}, so that
%$(\hat g_{hbb}-\hat g_{h\tau \tau})/(\hat g_{htt} - \hat g_{hbb}) 
%\simeq \Delta_b$
%$(1 - \sqrt{x})/(\sqrt{x} - \sqrt{y}) = \Delta_b$ 
%gives a direct measurement of $\Delta_b$.
The error in $\Delta_b$ is calculated
using the fractional uncertainties in the Higgs couplings given in 
%$\BR(b)$ and $\BR(\tau)$ 
%from the first line of Table~\ref{tab:BRmeas}~\cite{BattagliaDesch}
%and the fractional uncertainty in 
%$g^2_{hbb}$, $g^2_{h\tau \tau}$ and $g^2_{htt}$ from 
Table~\ref{tab:g^2}.  
In the two upper panels of Fig.~\ref{fig:dmb}, in the region 
outside of the dot-dashed contours 
%({\it e.g.}, toward the region of large $m_A$)
%converging near $\tan\beta = 40$ 
[and also within the smaller closed dot-dashed circles], the error on 
$\Delta_b$ is larger than $\Delta_b$ itself, and no meaningful 
value for $\Delta_b$ can be obtained.  Three other contour lines are
shown corresponding to a fractional uncertainty in $\Delta_b$ of
50\% [dotted], 10\% [dashed] and 5\% [solid].
%In the region between the 
%dot-dashed and the short-dashed contours, $\Delta_b$ can be measured with
%a relative error between 50 and 100\%.  Similarly, between the long-dashed
%and short-dashed [solid and long-dashed] contours, the relative error
%on $\Delta_b$ is between 10 and 50\% [5 and 10\%].  
Within the region bracketed only by solid contours, the 
relative error on $\Delta_b$ is less than 5\%.
%In the upper right
%panel of Fig.~\ref{fig:dmb}, the contours are labeled in the same way;
%here there is a region bracketed only by solid contours within which the 
%relative error on $\Delta_b$ is less than 5\%.
There is an approximate (distorted reflection) symmetry in these contours
about $m_A=m_h^{\rm max}= 119$~GeV.  This corresponds to the
interchange of roles between $h$ and $H$ at large $\tan\beta$ as noted
above.  Note also the convergence of the contours at $\tan\beta\sim
41$ (left panels) and  $\tan\beta\sim 33$ (right panels) corresponding
to the $\tan\beta$ values at which the $m_A$-independent decoupling occurs.

For positive $\mu$ (the upper left panel of Fig.~\ref{fig:dmb}),
$\Delta_b$ can only be distinguished from zero at the $2\sigma$ level
({\it i.e.}, a 50\% measurement)
for moderate to large $\tan\beta$ and $m_A \lsim 160$ GeV.
For such low values of $m_A$, the heavy Higgs bosons will be directly
observed at the LC.
In contrast, for negative $\mu$
(the upper right panel of Fig.~\ref{fig:dmb}; note the change in the
horizontal scale),
$\Delta_b$ can be determined with 10\% accuracy even for $m_A$ as
large as 500 GeV for large $\tan\beta$.
This significant difference in sensitivity to $\Delta_b$ between the  
positive and negative $\mu$ cases
is a combination of two effects.
First, for negative $\mu$, the bottom-squark--gluino
and top-squark--Higgsino contributions to $\Delta_b$ enter with
the same (negative) sign, while for positive $\mu$ they enter with
opposite signs (see Eq.~\ref{eq:Deltab}).  Thus for fixed values of
$|A_t|$ and $|\mu|$ (with $A_t\mu<0$), $|\Delta_b|$ is larger
for negative $\mu$ than for positive $\mu$.
Second, $\Delta_b$ enters the $b$ quark Yukawa coupling non-linearly
as $1/(1 + \Delta_b)$,
so that $\Delta_b<0$ has a more pronounced effect.

The upper right panels of Figs.~\ref{fig:widths1} and \ref{fig:widthsmuAt}
demonstrate that $\Gamma(c) \propto g^2_{hcc}$ is very close to its SM value
once $m_A\gsim 150$~GeV and $\tan\beta\gsim 5$.
In this region of parameter space, $|\cos(\beta-\alpha)|\ll 1$ and
\begin{equation}
	\hat g_{hcc} = \hat g_{htt}
	\simeq 1 + \cot\beta \cos(\beta-\alpha)
	+ \mathcal{O}\left(\cos^2(\beta-\alpha) \right)
	\simeq 1\,. \label{hatone}
\end{equation}
Thus a value of $\Delta_b$
%$(\Delta_b - \Delta_\tau)/(1 + \Delta_\tau)$ 
can be  extracted by taking $\hat g_{hcc}=
\hat g_{htt} = 1$ in Eq.~\ref{eq:ghatreprise}.
To the extent that this is a good approximation, there is no sensitivity
in the value of $\Delta_b$ 
%$(\Delta_b - \Delta_\tau)/(1 + \Delta_\tau)$
to the theoretical uncertainty in the charm or top quark Yukawa
couplings.  Likewise, for $m_A\lsim m_h^{\rm max}$ and $\tan\beta\gg
1$, $\sin(\beta-\alpha)$ is close to zero so that
\begin{equation}
	\hat g_{Hcc} = \hat g_{Htt}
	\simeq 1 - \cot\beta \sin(\beta-\alpha)
	+ \mathcal{O}\left(\sin^2(\beta-\alpha) \right)
	\simeq 1\,, \label{Hatone}
\end{equation}
and $\Delta_b$ can be extracted by taking  $\hat g_{Hcc}=
\hat g_{Htt} = 1$ in Eq.~\ref{eq:gHatreprise}.
Results are shown in the lower two panels of Fig.~\ref{fig:dmb}.
%We are then able to obtain $\Delta_b$ in the large $\mu$ and $A_t$
%scenarios with higher precision in the large region of MSSM parameter
%space where this assumption is valid, as shown in the lower two
%panels of Fig.~\ref{fig:dmb}.
With the assumption that the Higgs coupling to $t\bar t$ is given by
its SM value, 
the precision in $\Delta_b$ is improved somewhat, especially 
for large values of $\tan\beta$ and $m_A$ and $\mu < 0$ [the lower
right panel of Fig.~\ref{fig:dmb}].  
%For $\tan\beta \gsim 10$, the approximation $\hat g_{htt} = 1$ is valid
%for $m_A$ larger than about 150 GeV [see the upper right panels of 
%Figs.~\ref{fig:widths1} and \ref{fig:widthsmuAt}];
Note that in the region of the $m_A$---$\tan\beta$ plane in which
$\cos(\beta-\alpha)$ takes on some intermediate value between zero and
one, neither Eq.~\ref{hatone} nor Eq.~\ref{Hatone} is a good
approximation, and the
results in the lower two panels of Fig.~\ref{fig:dmb} are not
reliable.
% in this region of parameter space.

%------------------------------------------------------------------
\section{Conclusions}
\label{sec:conclusions}

There are significant regions of MSSM parameter space where the
Tevatron or LHC can discover a light Higgs boson which is
indistinguishable from the SM Higgs boson using only hadron collider
measurements.  Additional Higgs bosons may
be observed, but further measurements will be necessary to
constrain the underlying Higgs sector.  Also, there is a region
of MSSM parameter space
for moderate $\tan\beta$ and moderate to large $m_A$ where only a single
light Higgs boson with SM-like couplings would be observed in experiments
at the Tevatron, the LHC, and a $\sqrt{s}=
350$ or 500 GeV $e^+e^-$ linear collider.
In this case, precision measurements of the properties of the light
Higgs boson at the LC can be used to
distinguish between
the SM and MSSM, and further to extract or constrain the model parameters.
At the LC, the production cross sections and the most important branching
ratios of a light SM-like Higgs boson can be measured with precisions
between 2 and 15\%.
We have examined how these measurements can be used to distinguish an MSSM
Higgs boson from the SM Higgs boson.
We have also demonstrated the extraction of the important SUSY quantity
$\Delta_b$ that parameterizes the SUSY radiative corrections to the relation
between the $b$ quark mass and its Yukawa coupling.

In general, the couplings of the lightest CP-even MSSM Higgs boson $h$
to pairs of SM particles approach their SM values in the decoupling limit
of large $m_A$.
Thus, the deviation of $h$ partial widths
from their SM values goes to zero in the limit of large $m_A$, except
for the loop-induced couplings of $h$ to
photon or gluon pairs, which
approach their SM values in the limit that
both $m_A$ and the SUSY masses are taken large simultaneously.
As one moves away from the decoupling limit, the radiative corrections
to the CP-even Higgs mixing angle $\alpha$, the Yukawa vertex
corrections $\Delta_b$, and
contributions of SUSY particles to the loop induced
$h gg$ and $h \gamma\gamma$ couplings, all contribute to deviations of
the $h$ couplings from the corresponding SM values.  The sensitivity
of these couplings 
%the reach of the BR and coupling measurements
in the $m_A$---$\tan\beta$ plane for distinguishing the 
SM-like MSSM Higgs boson from
the SM Higgs boson depends strongly on the MSSM parameters.
In order to illustrate this strong dependence, we have chosen three
benchmark MSSM scenarios
that lead to very different behavior of the Higgs couplings.

The decoupling behavior of the $h$ partial widths at large $m_A$
and moderate to large $\tan\beta$ is summarized qualitatively as follows:
$\delta \Gamma(W) \sim m_Z^4 \cot^2\beta / m_A^4$,
$\delta \Gamma(c) \sim m_Z^2 \cot^2\beta / m_A^2$,
and $\delta \Gamma(b) \sim \delta \Gamma(\tau) \sim m_Z^2/m_A^2$.
Moreover, there is a contribution to $\delta\Gamma(b)$ proportional to
$\Delta_b$ [see Eq.~\ref{eq:g2bbdecoupling}],
which is enhanced at large $\tan\beta$ and can further slow
the approach to decoupling.
Thus we expect that $\Gamma(W)$ will decouple very quickly with increasing
$m_A$ (so that $\delta\BR(W) \simeq \delta \Gamma_{\rm tot}$ over most
of the parameter space),
$\Gamma(c)$ will decouple more slowly at low $\tan\beta$ but
quickly at large $\tan\beta$, and $\Gamma(b)$ and $\Gamma(\tau)$ will
decouple quite slowly, at all values of $\tan\beta$.  This expected
behavior is borne out in the no-mixing and maximal-mixing scenarios.
In the large $\mu$ and $A_t$ scenarios the decoupling behavior of these
partial widths is more complicated, because the radiative corrections
to the CP-even 
Higgs mixing angle $\alpha$ are quite significant in some regions
of the $m_A$---$\tan\beta$ plane.  For some (large)
values of $\tan\beta$, $\cos(\beta - \alpha)$ is driven to zero
by the radiative corrections to $\alpha$, leading to
``$m_A$-independent decoupling''.
In particular, $\Gamma(b)$ and
$\Gamma(\tau)$ attain their SM values at much lower values of $m_A$
than would be expected from the tree-level decoupling behavior.

The $m_A$-independent decoupling is not an exclusive feature of 
the specific large values of $\mu$ and $A_t$ chosen in our benchmark 
scenarios.
%rather it is a consequence of the relative values of $X_t$ and $M_S$.
Additional sets of MSSM parameters that exhibit $m_A$-independent
decoupling can be found by scaling $\mu$, $A_t$, $M_{\tilde g}$ and $M_S$ 
by a common factor; in this case the value of $\tan\beta$ at which
the $m_A$-independent decoupling occurs varies primarily due to the
$\log(M_S^2/m_t^2)$ dependence of $\delta \mathcal{M}^2_{ij}$
[see Eq.~\ref{eq:M12approx}].
In particular, we have found sets of MSSM parameters with relatively
small $M_S$ that exhibit $m_A$-independent decoupling 
even for $\tan\beta$ less than 10, within the parameter region in 
which the heavy MSSM Higgs bosons would be missed at the LHC.  
In this case, due to the relatively low value of $M_S$, 
deviations from the SM Higgs couplings may be observable in $h \to gg$
or $h \to \gamma \gamma$
due to top-squark and bottom-squark loop contributions.

In the case of the loop-induced Higgs couplings to gluon or photon pairs,
deviations from the SM widths can occur due to SUSY loop contributions
even in the limit of very large $m_A$.
These SUSY loops decouple at large SUSY particle masses, but the decoupling
can occur more slowly if $X_t$ is large (as in the maximal-mixing
scenario), resulting in several percent deviations of $\Gamma(g)$ from
its SM value even for TeV mass squarks.

Because $\Gamma(b)$ dominates the total width of a 120 GeV SM-like
Higgs boson over most of the MSSM parameter space, a change in 
$\Gamma(b)$ causes a corresponding change in $\Gamma_{\rm tot}$,
so that the resulting change in $\BR(b)$ is relatively small.
In contrast, branching ratios
for rarer Higgs decays such as $\BR(W)$ and $\BR(g)$ are
sensitive both to changes in $\Gamma(W)$ and $\Gamma(g)$,
respectively, and to changes in $\Gamma_{\rm tot}$.  Thus, although
$\BR(b)$ will be the best-measured BR and that $\Gamma(b)$
typically deviates significantly from its SM value, $\BR(W)$ and
$\BR(g)$ will in general be more sensitive to deviations from the SM
than $\BR(b)$, making their high-precision measurements a priority.

To demonstrate how LC measurements of Higgs properties can be used to 
distinguish a SM-like MSSM Higgs boson from the SM Higgs boson,
we studied the expected uncertainties in the branching
ratios of a 120 GeV SM Higgs boson to $b \bar b$, $WW^*$ and $gg$, and
performed a $\chi^2$ analysis using the anticipated
uncertainties in the squared couplings
of the SM Higgs boson to $b \bar b$, $\tau^+\tau^-$ and $gg$.
In particular,
we draw the following conclusions
for our three benchmark scenarios from the $\chi^2$ analysis
(a somewhat lower reach in $m_A$ is obtained using
the analysis of individual BR measurements):
\begin{itemize}
\item  In the no-mixing and maximal-mixing scenarios,
the MSSM Higgs boson can be distinguished
from the SM Higgs boson at the $95\%$ confidence level
for $m_A \lsim 600$ and $650$ GeV, respectively, for $\tan\beta\gsim 5$.
These limits become slightly lower for 
values of $\tan\beta \lsim 5$.

\item In the large $\mu$ and $A_t$ scenario, there are regions of parameter
space at large $\tan\beta$ in which
the MSSM Higgs boson cannot be distinguished
from the SM Higgs boson using BR and cross section measurements
for any value of $m_A$.
For $\tan\beta \lsim 20$, the MSSM Higgs boson can be distinguished
from the SM Higgs boson at the 95\% confidence level for $m_A \lsim
450$ GeV.
\end{itemize}
The precise reach in $m_A$ in the maximal-mixing scenario should
be viewed liberally, since the SM-like Higgs boson mass 
can reach values near 130 GeV, and some of the coupling
and BR measurements used in this study have poorer precision for
$m_h > 120$ GeV.

Finally, we have shown that when $\Delta_b$ is sufficiently large,
information on $\Delta_b$ can be obtained from the 
measurements of the couplings $g_{hbb}$, $g_{h\tau\tau}$ and
$g_{htt}$ [or $g_{hcc}$]
%$\BR(b)/\BR(\tau)$ and $g^2_{htt}/\BR(\tau)$ [or $\BR(c)/\BR(\tau)$], 
unless the MSSM parameters
are such that $m_A$-independent decoupling occurs.
The precision achievable in this measurement
depends on the size and sign of $\Delta_b$; these depend in part
on the signs of $\mu M_{\tilde g}$ and $\mu A_t$.
Combining precision measurements in the Higgs sector, such as
those leading to the determination of $\Delta_b$, with direct
measurements of the SUSY spectrum will be of great value in 
deciphering the underlying supersymmetric structure of the theory.

\vskip1cm
%---------------------------
\noindent
{\Large \bf Acknowledgments}

\vskip0.5cm \noindent
We thank the organizers of the 5th International
Linear Collider Workshop (LCWS 2000)
at Fermilab, where portions of this work were presented.
We are also grateful to S.~Dawson, D.~Rainwater, M.~Schmitt, M.~Spira 
and especially C.~Wagner for useful comments and discussions.
H.E.H. and H.E.L. thank the Aspen Center for Physics for its hospitality
while this work was being completed.
Fermilab is operated by Universities Research Association Inc.\
under contract no.~DE-AC02-76CH03000 with the U.S. Department of
Energy.  H.E.H. is supported in part by the U.S. Department of Energy.
S.M. is supported by the Department of Energy and by the Davis Institute
for High Energy Physics.

\clearpage

%-------------------------------------------------------

\newpage

\end{document}